\documentclass[10pt]{spieman}  
\usepackage{amsmath,amsfonts,amssymb}
\usepackage{graphicx}
\usepackage{setspace}
\usepackage{tocloft}
\usepackage[table,x11names]{xcolor}
\usepackage{pdflscape}
\usepackage[flushleft]{threeparttable}
\usepackage{wrapfig}

\usepackage{natbib}
 \setcitestyle{super,none,aysep={},yysep={,}} 
\def\deg{\hbox{$^\circ$}}
\def\arcmin{\hbox{$^\prime$}}
\def\arcsec{\hbox{$^{\prime\prime}$}}

\newcommand\aj{AJ}%                                % Astronomical Journal
\newcommand\araa{ARA\&A}%                          % Annual Review of Astron and Astrophys
\newcommand\apj{ApJ}%                              % Astrophysical Journal
\newcommand\apjl{ApJ}%                             % Astrophysical Journal, Letters
\newcommand\apjs{ApJS}%                            % Astrophysical Journal, Supplement
\newcommand\ao{Appl.~Opt.}%                        % Applied Optics
\newcommand\aap{A\&A}%                             % Astronomy and Astrophysics
\newcommand\aaps{A\&AS}%                           % Astronomy and Astrophysics, Supplement
\newcommand\jcap{J. Cosmology Astropart. Phys.}%   % Journal of Cosmology and Astroparticle Physics
\newcommand\mnras{MNRAS}%                          % Monthly Notices of the RAS
\newcommand\prb{Phys.~Rev.~B}%                     % Physical Review B: Solid State
\newcommand\pasp{PASP}%                            % Publications of the ASP
\newcommand\pasj{PASJ}%                            % Publications of the ASJ
\newcommand\ssr{Space~Sci.~Rev.}%                  % Space Science Reviews
\newcommand\nat{Nature}%                           % Nature
\newcommand\physrep{Phys.~Rep.}%                   % Physics Reports
\newcommand\procspie{Proc.~SPIE}%                  % Proceedings of the SPIE

\title{Review: Far-Infrared Instrumentation and Technology Development for the Next Decade}

 \author[a,b,*]{Duncan Farrah}          %% D
 \author[c]{Kimberly Ennico Smith}    %% NASA Ames 
 \author[d]{David Ardila}             %% JPL
 \author[d,e]{Charles M. Bradford}    %% caltech and jpl
 \author[f]{Michael Dipirro}          %% GSFC
 \author[g]{Carl Ferkinhoff}          %% winona state
 \author[h]{Jason Glenn}              %% colorado
 \author[d]{Paul Goldsmith}           %% jpl
 \author[f]{David Leisawitz}          %% NASA Goddard
 \author[i]{Thomas Nikola}            %% Cornell
 \author[c,j]{Naseem Rangwala}        %% NASA Ames AND Bay area enviro
 \author[f]{Stephen A. Rinehart}      %% NASA Goddard 
 \author[k,f]{Johannes Staguhn}       %% JHU and GSFC 
 \author[l,d]{Michael Zemcov}         %% RIT AND JPL
 \author[e]{Jonas Zmuidzinas}         %% caltech
 \author[d]{James Bartlett}           %% JPL 
 \author[m]{Sean Carey}               %% ipac
 \author[n]{William J. Fischer}       %% stsci
 \author[o]{Julia Kamenetzky}         %% westminster
 \author[l]{Jeyhan Kartaltepe}        %% rit
 \author[p]{Mark Lacy}                %% NRAO
 \author[q,e]{Dariusz C. Lis}         %% caltech
 \author[p,r]{Lisa Locke}             %% 
 \author[c]{Enrique Lopez-Rodriguez}  %% caltech
 \author[s]{Meredith MacGregor}       %% carnegie dtm
 \author[t]{Elisabeth Mills}          %% Boston
 \author[f]{S. Harvey Moseley}        %% NASA Goddard
 \author[o]{Eric J. Murphy}           %% NRAO
 \author[c]{Alan Rhodes}              %% Ames
 \author[u]{Matt Richter}             %% davis
 \author[v,w]{Dimitra Rigopoulou}     %% Oxford and RAL
 \author[b]{David Sanders}            %% hawaii
 \author[n,c]{Ravi Sankrit}           %% stsci, ames 
 \author[x]{Giorgio Savini}           %% UCL
 \author[y]{John-David Smith}         %% toledo
 \author[m]{Sabrina Stierwalt}        %% ipac

\affil[a]{Department of Physics and Astronomy, University of Hawaii, 2505 Correa Road, Honolulu, HI 96822, USA}
\affil[b]{Institute for Astronomy, 2680 Woodlawn Drive, University of Hawaii, Honolulu, HI 96822}
\affil[c]{NASA Ames Research Center, Moffet Field, CA 94035, USA}
\affil[d]{Jet Propulsion Laboratory, Pasadena, CA 91109, USA}
\affil[e]{Division of Physics, Mathematics, and Astronomy, California Institute of Technology, Pasadena, CA 91125, USA}
\affil[f]{NASA Goddard Space Flight Center, Greenbelt, MD 20771, USA}
\affil[g]{Department of Physics, Winona State University, Winona, MN 55987, USA}
\affil[h]{Department of Astrophysical and Planetary Sciences, University of Colorado, Boulder, CO 80309, USA}
\affil[i]{Cornell Center for Astrophysics and Planetary Sciences, Cornell University, Ithaca, NY 14853, USA}
\affil[j]{Bay Area Environmental Research Institute, NASA Research Park, Moffet Field, CA 94035, USA}
\affil[k]{The Henry A. Rowland Department of Physics and Astronomy, Johns Hopkins University, Baltimore, MD 21218, USA}
\affil[l]{School of Physics and Astronomy, Rochester Institute of Technology, Rochester, NY 14623, USA}
\affil[m]{Infrared Processing Analysis Center, California Institute of Technology, Pasadena, CA 91125, USA}
\affil[n]{Space Telescope Science Institute, Baltimore, MD 21218, USA}
\affil[o]{Westminster College, Salt Lake City, UT 84105, USA}
\affil[p]{National Radio Astronomy Observatory, Charlottesville, VA 22903, USA}
\affil[q]{Sorbonne Universite, Observatoire de Paris, Universite PSL, CNRS, LERMA, F-75014, Paris, France}
\affil[r]{National Research Council Canada, Herzberg Astronomy and Astrophysics, Victoria, BC, Canada}
\affil[s]{Department of Terrestrial Magnetism, Carnegie Institution of Washington, Washington, DC 20015}
\affil[t]{Department of Astronomy, Boston University, Boston, MA 02215, USA}
\affil[u]{Department of Physics, University of California at Davis, Davis, CA 95616, USA}
\affil[v]{Department of Physics, University of Oxford, Oxford OX1 3RH, UK} 
\affil[w]{RAL Space, Science, and Technology Facilities Council, Rutherford Appleton Laboratory, Didcot OX11 0QX, UK}
\affil[x]{Physics \& Astronomy Department, University College London, Gower Street, London WC1E 6BT, UK}
\affil[y]{Department of Physics and Astronomy, The University of Toledo, 2801 W. Bancroft, Toledo, OH 43606}

\cftpagenumbersoff{figure}
\cftpagenumbersoff{table}

\begin{document}

\maketitle

\begin{abstract}
Far-infrared astronomy has advanced rapidly since its inception in the late 1950's, driven by a maturing technology base and an expanding community of researchers. This advancement has shown that observations at far-infrared wavelengths are important in nearly all areas of astrophysics, from the search for habitable planets and the origin of life, to the earliest stages of galaxy assembly in the first few hundred million years of cosmic history. The combination of a still developing portfolio of technologies, particularly in the field of detectors, and a widening ensemble of platforms within which these technologies can be deployed, means that far-infrared astronomy holds the potential for paradigm-shifting advances over the next decade. In this review, we examine current and future far-infrared observing platforms, including ground-based, sub-orbital, and space-based facilities, and discuss the technology development pathways that will enable and enhance these platforms to best address the challenges facing far-infrared astronomy in the 21st century. 
\end{abstract}


{\noindent \footnotesize\textbf{*}Duncan Farrah, \linkable{dfarrah@hawaii.edu} }

\pagebreak

\tableofcontents

\twocolumn

\section{Introduction}\label{sect:intro} 
Far-infrared astronomy, defined broadly as encompassing science at wavelengths of $30-1000\,\mu$m, is an invaluable tool in understanding all aspects of our cosmic origins. Tracing its roots to the late 1950's, with the advent of infrared detectors sensitive enough for astronomical applications, far-infrared astronomy has developed from a niche science, pursued by only a few teams of investigators, to a concerted worldwide effort pursued by hundreds of astronomers, targeting areas ranging from the origins of our Solar System to the ultimate origin of the Universe. 

By their nature, far-infrared observations study processes that are mostly invisible at other wavelengths, such as young stars still embedded in their natal dust clouds, or the obscured, rapid assembly episodes of supermassive black holes. Moreover, the $30-1000\,\mu$m wavelength range includes a rich and diverse assembly of diagnostic features. The most prominent of these are:

\begin{itemize}

\item Continuum absorption and emission from dust grains with equilibrium temperatures approximately in the range 15-100\,K. The dust is heated by any source of radiation at shorter wavelengths, and cools via thermal emission. 

\item Line emission and absorption from atomic gas, the most prominent lines including [O I], [N II], [C I], [C II], as well as several hydrogen recombination lines. 

\item A plethora of molecular gas features, including, but not limited to: CO, H$_2$O, H$_2$CO, HCN, HCO$^+$, CS, NH$_3$, CH$_3$CH$_2$OH, CH$_3$OH, HNCO, HNC, N$_2$H$^+$, H$_3$O$^+$, their isotopologues (e.g. $^{13}$CO, H$_2$$^{18}$O), and deuterated species (e.g. HD, HDO, DCN).

\item Amorphous absorption and emission features arising from pristine and processed ices, and crystalline silicates.

\end{itemize}

\noindent This profusion and diversity of diagnostics allows for advances across a wide range of disciplines. We briefly describe four examples:
\vspace{0.2cm}

\noindent {\bfseries Planetary systems and the search for life:} Far-infrared continuum observations in multiple bands over $50-200\,\mu$m measure the size distributions, distances, and orbits of both Trans-Neptunian Objects \citep{backman95,santos12,lebret12,eiroa13} and of zodiacal dust \citep{nesvor10}, which gives powerful constraints on the early formation stages of our Solar System, and of others. Molecular and water features determine the composition of these small bodies, provide the first view of how water pervaded the early Solar System via deuterated species ratios, and constrain how water first arrived on Earth \citep{morb00,mumma11,harto11}. Far-infrared observations are also important for characterizing the atmospheric structure and composition of the gas giant planets and their satellites, especially Titan. 

Far-infrared continuum observations also give a direct view of the dynamics and evolution of protoplanetary disks, thus constraining the early formation stages of other solar systems \citep{holland98,andwil05,bryden06,wyatt08}. Deuterated species can be used to measure disk masses, ice features and water lines give a census of water content and thus the earliest seeds for life \citep{hoger11}, while the water lines and other molecular features act as bio-markers, providing the primary tool in the search for life beyond Earth \citep{kalt07,hedelt13}. 
\vspace{0.1cm}

\noindent {\bfseries The early lives of stars:} The cold, obscured early stages of star formation make them especially amenable to study at far-infrared wavelengths. Far-infrared continuum observations are sensitive to the cold dust in star forming regions, from the filamentary structures in molecular clouds \citep{andre10} to the envelopes and disks that surround individual pre-main-sequence stars \citep{brog15}. They thus trace the luminosities of young stellar objects and can constrain the masses of circumstellar structures. Conversely, line observations such as [O I], CO, and H$_2$O probe the gas phase, including accretion flows, outflows, jets, and associated shocks \citep{motte98,evans09,schul09,kristen12,manoj13,vandish11,watson16}. 

For protostars, since their SEDs peak in the far-infrared, photometry in this regime is required to refine estimates of their luminosities and evolutionary states \citep{dunham08,furl16,fisch17}, and can break the degeneracy between inclination angle and evolutionary state\footnote{At mid-infrared and shorter wavelengths a more evolved protostar seen through its edge-on disk has an SED similar to a deeply embedded protostar viewed from an intermediate angle \citep{whit03}.}. With {\itshape Herschel}, it became possible to measure temperatures deep within starless cores \citep{launh13}, and young protostars were discovered that were only visible at far-infrared and longer wavelengths \citep{stutz13}. These protostars have ages of $\sim25,000$\,yr, only 5\% of the estimated protostellar lifetime.

In the T Tauri phase, where the circumstellar envelope has dispersed, far-infrared observations probe the circumstellar disk \citep{howard13}. At later phases, the far-infrared traces extrasolar analogs of the Kuiper belt in stars such as Fomalhaut \citep{acke12}.

Future far-infrared observations hold the promise to understand the photometric variability of protostars. {\itshape Herschel} showed that the far-infrared emission from embedded protostars in Orion could vary by as much as 20\% over a time scale of weeks \citep{billot12}, but such studies were limited by the $<4$ year lifetime of {\itshape Herschel}. Future observatories will allow for sensitive mapping of entire star-forming regions several times over the durations of their missions. This will enable a resolution to the long-running question of whether protostellar mass accretion happens gradually over a few hundred thousand years, or more stochastically as a series of short, episodic bursts \citep{kenyon90}.
\vspace{0.1cm}

\noindent {\bfseries The physics and assembly history of galaxies:} The shape of the mid/far-infrared dust continuum is a sensitive diagnostic of the dust grain size distribution in the ISM of our Milky Way, and nearby galaxies, which in turn diagnoses energy balance in the ISM\citep{cal00,li01,dale01,moli10}. Emission and absorption features measure star formation, metallicity gradients, gas-phase abundances and ionization conditions, and gas masses, all independently of extinction, providing a valuable perspective on how our Milky Way, and other nearby galaxies, formed and evolved\citep{craw85,panu10,fisch10,diaz13,far13}. Continuum and line surveys at far-infrared wavelengths measure both obscured star formation rates and black hole accretion rates over the whole epoch of galaxy assembly, up to $z\gtrsim7$, and are essential to understand why the comoving rates of star formation and supermassive black hole accretion both peaked at redshifts of $z = 2 - 3$, when the Universe was only 2 or 3 billion years old, and have declined strongly since then \citep{lag05,madau14}. 

Of particular relevance in this context are the infrared-luminous galaxies, in which star formation occurs embedded in molecular clouds, hindering the escape of optical and ultraviolet radiation; however, the radiation heats dust, which reradiates infrared light, enabling star-forming galaxies to be identified and their star formation rates to be inferred. These infrared-luminous galaxies may dominate the comoving star formation rate density at $z>1$ and are most optimally studied via infrared observations \citep{lon06,rodig11,lutz11,oliver12,cas14}. Furthermore, far-infrared telescopes can study key processes in understanding stellar and black hole mass assembly, whether or not they depend directly on each other, and how they depend on environment, redshift, and stellar mass \citep{gen10,fabian12,far12}. 
\vspace{0.1cm}

\noindent {\bfseries The origins of the Universe:} Millimeter-wavelength investigations of primordial B- and E-modes in the cosmic microwave background provide the most powerful observational constraints on the very early Universe, at least until the advent of space-based gravitational-wave observatories\citep{page07,planck16}. However, polarized dusty foregrounds are a pernicious barrier to such observations, as they limit the ability to measure B-modes produced by primordial gravitational waves, and thus to probe epochs up to $10^{-30}$ seconds after the Big Bang. Observations at far-infrared wavelengths are the only way to isolate and remove these foregrounds. CMB instruments that also include far-infrared channels thus allow for internally consistent component separation and foreground subtraction. 
\vspace{0.2cm}

The maturation of far-infrared astronomy as a discipline has been relatively recent, in large part catalyzed by the advent of truly sensitive infrared detectors in the early 1990s. Moreover, the trajectory of this development over the last two decades has been steep, going from one dedicated satellite and a small number of other observatories by the mid-1980's, to at least eight launched infrared-capable satellites, three airborne facilities, and several balloon/sub-orbital and dedicated ground based observatories by 2018. New detector technologies are under development, and advances in areas such as mechanical coolers enable those detectors to be deployed within an expanding range of observing platforms. Even greater returns are possible in the future, as far-infrared instrumentation capabilities remain far from the fundamental sensitivity limits of a given aperture.

This recent, rapid development of the far-infrared is reminiscent of the advances in optical and near-infrared astronomy from the 1940s to the 1990s. Optical astronomy benefited greatly from developments in sensor, computing, and related technologies that were driven in large part by commercial and other applications, and which by now are fairly mature. Far-infrared astronomers have only recently started to benefit from comparable advances in capability. The succession of rapid technological breakthroughs, coupled with a wider range of observing platforms, means that far-infrared astronomy holds the potential for paradigm-shifting advances in several areas of astrophysics over the next decade. 

We here review the technologies and observing platforms for far-infrared astronomy, and discuss potential technological developments for those platforms, including in detectors and readout systems, optics, telescope and detector cooling, platform infrastructure on the ground, sub-orbital, and in space, and software development and community cohesion. We aim to identify the technologies needed to address the most important science goals accessible in the far-infrared. We do not review the history of infrared astronomy, as informative historical reviews can be found elsewhere \citep{soipi78,lrg07,rie07,siegel07,price2008history,price09,rieke09,leq09,rowan2013night,cle14}. We focus on the $30-1000\,\mu$m wavelength range, though we do consider science down to $\sim10\,\mu$m, and into the millimeter range, as well. We primarily address the US mid/far-infrared astronomy community; there also exist roadmaps for both European\citep{rigo17} and Canadian\citep{webb2013roadmap} far-infrared astronomy, and for THz technology covering a range of applications\citep{walker2015terahertz,lee2009principles,dhillon17}.

\begin{figure*}
\centering
\includegraphics[width=14cm,angle=0]{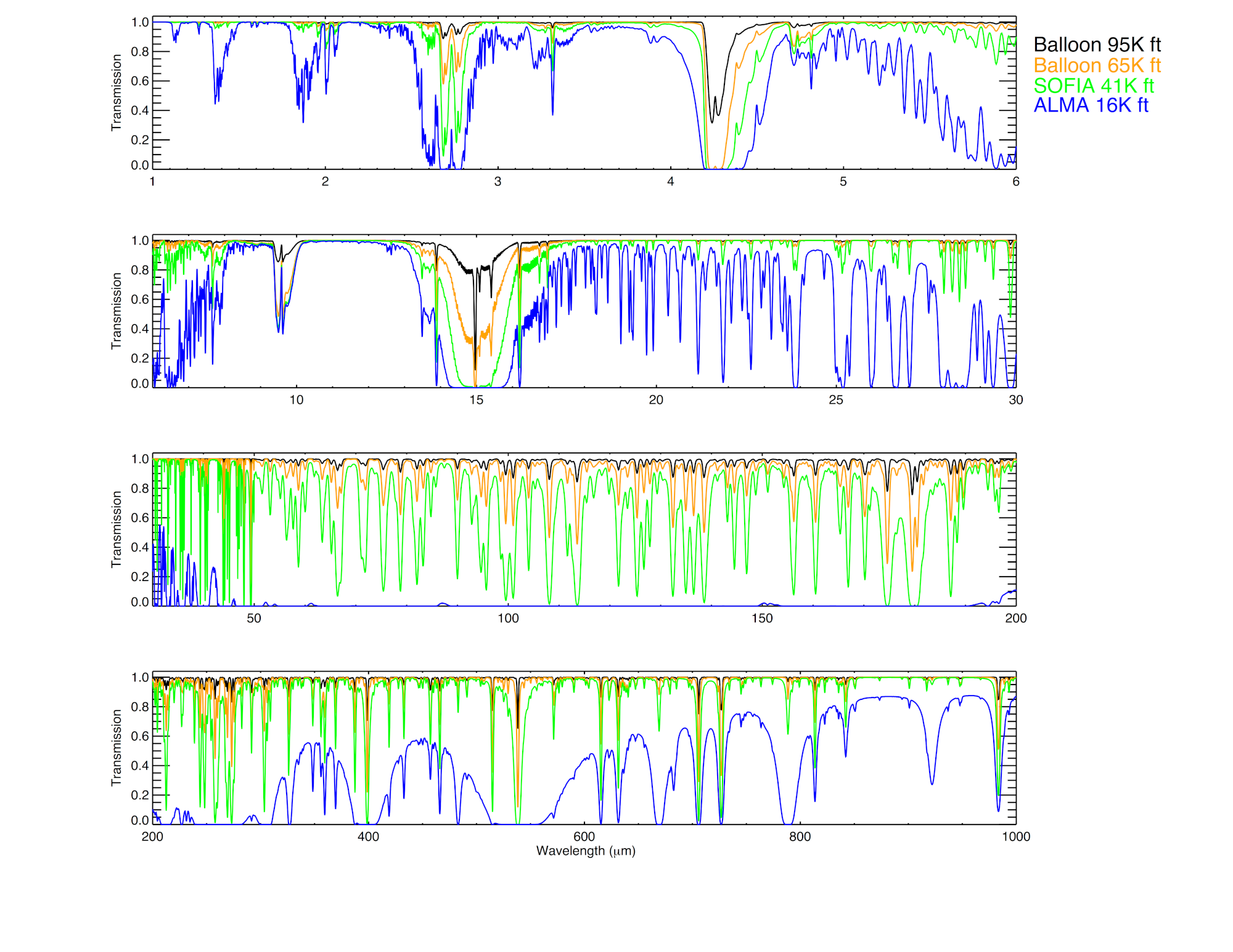}
\vspace{-1cm}
\caption[Atmospheric transmission]{Atmospheric transmission over $1-1000\,\mu$m\citep{lord92}. The curves for ALMA and SOFIA were computed with a 35$^{\circ}$ telescope zenith angle. The two balloon profiles were computed with a 10$^{\circ}$ telescope zenith angle. The precipitable water vapor for ALMA (5060\,m), SOFIA (12,500\,m), and the 19,800\,m and 29,000\,m altitudes are 500, 7.3, 1.1, and 0.2\,$\mu$m, respectively. The data were smoothed to a resolution of $R = 2000$.} 
\label{fig:atmos} 
\end{figure*}

\section{Observatories: Atmosphere-Based}

\subsection{Ground-Based}\label{ssect:ground}
Far-infrared astronomy from the ground faces the fundamental limitation of absorption in Earth's atmosphere, primarily by water vapor. The atmosphere is mostly opaque in the mid- through far-infrared, with only a few narrow wavelength ranges with modest transmission. This behavior can be seen in Figure \ref{fig:atmos}, which compares atmospheric transmission for ground-based observing, observing from SOFIA (\S\ref{ssect:sofia}), and two higher altitudes that are accessible by balloon-based platforms. The difficulties of observing from the ground at infrared wavelengths are evident. Moreover, the transmissivity and widths of these windows are heavily weather dependent. Nevertheless, there do exist spectral windows at $34\,\mu$m, $350\,\mu$m, $450\,\mu$m, $650\,\mu$m and $850\,\mu$m with good, albeit weather-dependent transmission at dry, high-altitude sites, with a general improvement towards longer wavelengths. At wavelengths longward of about 1\,mm there are large bands with good transmission. These windows have enabled an extensive program of ground-based far-infrared astronomy, using both single-dish and interferometer facilities.

\subsubsection{Single-dish facilities}\label{ssect:singdish}
Single-dish telescopes dedicated to far-infrared through millimeter astronomy have been operated for over 30 years. Examples include the 15-m James Clerk Maxwell Telescope (JCMT), the 12-m Caltech Submillimeter Observatory (CSO, closed September 2015), the 30-m telescope operated by the Institut de Radioastronomie Millimetrique (IRAM), the 12-m Atacama Pathfinder Experiment (APEX), the 50-m Large Millimeter Telescope (LMT) in Mexico, the 10-m Submillimeter Telescope (SMT, formerly the Heinrich Hertz Submillimeter Telescope) in Arizona, and the 10-m South Pole Telescope (SPT). These facilities have made major scientific discoveries in almost every field of astronomy, from planet formation to high-redshift galaxies. They have also provided stable development platforms, resulting in key advances in detector technology, and pioneering techniques that subsequently found applications in balloon-borne and space missions.

There is an active program of ground-based single-dish far-infrared astronomy, with current and near-future single-dish telescopes undertaking a range of observation types, from wide-field mapping to multi-object wideband spectroscopy. This in turn drives a complementary program of technology development. In general, many applications for single-dish facilities motivate development of detector technologies capable of very large pixel counts (\S\ref{directdetect}). Similarly large pixel counts are envisioned for planned space-based far-infrared observatories, including the Origins Space Telescope (OST, \S\ref{origins}). Since far-infrared detector arrays have few commercial applications, they must be built and deployed by the science community itself. Thus, ground-based instruments represent a vital first step toward achieving NASA's long-term far-infrared goals.

We here briefly describe two new ground-based facilities; CCAT-prime (CCAT-p), and the Large Millimeter Telescope (LMT):

\noindent {\bfseries CCAT-p}: will be a 6\,m telescope at 5600\,m altitude, near the summit of Cerro Chajnantor in Chile \citep{koop17}. CCAT-p is being built by Cornell University and a German consortium that includes the universities of Cologne and Bonn, and in joint venture with the Canadian Atacama Telescope Corporation. In addition, CCAT-p collaborates with CONICYT and several Chilean universities. The project is funded by a private donor and by the collaborating institutions, and is expected to achieve first light in 2021.

The design of CCAT-p is an optimized crossed-Dragone \citep{niemack16} system that delivers an $8\deg$ field of view (FoV) with a nearly flat image plane. At 350\,$\mu$m the FoV with adequate Strehl ratio reduces to about $4\deg$. The wavelength coverage of the anticipated instruments will span wavelengths of 350\,$\mu$m to 1.3\,mm. With the large FoV and a telescope surface RMS of below 10.7\,$\mu$m, CCAT-p is an exceptional observatory for survey observations. Since the 200\,$\mu$m zenith transmission is $\geq10$\% in the first quartile at the CCAT-p site \citep{radford16}, a 200\,$\mu$m observing capability will be added in a second generation upgrade.

The primary science drivers for CCAT-p are 1) tracing the epoch or reionization via [CII] intensity mapping, 2) studying the evolution of galaxies at high redshifts, 3) investigating dark energy, gravity, and neutrino masses via measurements of the Sunyaev-Zel'dovich effect, and 4) studying the dynamics of the interstellar medium in the Milky Way and nearby galaxies via high spectral resolution mapping of fine-structure and molecular lines. 

CCAT-p will host two facility instruments, the CCAT Heterodyne Array Instrument (CHAI), and the direct detection instrument Prime-Cam (P-Cam). CHAI is being built by the University of Cologne and will have two focal plane arrays that simultaneously cover the 370\,$\mu$m and 610\,$\mu$m bands. The arrays will initially have $8\times8$ elements, with a planned expansion to 128 elements each. The direct detection instrument P-Cam, which will be built at Cornell University, will encompass seven individual optics-tubes. Each tube has a FoV of about $1.3\deg$. For first light, three tubes will be available, 1) a four-color, polarization sensitive camera with 9000 pixels that simultaneously covers the 1400, 1100, 850, and 730\,$\mu$m bands, 2) a 6000 pixel Fabry-Perot spectrometer, and 3) a 18,000 pixel camera for the 350\,$\mu$m band.
\vspace{0.1cm}

\noindent {\bfseries LMT}: The LMT is a 50-m diameter telescope sited at 4600\,m on Sierra Negra in Mexico. The LMT has a FoV of $4\,\arcmin$ and is optimized for maximum sensitivity and small beamsize at far-infrared and millimeter wavelengths. It too will benefit from large-format new instrumentation in the coming years. A notable example is TolTEC, a wide-field imager operating at 1.1\,mm, 1.4\,mm, and 2.1\,mm, and with an anticipated mapping speed at 1.1\,mm of 14 deg$^2$ my$^{-2}$ hr$^{-1}$ (Table \ref{tbl:requirements}). At 1.1\,mm, the TolTEC beam size is anticipated to be $\sim5\arcsec$, which is smaller than the $6\arcsec$ beamsize of the 24\,$\mu$m {\itshape Spitzer} extragalactic survey maps. As a result, the LMT confusion limit at 1.1\,mm is predicted to be $\sim0.1$\,mJy, making LMT capable of detecting sources with star formation rates below 100\,M$_{\odot}$yr$^{-1}$ at $z\sim6$. This makes TolTEC an excellent ``discovery machine'' for high-redshift obscured galaxy populations. As a more general example of the power of new instruments mounted on single-aperture ground-based telescopes, a $\sim$100-object steered-beam multi-object spectrometer mounted on LMT would exceed the abilities of any current ground-based facility, including ALMA, for survey spectroscopy of galaxies, and would require an array of $\sim10^{5.5}$ pixels.

\begin{figure*}
\includegraphics[width=16cm,angle=0]{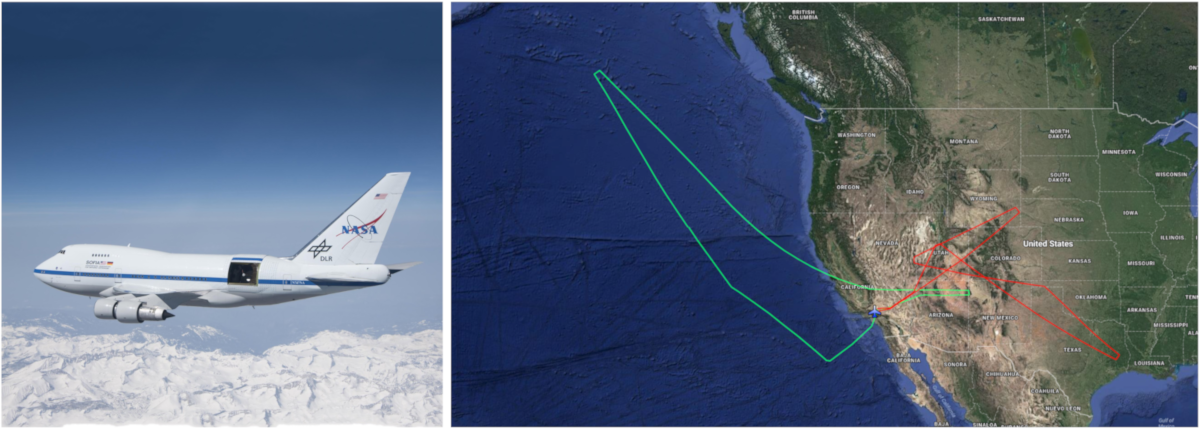}
\caption[The SOFIA observatory, and example flight paths]{{\itshape Left:} The world's largest flying infrared astronomical observatory, SOFIA (\S\ref{ssect:sofia}). {\itshape Right:}  Two flight plans, originating from SOFIA's prime base in Palmdale, California. In a typical 8-10 hour flight, SOFIA can observe 1-5 targets.} 
\label{fig:sofflight} 
\end{figure*}

\subsubsection{Interferometry}\label{ssect:gbinter}
Interferometry at far-infrared wavelengths is now routinely possible from the ground, and has provided order of magnitude improvements in spatial resolution and sensitivity over single-dish facilities. Three major ground-based far-infrared/millimeter interferometers are currently operational. The NOEMA array (the successor to the IRAM Plateau de Bure interferometer) consists of nine 15-m dishes at 2550\,m elevation in the French Alps. The Submillimeter Array (SMA) consists of eight 6-m dishes on the summit of Mauna Kea in Hawaii (4200\,m elevation). Both NOEMA and the SMA are equipped with heterodyne receivers. NOEMA has up to 16\,GHz instantaneous bandwidth, while the SMA has up to 32\,GHz of instantaneous bandwidth (or 16\,GHz with dual polarization) with 140\,KHz uniform spectral resolution.

Finally, the Atacama Large Millimeter/submillimeter Array (ALMA) is sited on the Chajnantor Plateau in Chile at an elevation of 5000\,m. It operates from 310\,$\mu$m to 3600\,$\mu$m in eight bands covering the primary atmospheric windows. ALMA uses heterodyne receivers based on Superconductor-Insulator-Superconductor (SIS) mixers in all bands, with 16\,GHz maximum instantaneous bandwidth split across 2 polarizations and 4 basebands. ALMA consists of two arrays: the main array of fifty 12-m dishes (of which typically 43 are in use at any one time), and the Morita array (also known as the Atacama Compact Array), which consists of up to twelve 7-m dishes and up to four 12-m dishes equipped as single dish telescopes. 

At the ALMA site (which is the best of the three ground-based interferometer sites), the Precipitable Water Vapor (PWV) is below 0.5\,mm for 25\% of the time during the five best observing months (May-September). This corresponds to a transmission of about 50\% in the best part of the 900-GHz window (ALMA Band 10). In more typical weather (PWV=1\,mm) the transmission at 900-GHz is 25\%.

There are plans to enhance the abilities of ALMA over the next decade, by (1) increasing bandwidth, (2) achieving finer angular resolutions, (3) improving wide-area mapping speeds, and (4) improving the data archive. The primary improvement in bandwidth is expected to come from an upgrade to the ALMA correlator, which will effectively double the instantaneous bandwidth, and increase the number of spectral points by a factor of eight. This will improve ALMA's continuum sensitivity by a factor $\sqrt{2}$, as well as making ALMA more efficient at line surveys. Further bandwidth improvements include the addition of a receiver covering 35-50\,GHz (ALMA Band 1, expected in 2022), and 67-90\,GHz (ALMA Band 2). To improve angular resolution, studies are underway to explore the optimal number and placement of antennas for baseline lengths of up to tens of km. Other possible improvements include increasing the number of antennas in the main array to 64, the incorporation of focal-plane arrays to enable wider field imaging, and improvements in the incorporation of ALMA into the global Very Long Baseline Interferometry (VLBI) network.

\subsection{Stratospheric Observatory for Infrared Astronomy}\label{ssect:sofia}
The Stratospheric Observatory for Infrared Astronomy (SOFIA\citep{temi14}) is a 2.5-m effective diameter telescope mounted within a Boeing 747SP aircraft that flies to altitudes of 13,700\,m to get above over 99.9\% of the Earth's atmospheric water vapor. The successor to the Learjet observatory and NASA's Kuiper Airborne Observatory (KAO), SOFIA saw first light in May 2010, began prime operations in May 2014, and offers approximately 600 hours per year for community science observations\citep{young2012early}. SOFIA is the only existing public observatory with access to far-infrared wavelengths inaccessible from the ground, though CMB polarization studies at millimeter wavelengths have also been proposed from platforms at similar altitudes to SOFIA\citep{feen17}.

SOFIA's instrument suite can be changed after each flight, and is evolvable with new or upgraded instruments as capabilities improve. SOFIA is also a versatile platform, allowing for (1) continuous observations of single targets of up to five hours, (2) repeated observations over multiple flights in a year, and (3) in principle, observations in the visible though millimeter wavelength range. Example flight paths for SOFIA are shown in Figure \ref{fig:sofflight}. Each flight path optimizes observing conditions (e.g., elevation, percentage water vapor, maximal on-target integration time). SOFIA can be positioned to where the science needs, enabling all-sky access. Annually, SOFIA flies from Christchurch, New Zealand to enable southern hemisphere observations.

\begin{figure*}
\begin{center}
\includegraphics[width=16cm,angle=0]{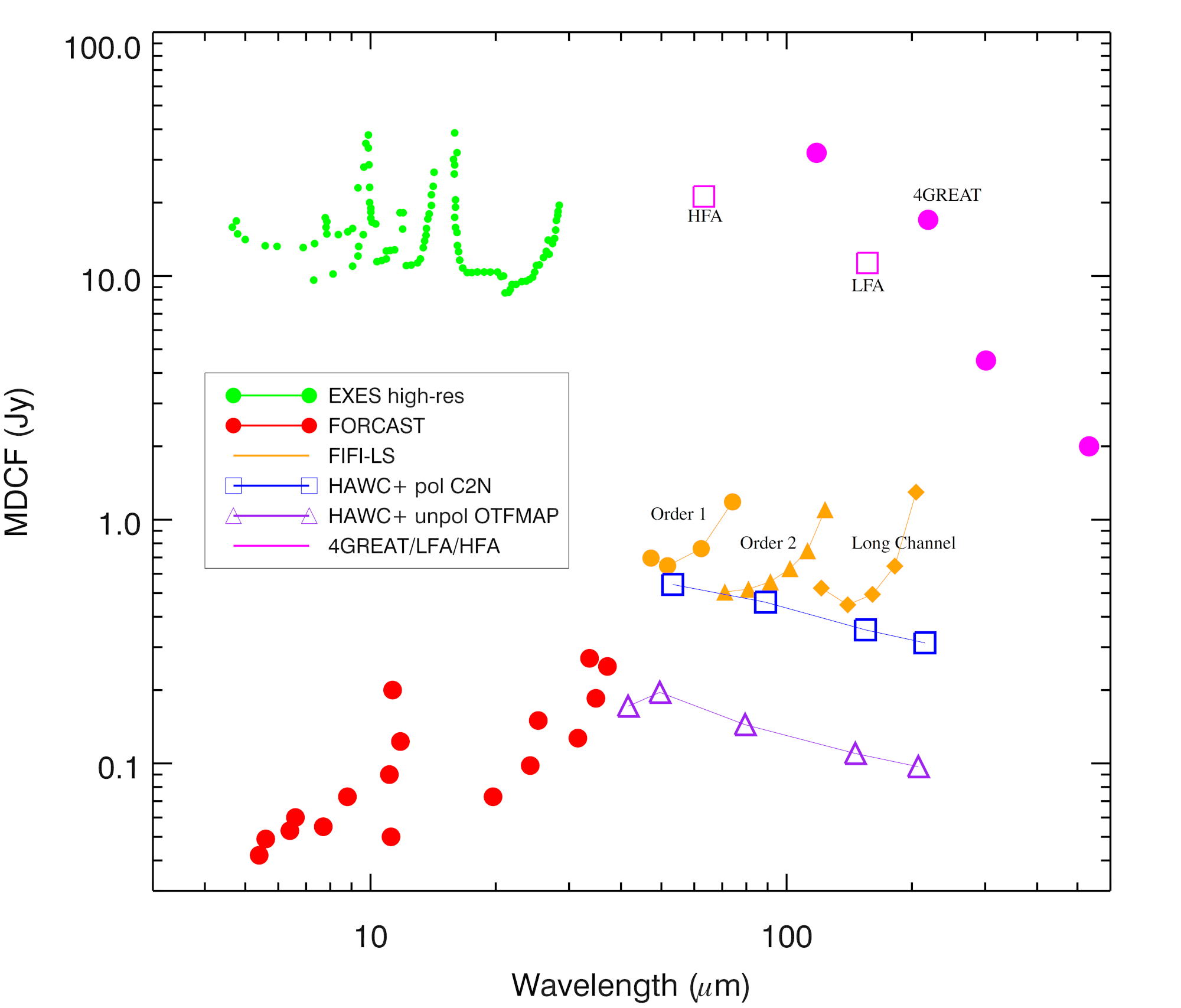}
\caption[SOFIA instrument sensitivities as a function of wavelength]{The continuum sensitivities, as a function of wavelength, of SOFIA's mid- to far-infrared instrument suite. Shown are the 4$\sigma$ Minimum Detectable Continuum Flux (MDCF) densities for point sources in Janskys for 900\,s of integration time.} 
\label{fig:sofia}
\end{center} 
\end{figure*}

\begin{figure}
\begin{center}
\includegraphics[width=8cm,angle=0]{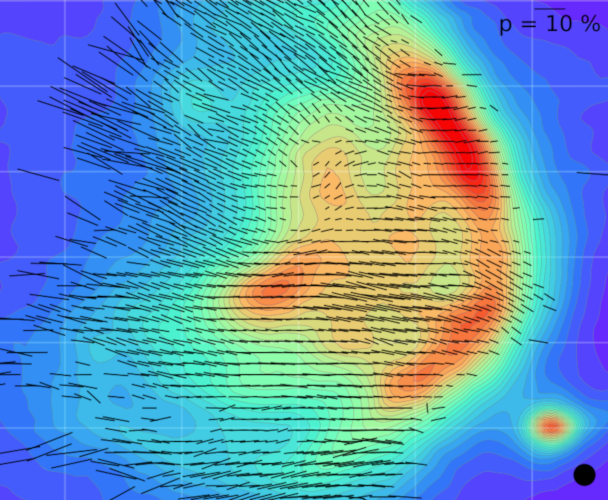}
\caption[An example of a HAWC+ polarimetric image from SOFIA]{The 89\,$\mu$m image (intensity represented by color) with polarization measurements at the same wavelength (black lines), taken using HAWC+ on SOFIA, of $\rho$-ophiucus (courtesy of Fabio Santos, Northwestern University). The length of the line is the degree of polarization. The SOFIA beam size is 7.8$^{\prime\prime}$ indicated by the black circle, lower right.} 
\label{fig:sofiascience}
\end{center} 
\end{figure}

SOFIA's instruments include the $5-40\,\mu$m camera and grism spectrometer FORCAST\citep{adams2010forcast,herter2012first}, the high-resolution (up to $R=\lambda/\Delta\lambda=100,000$) $4.5-28.3\,\mu$m spectrometer EXES\citep{richter2003high}), the $51-203\,\mu$m integral field spectrometer FIFI-LS\citep{colditz2012sofia}), the $50-203\,\mu$m camera and polarimeter HAWC\citep{harper2000hawc}, and the $R\sim10^8$ heterodyne spectrometer GREAT\citep{heyminck2012great,klein12}. The first-generation HIPO\citep{dunham2004hipo} and FLITECAM\citep{mclean2006flitecam} instruments were retired in early 2018. The sensitivities of these instruments as a function of wavelength are presented in Fig \ref{fig:sofia}. Upgrades to instruments over the last few years have led to new science capabilities, such as adding a polarimetry channel to HAWC (HAWC+\citep{dowell2013hawc+}), and including larger arrays and simultaneous channels on GREAT (upGREAT\citep{risacher2016upgreat} \& 4GREAT, commissioned in 2017), making it into an efficient mapping instrument. Figure \ref{fig:sofiascience} shows early polarimetry measurements from HAWC+.

Given the versatility and long-term nature of SOFIA, there is a continuous need for more capable instruments throughout SOFIA's wavelength range. However, the unique niche of SOFIA, given its warm telescope and atmosphere, the imminent era of the James Webb Space Telescope (JWST), and ever more capable ground-based platforms, is high-resolution spectroscopy. This is presently realized with two instruments (GREAT and EXES). An instrument under development, the HIgh Resolution Mid InfrarEd Spectrometer (HIRMES), scheduled for commissioning in 2019, will enhance SOFIA's high resolution spectroscopy capabilities. HIRMES covers $25-122\,\mu$m, with three spectroscopic modes ($R = 600$, $R=10,000$, and $R=100,000$), and an imaging spectroscopy mode ($R=2,000$). 

As SOFIA can renew itself with new instruments, it provides both new scientific opportunities and maturation of technology to enable future far-infrared space missions. SOFIA offers a 20\,kVA cryo-cooler with two compressors able to service two cold heads. The heads can be configured to operate two cryostats or in parallel within one cryostat to increase heat pumping capacity, with 2nd stage cooling capacity Q2$\geq$800\,mW at 4.2\,K and 1st stage cooling capacity Q1$\geq$15\,W at 70\,K. Instruments aboard SOFIA can weigh up to 600\,kg excluding the instrument electronics in the counter-weight rack and PI Rack(s) and draw power up to 6.5\,kW. Their volume is limited by the aircraft's door-access and must fit within the telescope assembly constraints.

Capabilities that would be invaluable in a next-generation SOFIA instrument include, but are not limited to:

\begin{itemize}

\item Instruments with $\geq100$ beams that enable low- to high-resolution spectroscopy (up to sub-km s$^{-1}$ velocity resolution) from 30 to 600\,$\mu$m. This would enable large-area, velocity-resolved spectral line maps of key fine-structure transitions in giant molecular clouds, and complement the wavelengths accessible by JWST and ground-based telescopes. The current state of the art on SOFIA is upGREAT LFA: 14 beams, 44\,kHz channel spacing.

\item Medium to wide-band imaging and imaging polarimetry from 30 to 600\,$\mu$m, with $10^{4-5}$ pixels and FoV's of tens of arcminutes. The current state of the art on SOFIA is HAWC+, with a $64\times40$ pixel array and a largest possible FoV of $8.0'\times6.1'$.

\item High spectral resolution (R=4,000-100,000) 5-30\,$\mu$m mapping spectroscopy with factor $\geq3$ greater observation efficiency and sensitivity than EXES. This would complement JWST, which observes in the same wavelength range but at $R<3300$ with the mid-infrared instrument (MIRI). Such an instrument on SOFIA could then identify the molecular lines that JWST may detect but not spectrally resolve. The current state of the art on SOFIA is EXES, with R$\sim$100,000 and sensitivities ($10\sigma, 100s$) of 10\,Jy at 10\,$\mu$m; 20 Jy at 20\,$\mu$m (NELB, $10\sigma, 100s$: $1.4\times10^{-6}\,$W m$^{-2}$ sr$^{-1}$ at 10\,$\mu$m; $7.0\times10^{-7}\,$W m$^{-2}$ sr$^{-1}$ at $20\,\mu$m).

\item High resolution (R$\sim$100,000) spectroscopy at $2.5-8\,\mu$m, to identify several gas-phase molecules. These molecules are not readily accessible from the ground (Fig \ref{fig:atmos}), and cannot be reliably identified by JWST as its near-infrared spectrometer NIRSpec only goes up to $R=2700$. Currently, SOFIA does not have such an instrument.

\item A wide-field, high-resolution integral-field spectrometer covering $30-600$\,$\mu$m. This would allow rapid, large-area spectrally-resolved mapping of fine structure lines in the Milky Way, and integral field-spectroscopy of nearby galaxies. The current state of the art on SOFIA is FIFI-LS, with FoV $12\arcsec$ over 115-203$\,\mu$m) and $6\arcsec$ over 51-120$\,\mu$m.

\item A broadband, wide-field, multi-object spectrograph, with resolution $R=10^3 - 10^4$ and up to $1000$ beams, over $30-300\,\mu$m. Such an instrument could map velocity fields in galaxies or star-forming regions, with enough beams to allow mapping of complex regions. SOFIA currently does not have any multi-object spectroscopic capability.

\item An instrument to characterize exoplanet atmospheres: an ultra-precise spectro-imager optimized for bands not available from the ground and with sufficient FoV to capture simultaneous reference stars to decorrelate time-variable effects. JWST and ESA's ARIEL mission will both also contribute to this science. SOFIA currently does not have this capability. However, during Early Science with first-generation instruments, SOFIA demonstrated it could measure atmospheres with transiting exoplanets with performance similar to existing ground assets. 

\item A mid/far-infrared spectropolarimeter. Spectropolarimetric observations of the relatively unexplored $20\,\mu$m silicate feature with SOFIA would be a unique capability, and allow for e.g. new diagnostics of the chemistry and composition of protoplanetary disks. SOFIA currently does not have polarimetry shortward of $50\,\mu$m.

\end{itemize}

Other possible improvements to the SOFIA instrument suite include: (1) upgrading existing instruments (e.g. replacing the FIFI-LS germanium photoconductors to achieve finer spatial sampling through higher multiplexing factors), and (2) instruments that observe in current gaps in SOFIA wavelength coverage (e.g., $1-5\,\mu$m, $90-150\,\mu$m, $210-310\,\mu$m).

More general improvements include the ability to swap instruments faster than a two-day timescale, or the ability to mount multiple instruments. Mounting multiple instruments improves observing efficiency if both instruments can be used on the same source, covering different wavelengths or capabilities. This would also allow for flexibility to respond to targets of opportunity, time domain or transient phenomena, and increase flexibility as a development platform to raise Technology Readiness Levels (TRLs\citep{mankins1995technology,mankins2009technology}) of key technologies.

\begin{figure*}
\centering
\includegraphics[width=16cm,angle=0]{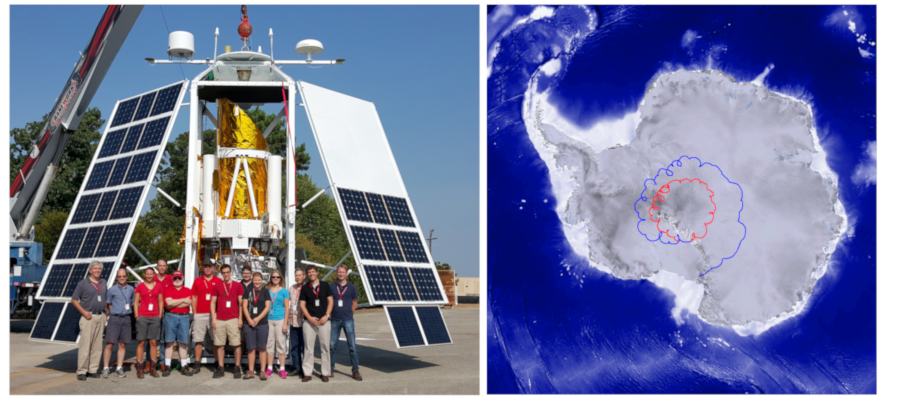}
\caption[The balloon-based Stratospheric Terahertz Observatory]{{\itshape Left:} The STO balloon observatory and science team, after the successful hang test in the Columbia Scientific Balloon Facility in Palestine, Texas, in August 2015. This image originally appeared on the SRON STO website. {\itshape Right:} The second science flight of the STO took place from from McMurdo in Antarctica on December 8th, 2016, with a flight time of just under 22 days. This image was taken from https://www.csbf.nasa.gov/antarctica/payloads.htm.} 
\label{fig:stoii} 
\end{figure*}

\subsection{Scientific Ballooning}\label{ssect:uldb}
Balloon-based observatories allow for observations at altitudes of up to $\sim40,000$\,m ($130,000$\,ft). At these altitudes, less than 1\% of the atmosphere remains above the instrument, with negligible water vapor. Scientific balloons thus give access, relatively cheaply, to infrared discovery space that is inaccessible to any ground-based platform, and in some cases even to SOFIA. For example, several key infrared features are inaccessible even at aircraft altitudes (Figure \ref{fig:atmos}), including low-energy water lines and the [N II]122\,$\mu$m line. Scientific ballooning is thus a valuable resource for infrared astronomy. Both standard balloons, with flight times of $\lesssim24$ hours, and Long Duration Balloons (LDBs) with typical flight times of $7-15$ days (though flights have lasted as long as 55 days) have been used. Balloon projects include the Balloon-borne Large Aperture Submillimeter Telescopes (BLAST\citep{tuck04,fiss10,galit14}), PILOT \citep{bern16}, the Stratospheric Terahertz Observatory\citep{walk10} (Figure \ref{fig:stoii}), and FITE\citep{shib10} \& BETTII\citep{rine14}, both described in \S\ref{ssect:firint}. Approved future missions include GUSTO, scheduled for launch in 2021. With the development of Ultra-Long Duration Balloons (ULDB), with potential flight times of over 100 days, new possibilities for far-infrared observations become available.

A further advantage of ballooning, in a conceptually similar manner to SOFIA, is that the payloads are typically recovered and available to refly on $\sim$ one year timescales, meaning that balloons are a vital platform for technology development and TRL raising. For example, far-infrared direct detector technology shares many common elements (detection approaches, materials, readouts) with CMB experiments, which are being conducted on the ground\citep{bicep14,flauger14,bikec15}, from balloons\citep{kogut12,gand16,ebex17}, and in space. These platforms have been useful for developing bolometer and readout technology applicable to the far-infrared.

All balloon projects face challenges, as the payload must include both the instrument and all of the ancillary equipment needed to obtain scientific data. For ULDBs, however, there are two additional challenges: 
\vspace{0.1cm}

\noindent {\bfseries Payload mass}: While zero-pressure balloons (including LDBs) can lift up to about 2,700\,kg, ULDBs have a mass limit of about 1,800\,kg. Designing a payload to this mass limit is non-trivial, as science payloads can have masses in excess of 2,500\,kg. For example, the total mass of the GUSTO gondola is estimated to be 2,700\,kg.
\vspace{0.1cm}

\noindent {\bfseries Cooling}: All far-infrared instruments must operate at cryogenic temperatures. Liquid cryogens have been used for instruments on both standard and LDB balloons, with additional refrigerators (e.g. $^3$He, adiabatic demagnetization) to bring detector arrays down to the required operating temperatures, which can be as low as $100\,$mK. These cooling solutions typically operate on timescales commensurate with LDB flights. For the ULDB flights however it is not currently possible to achieve the necessary cryogenic hold times. Use of mechanical coolers to provide first-stage cooling would solve this problem, but current technology does not satisfy the needs of balloon missions. Low-cost cryocoolers for use on the ground are available, but have power requirements that are hard to meet on balloons, which currently offer total power of up to about $2.5$\,kW. Low-power cryocoolers exist for space use, but their cost (typically $\gtrsim\$1$M) does not fit within typical balloon budgets. Cryocoolers are discussed in detail in \S\ref{ssect:cryoc}. 
\vspace{0.1cm}

In addition to addressing the challenges described above, there exist several avenues of development that would enhance many balloon experiments. Three examples are: 

\begin{itemize}

\item Large aperture, lightweight mirrors for $50-1000\,\mu$m observing (see also \S\ref{ssect:general}).

\item Common design templates for certain subsystems such as star cameras, though attempting to standardize on {\itshape gondola} designs would be prohibitively expensive since most systems are still best implemented uniquely for each payload.
  
\item Frameworks to enhance the sharing of information, techniques and approaches. While balloon experiments are in general more ``PI driven'' than facility class observatories (since much of the hardware is custom-built for each flight), there does exist a thriving user community in ballooning, within which information and ideas are shared. Nurturing the sharing of information would help in developing this community further. The PI-driven balloon missions also serve as pathfinders for larger facilities, as was the case for BLAST and {\itshape Herschel}, and thus may lay the groundwork for a future ``General Observatory'' class balloon mission. 

\end{itemize}

\begin{figure}
\includegraphics[width=8cm,angle=0]{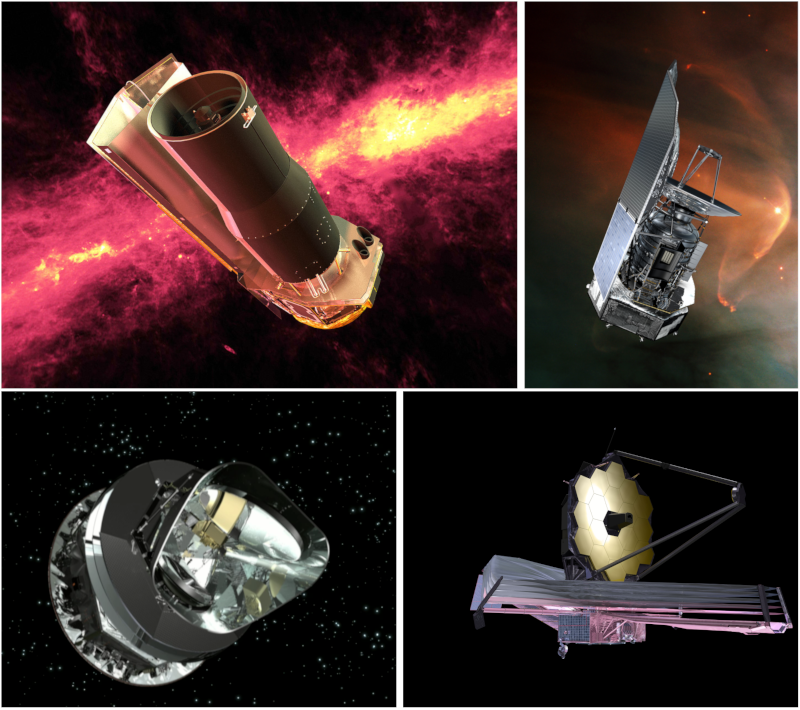}
\caption[Four examples of infrared satellites.]{Four examples of satellites that observe at mid/or far-infrared wavelengths (\S\ref{spaceobs}). {\itshape Top row:} {\itshape Spitzer} and {\itshape Herschel}. {\itshape Bottom row:} {\itshape Planck} and JWST, which also use V-groove radiators (thermal shields) to achieve passive cooling to $< 40$\,K.} 
\label{fig:launchir} 
\end{figure}

\subsection{Short Duration Rocket Flights}\label{ssect:rockets}
Sounding rockets inhabit a niche between high-altitude balloons and fully orbital platforms, providing $5{-}10$ minutes of observation time above the Earth's atmosphere, at altitudes of $50$\,km to $\sim1500$\,km. They have been used for a wide range of astrophysical studies, with a heritage in infrared astronomy stretching back to the 1960's \citep{shiv68,houck69,afgl76,seibert2006history}. 

Though an attractive way to access space for short periods, the mechanical constraints of sounding rockets are limiting in terms of the size and capability of instruments. However, sounding rockets observing in the infrared are flown regularly\citep{zemc13}, and rockets are a viable platform for both technology maturation and certain observations in the far-infrared. In particular, measurements of the absolute brightness of the far-infrared sky, intensity mapping, and development of ultra-low noise far-infrared detector arrays are attractive applications of this platform. 

Regular access to sounding rockets is now a reality, with the advent of larger, more capable Black Brant XI vehicles launched from southern Australia via the planned Australian NASA deployment in 2019-2020. Similarly, there are plans for recovered flights from Kwajalein Atoll using the recently tested NFORCE water recovery system. Long-duration sounding rockets capable of providing limited access to orbital trajectories and $> 30 \,$min observation times have been studied \citep{NAP12862}, and NASA is continuing to investigate this possibility. However, no missions using this platform are currently planned, and as a result the associated technology development is moving slowly.

\begin{figure*}
\includegraphics[width=16cm,angle=0]{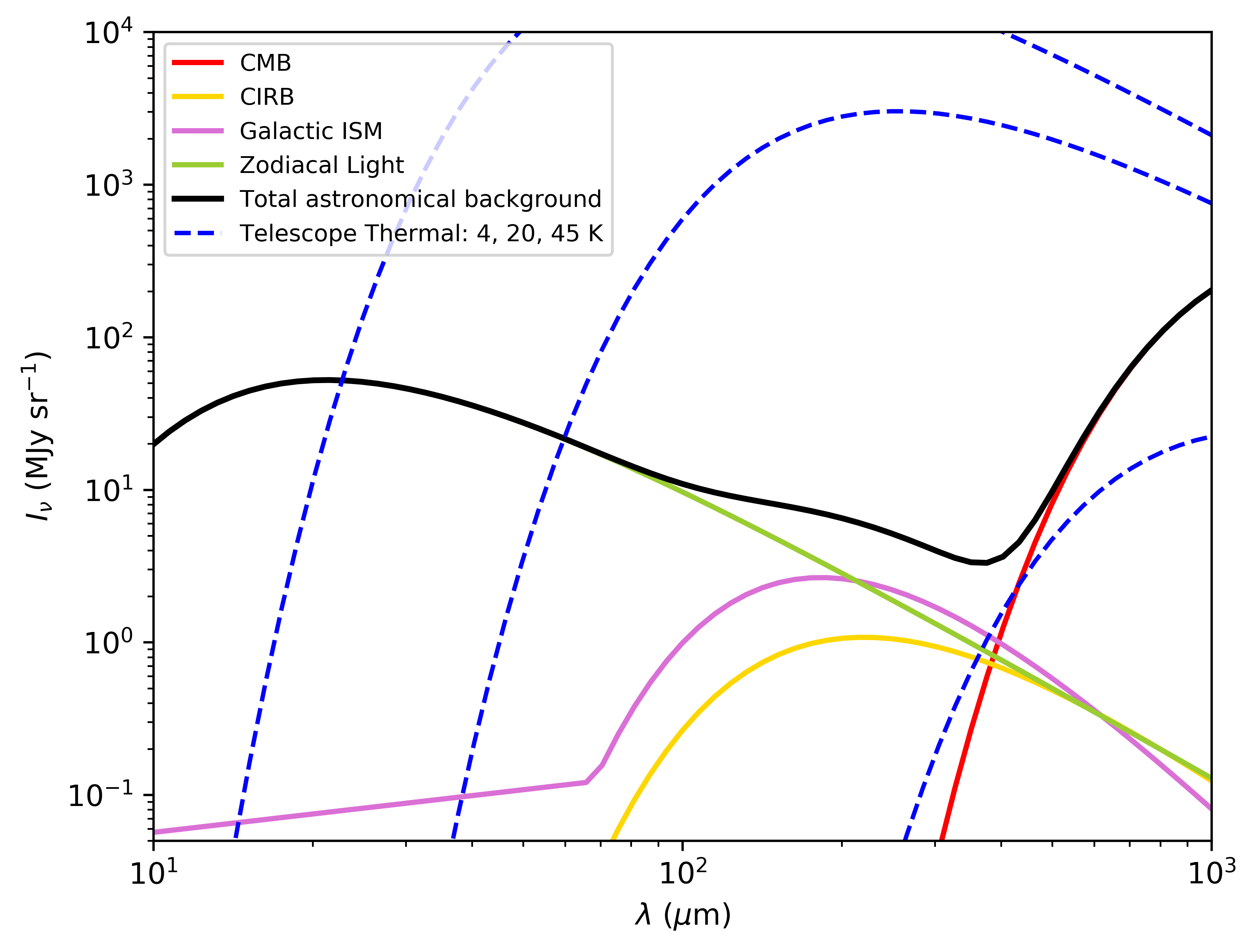}
\caption[A comparison between astrophysical infrared background, and thermal backgrounds from telescope optics]{A comparison between the primary astrophysical continuum backgrounds at infrared wavelengths (the Cosmic Microwave Background\citep{fixs09}, the Cosmic Infrared Background\citep{cobefiras}, Galactic ISM emission\citep{parad11}, and Zodiacal emission from interplanetary dust\citep{lein98}) and representative thermal emission from telescope optics at three temperatures, assuming uniform thermal emissivity of 4\%. The astrophysical backgrounds assume observations outside the atmosphere towards high ecliptic and galactic latitudes, and at a distance of 1\,AU from the Sun. The advantages of ``cold'' telescope optics are apparent; at $300\,\mu$m the thermal emission from a 4\,K telescope is five orders of magnitude lower than for a telescope at 45\,K, and enables the detection of the CIRB, Galactic ISM, and zodiacal light.}
\label{fig:backgr} 
\end{figure*}

\section{Observatories: Space-Based}\label{spaceobs}
All atmospheric-based observing platforms, including SOFIA and balloons, suffer from photon noise from atmospheric emission. Even at balloon altitudes, of order 1\% emissivity on average through the far-infrared remains from residual water vapor, which can contaminate astrophysical water lines unless they are shifted by velocities of at least a few tens of km s$^{-1}$. The telescope optics are another source of loading, with an unavoidable $2-4$\%  emissivity. Though the total emissivity can be less than 5\%, these ambient-temperature ($\sim$250\,K) background sources dominate that of the zodiacal and galactic dust. Space-based platforms are thus, for several paths of inquiry, the only way to perform competitive infrared observations.

There exists a rich history of space-based mid/far-infrared observatories (Figure \ref{fig:launchir}), including IRAS\citep{neugebauer1984infrared}, MSX \citep{mill94}, the IRTS\citep{mura96}, ISO\citep{kessler1996infrared}, SWAS\citep{meln00}, {\itshape Odin}\citep{nordh03}, AKARI \citep{murakami2007infrared}, {\itshape Herschel}\citep{pilb10}, WISE\citep{wright10}, and {\itshape Spitzer}\citep{wer04}. Far-infrared detector arrays are also used on space-based CMB missions, with past examples including {\itshape Planck} \citep{plan11}, WMAP\citep{benn03}, and COBE\citep{bogg92,cobefiras}, as well as concepts such as PIXIE\citep{pixie16}, LiteBIRD\citep{litebird14}, and CORE\citep{core17}.

\begin{figure*}
\includegraphics[width=16cm,angle=0]{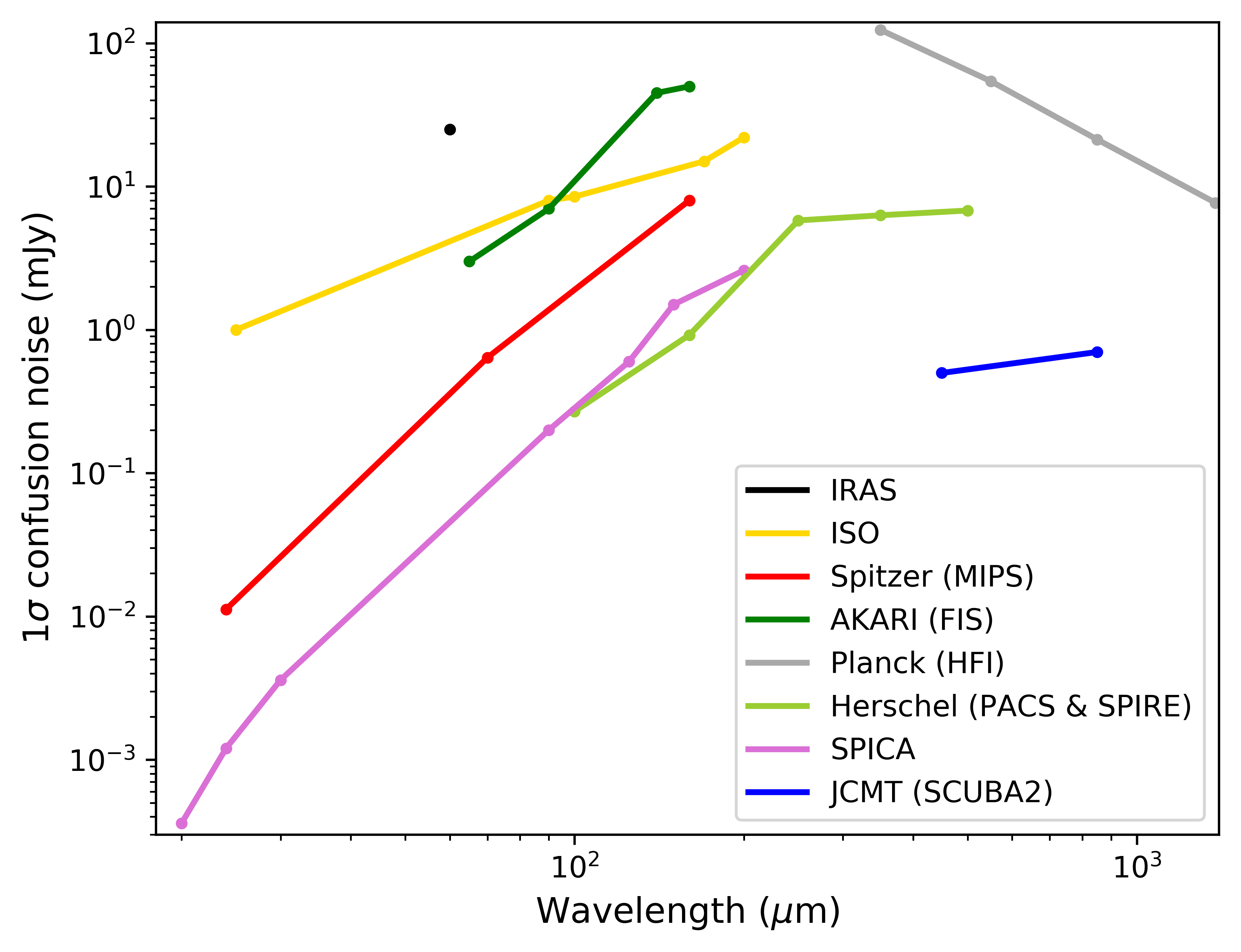}
\caption[A summary of confusion noise levels for selected telescopes]{A summary of literature estimates of confusion noise levels for selected telescopes. The confusion levels are not calculated with a uniform set of assumptions, but are comparable in that they are applicable to regions of sky away from the galactic plane, and with low galactic cirrus emission. Shown are estimates for IRAS at 60\,$\mu$m\citep{hahou87}, ISO\citep{kiss05}, {\itshape Spitzer}\citep{dole04}, {\itshape Herschel}\citep{nguyen10}, {\itshape Planck}\citep{negrel04,fernan08}, AKARI, SPICA\citep{takeuchi2018estimation}, and JCMT. The confusion limits for interferometers such as ALMA at the SMA are all below $10^{-6}$\,mJy.} 
\label{fig:confnoise} 
\end{figure*}

\begin{figure*}
\includegraphics[width=16cm,angle=0]{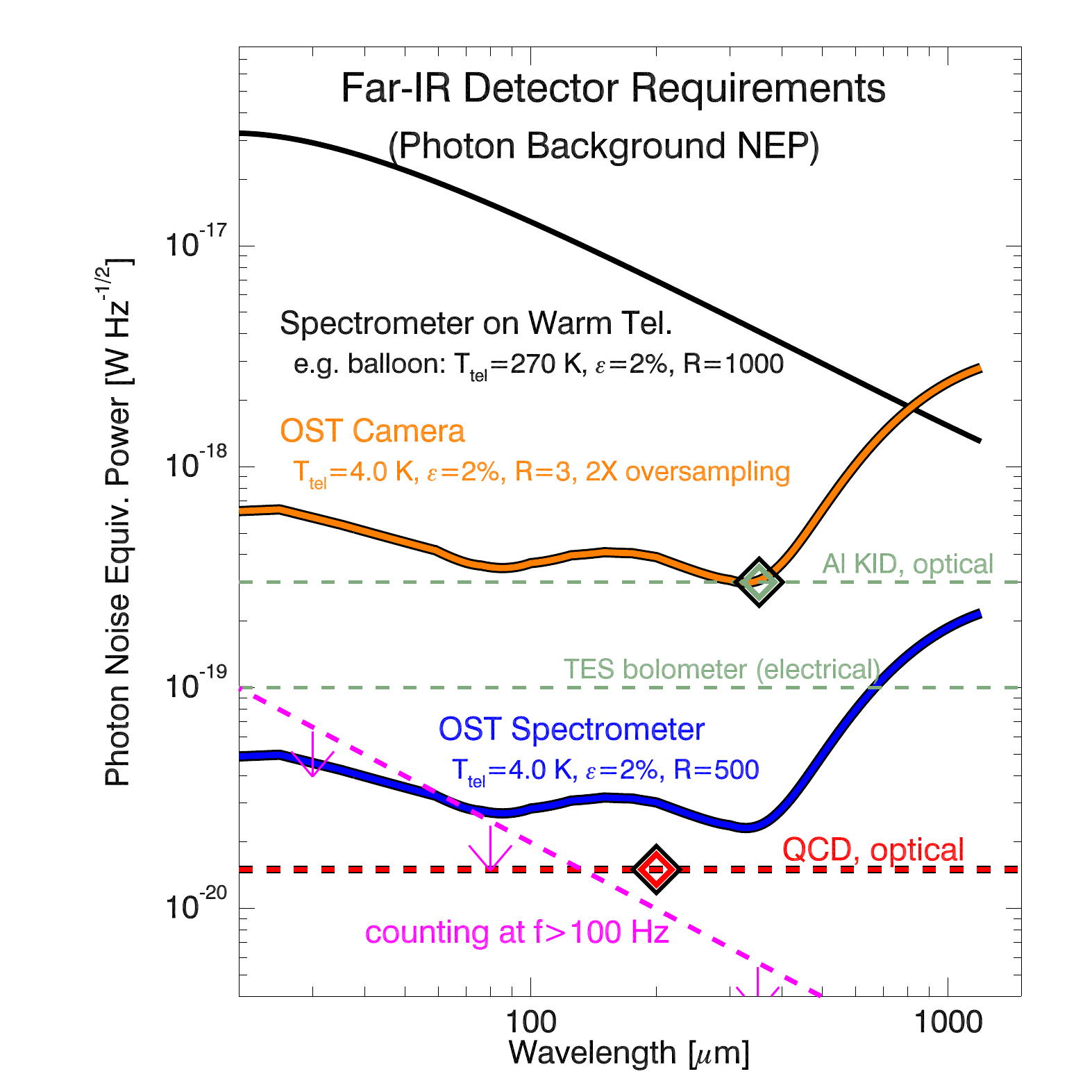}
\caption[Sensitivity requirements to meet photon background levels in the far-infrared]{Detector sensitivity requirements to meet photon background levels in the far-infrared. With a cryogenic space telescope, the fundamental limits are the zodiacal dust and galactic cirrus emission, and the photon noise level scales as the square root of bandwidth. Of particular interest is the requirement for moderate-resolution dispersive spectroscopy (blue). Also shown are detector sensitivity measurements for the TES, KIDS and QCD technologies described in \S\ref{directdetect}. The magenta dotted line shows the photon counting threshold at 100\,Hz: a device which can identify individual photons at this rate (photon counting) at high efficiency is limited by the dark counts rate rather than classical NEP.} 
\label{fig:detnep} 
\end{figure*}

\begin{figure*}
\includegraphics[width=16cm,angle=0]{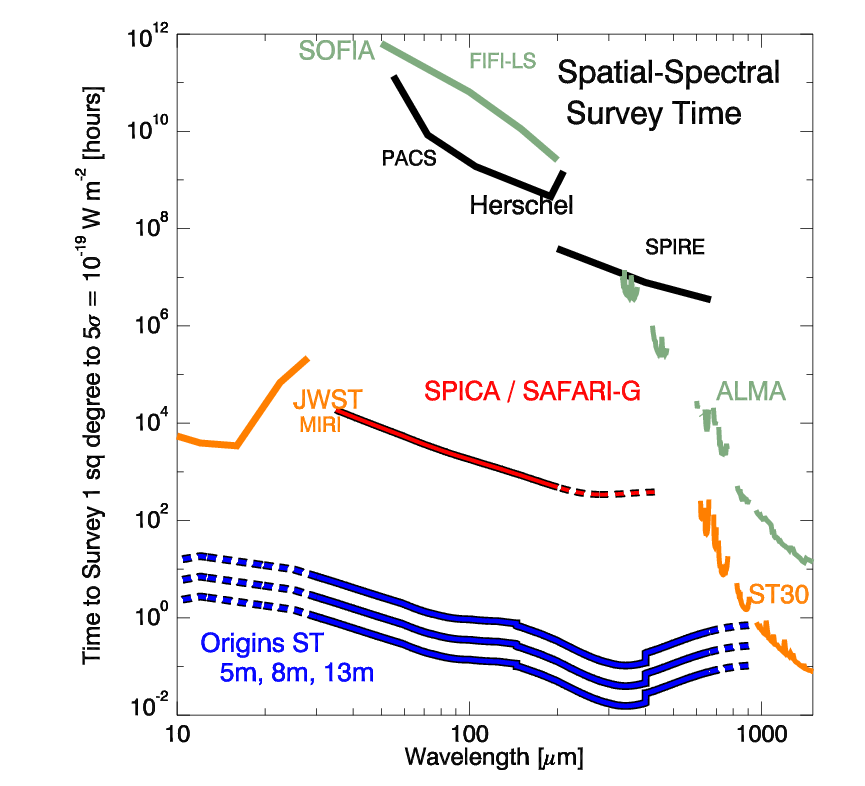}
\caption[A comparison of far-infrared discovery speeds as a function of wavelength]{A comparison of the times required to perform a blank-field spatial-spectral survey reaching a depth of 10$^{-19}\,\rm W\,m^{-2}$ over one square degree, as a function of wavelength, for various facilities. This figure uses current estimates for sensitivity, instantaneous bandwidth covered, telescope overheads, and instantaneous spatial coverage on the sky. The OST curves assume $R=500$ grating spectrometers with $60-200$ beams (depending on wavelength), 1:1.5 instantaneous bandwidth. Pixels are assumed to operate with a NEP of $2\times10^{-20}\,$W\,Hz$^{-1/2}$. The SPICA/SAFARI-G curve is for a 2.5-m telescope with $R=300$ grating spectrometer modules with 4 spatial beams, and detector arrays with a NEP of $2\times10^{-19}\,$W\,Hz$^{-1/2}$. ST30 is a ground-based 30-m telescope with 100 spectrometer beams, each with 1:1.5 bandwidth, ALMA band-averaged sensitivity, and survey speed based on 16\,GHz bandwidth in the primary beam.} 
\label{fig:speed} 
\end{figure*}

It is notable, however, that the performance of many past and present facilities is limited by thermal emission from telescope optics (Figure \ref{fig:backgr}). The comparison between infrared telescopes operating at 270\,K vs. temperatures of a few kelvins is analogous to the comparison between the sky brightness during the day and at night in the optical. Even with {\itshape Herschel} and its $\sim$85\,K telescope, the telescope emission was the dominant noise term for both its Photodetector Array Camera and Spectrometer (PACS\citep{pog10}) and Spectral and Photometric Imaging REceiver (SPIRE\citep{gri10}). Thus, the ultimate scientific promise of the far-infrared is in orbital missions with actively-cooled telescopes {\itshape and} instruments. Cooling the telescope to a few kelvins effectively eliminates its emission through most of the far-infrared band. When combined with appropriate optics and instrumentation, this results in orders-of-magnitude improvement in discovery speed over what is achievable from atmospheric-based platforms (Figures \ref{fig:detnep} \& \ref{fig:speed}). A ``cold'' telescope can bring sensitivities at observed-frame $30-500\,\mu$m into parity with those at shorter (JWST) and longer (ALMA) wavelengths.

A further limiting factor is source confusion - the fluctuation level in image backgrounds below which individual sources can no longer be detected. Unlike instrument noise, confusion noise cannot be reduced by increasing integration time. Source confusion can arise from both smooth diffuse emission and fluctuations on scales smaller than the beamsize of the telescope. Outside of the plane of the Milky Way, the primary contributors to source confusion are structure in Milky Way dust emission, individually undetected extragalactic sources within the telescope beam, and individually detected sources that are blended with the primary source. Source confusion is thus a strong function of the location on the sky of the observations, the telescope aperture, and observed wavelength. Source confusion is a concern for all previous and current single-aperture infrared telescopes, especially space-based facilities whose apertures are modest compared to ground-based facilities. A summary of the confusion limits of some previous infrared telescopes is given in Figure \ref{fig:confnoise}. 

A related concept is line confusion, caused by the blending and overlapping of individual lines in spectral line surveys. While this is barely an issue in e.g. H\,I surveys as the 21\,cm H\,I line is bright and isolated\citep{jonesmg16}, it is potentially a pernicious source of uncertainty at far-infrared wavelengths, where there are a large number of bright spectral features. This is true in galactic studies\citep{terce10} and in extragalactic surveys. Carefully chosen spatial and spectral resolutions are required to minimize line confusion effects\citep{kogut15}.

Several approaches have been adopted to extract information on sources below the standard confusion limit. They include; novel detection methods applied to single-band maps\citep{knud06}, the use of prior positional information from higher spatial resolution images to deconvolve single far-infrared sources\citep{rose10,safar15}, and combining priors on positions with priors from spectral energy distribution modelling\citep{macken16,hurl17}. Finally, the spatial-spectral surveys from upcoming facilities such as SAFARI on SPICA or the OSS on the OST should push significantly below the classical confusion limit by including spectral information to break degeneracies in the third spatial dimension\citep{raym10}. 

There are two further challenges that confront space-based far-infrared observatories, which are unfamiliar to sub-orbital platforms:
\vspace{0.1cm}

\noindent {\bfseries Dynamic range}: In moving to ``cold'' telescopes, sensitivity is limited only by the far-infrared sky background. We enter a regime where the dominant emission arises from the sources under study, and the sky has genuinely high contrast. This imposes a new requirement on the detector system - to observe the full range of source brightnesses - that is simple from sub-orbital platforms but challenging for cooled space-based platforms, since the saturation powers of currently proposed high-resolution detector arrays are within $\sim2$ orders of magnitude of their Noise Equivalent Powers (NEP \footnote{The Noise Equivalent Power (NEP) is, briefly, the input signal power that results in a signal-to-noise ratio of unity in a 1\,Hz bandwidth - the minimum detectable power per square root of bandwidth. Thus, a lower NEP is better. In-depth discussions of the concept of NEP can be found in \citep{lamarre86,richards94bol,benford98}.}). This would limit observations to relatively faint sources. Dynamic range limitations were even apparent for previous-generation instruments such as the Multiband Imaging Photometer onboard {\itshape Spitzer} and PACS onboard {\itshape Herschel}, with saturation limits at 70\,$\mu$m of 57\,Jy and 220\,Jy, respectively. Thus, we must either design detector arrays with higher dynamic range, or populate the focal plane with detector arrays, each suited to part of the range of intensities.  
\vspace{0.1cm}

\noindent {\bfseries Interference}: The susceptibility of cooled detector arrays to interference from ionizing radiation in space was noted in the development of microcalorimeter arrays for X-ray telescopes \citep{stahl99,stahl04,saab04}. Moreover, this susceptibility was clearly demonstrated by the bolometers on {\itshape Planck}. An unexpectedly high rate and magnitude of ionizing radiation events were a major nuisance for this mission, requiring corrections to be applied to nearly all of the data. Had this interference been a factor of $\sim2$ worse, it would have caused significant loss of science return from {\itshape Planck}. Techniques are being developed and demonstrated to mitigate this interference for X-ray microcalorimeters by the addition of a few micron thick layer of gold on the back of the detector frame. It is likely that a similar approach can mitigate interference in high-resolution far-infrared detector arrays as well. Moreover, work on reducing interference in far-infrared detector arrays is being undertaken in the SPACEKIDS program (\S\ref{ssect:kids}).
\vspace{0.1cm}

NASA, the European Space Agency (ESA), and the Japan Aerospace Exploration Agency (JAXA), in collaboration with astronomers and technologists around the world, are studying various options for cryogenic space observatories for the far-infrared. There are also opportunities to broaden the far-infrared astrophysics domain to new observing platforms. We give an overview of these space-based observing platforms in the following sections. We do not address the James Webb Space Telescope, as comprehensive overviews of this facility are given elsewhere\citep{gard06}. We also do not review non-US/EU projects such as Millimetron/Spektr-M\citep{smirn12,kard14}.

\begin{figure}
\begin{center}
\includegraphics[width=8cm,angle=0]{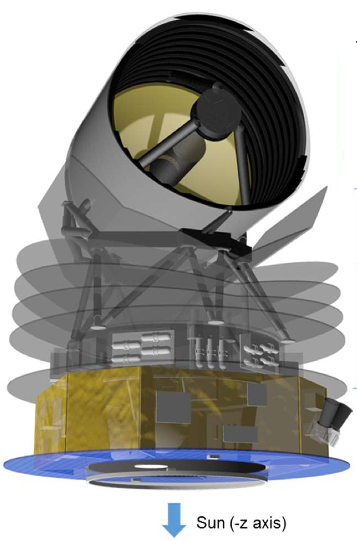}
\caption[Satellite concept - SPICA]{A concept image for the proposed SPICA satellite (\S\ref{sec:spica}).} 
\label{fig:satconceptspica}
\end{center} 
\end{figure}

\subsection{The Space Infrared Telescope for Cosmology and Astrophysics}\label{sec:spica}

First proposed by JAXA scientists in 1998, the Space Infrared Telescope for Cosmology and Astrophysics (SPICA\citep{naka98,nakagawa2004spica,swinyard2009space,naka17,sibth15}) garnered worldwide interest due to its sensitivity in the mid- and far-infrared, enabled by the combination of the actively-cooled telescope and the sensitive far-infrared detector arrays. Both ESA and JAXA have invested in a concurrent study, and an ESA-JAXA collaboration structure has gelled. ESA will provide the 2.5-m telescope, science instrument assembly, satellite integration and testing, and the spacecraft bus. JAXA will provide the passive and active cooling system (supporting a telescope cooled to below 8\,K), cryogenic payload integration, and launch vehicle.  JAXA has indicated commitment to their portion of the collaboration, and the ESA selected SPICA as one of the 3 candidates for the Cosmic Visions M5 mission. The ESA phase-A study is underway now, and the downselect among the 3 missions will occur in 2021.  Launch is envisioned for 2031. An example concept design for SPICA is shown in Figure \ref{fig:satconceptspica}.

SPICA will have three :instruments. JAXA’s SPICA mid-infrared instrument (SMI) will offer imaging and spectroscopy from $12$ to $38\,\mu$m. It is designed to complement JWST-MIRI with wide-field mapping (broad-band and spectroscopic), R$\sim$30,000 spectroscopy with an immersion grating, and an extension to $38\,\mu$m with antimony-doped silicon detector arrays.  A polarimeter from a French-led consortium will provide dual-polarization imaging in 2-3 bands using high-impedance semiconductor bolometers similar to those developed for {\itshape Herschel}-PACS, but modified for the lower background and to provide differential polarization. A sensitive far-infrared spectrometer, SAFARI, is being provided by an SRON-led consortium \citep{roelf14,past16}. It will provide full-band instantaneous coverage over $35-230\,\mu$m, with a longer wavelength extension under study, using four $R=300$ grating modules. A Fourier-transform module which can be engaged in front of the grating modules will offer a boost to the resolving power, up to R=3000. A US team is working in collaboration with the European team and aims to contribute detector arrays and spectrometer modules to SAFARI\citep{brad17} through a NASA Mission of Opportunity.

\begin{figure*}
\centering
\includegraphics[width=14cm,angle=0]{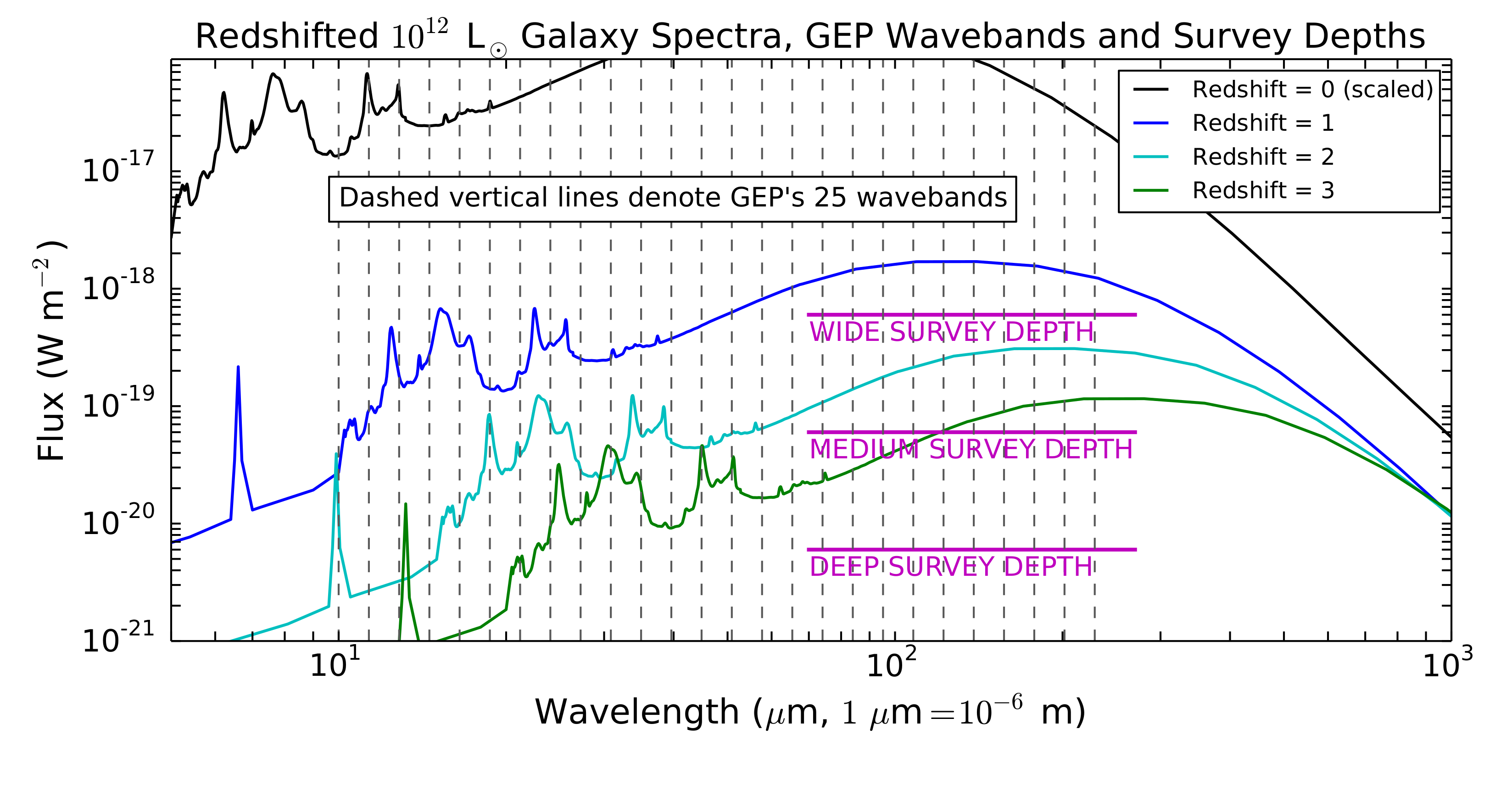}
\caption[Photometric redshifts in the infrared from the Galaxy Evolution Probe]{A mid/far-infrared galaxy spectrum, the GEP photometric bands, and notional survey depths.  The spectrum is a model of a star-forming galaxy\citep{dale02} exhibiting strong PAH features and far-infrared dust continuum emission. The black spectrum is the galaxy at a redshift of $z = 0$, but scaled vertically by a luminosity distance corresponding to $z = 0.1$ to reduce the plot range. The same spectrum is shown at redshifts $z = 1$, 2, and 3.  The vertical dashed lines mark the GEP photometric bands. As the galaxy spectrum is redshifted, the PAH features move through the bands, enabling photometric redshift measurements. This figure does not include the effects of confusion noise.} 
\label{fig:probegep} 
\end{figure*}

\subsection{Probe-class Missions}
Recognizing the need for astronomical observatories beyond the scope of Explorer class missions but with a cadence more rapid than flagship observatories such as the Hubble Space Telescope (HST), JWST, and the Wide Field Infrared Survey Telescope (WFIRST), NASA announced a call for Astrophysics Probe concept studies in 2017. Ten Probe concepts were selected in Spring 2017 for 18-month studies. Probe study reports will be submitted to NASA and to the Astro 2020 Decadal Survey to advocate for the creation of a Probe observatory line, with budgets of \$400M to \$1B.  

Among the Probe concepts under development is the far-infrared Galaxy Evolution Probe (GEP), led by the University of Colorado Boulder and the Jet Propulsion Laboratory. The GEP concept is a two-meter-class, mid/far-infrared observatory with both wide-area imaging and followup spectroscopy capabilities. The primary aim of the GEP is to understand the roles of star formation and black hole accretion in regulating the growth of stellar and black hole mass. In the first year of the GEP mission, it will detect $\geq10^{6}$ galaxies, including $\gtrsim10^5$ galaxies at $z>3$, beyond the peak in redshift of cosmic star formation, by surveying several hundred square degrees of the sky. A unique and defining aspect of the GEP is that it will detect galaxies by bands of rest-frame mid-infrared emission from polycyclic aromatic hydrocarbons (PAHs), which are indicators of star formation, while also using the PAH emission bands and silicate absorption bands to measure photometric redshifts. 

The GEP will achieve these goals with an imager using approximately 25 photometric bands spanning $10\,\mu$m to at least $230\,\mu$m, giving a {\itshape spectral} resolution of $R \simeq 8$ (Figure \ref{fig:probegep}). Traditionally, an imager operating at these wavelengths on a 2-m telescope would be significantly confusion-limited, especially at the longer wavelengths (see e.g. the discussion in the introduction to \S\ref{spaceobs}. However, the combination of many infrared photometric bands, and advanced multi-wavelength source extraction techniques, will allow the GEP to push significantly below typical confusion limits. The GEP team is currently simulating the effects of confusion on their surveys, with results expected in early 2019. The imaging surveys from the GEP will thus enable new insights into the roles of redshift, environment, luminosity and stellar mass in driving obscured star formation and black hole accretion, over most of the cosmic history of galaxy assembly. 

In the second year of the GEP survey, a grating spectrometer will observe a sample of galaxies from the first-year survey to identify embedded AGN. The current concept for the spectrometer includes four or five diffraction gratings with $R \simeq 250$, and spectral coverage from $23\,\mu$m to at least $190\,\mu$m. The spectral coverage is chosen to enable detection of the high-excitation [NeV] $24.2\,\mu$m line, which is an AGN indicator, over $0<z<3.3$, and the [OI] $63.2\,\mu$m line, which is predominantly a star formation indicator, over $0 < z \lesssim 2$.

Recent advances in far-infrared detector array technology have made an observatory like the GEP feasible. It is now possible to fabricate large arrays of sensitive kinetic inductance detectors (KIDs, see \S\ref{ssect:kids}) that have a high frequency multiplex factor. The GEP concept likely will employ Si BIB arrays (similar to those used on JWST-MIRI) for wavelengths from $10\,\mu$m to $24\,\mu$m and KIDs at wavelengths longer than $24\,\mu$m.  Coupled with a cold ($\sim4\,$K) telescope, such that the GEP’s sensitivity would be photon-limited by astrophysical backgrounds (Figure \ref{fig:backgr}), the GEP will detect the progenitors of Milky Way-type galaxies at $z = 2$ ($\geq10^{12}$\,L$_{\odot}$). Far-infrared KID sensitivities have reached the NEPs required for the GEP imaging to be background limited ($3\times10^{-19}$ W / Hz$^{-0.5}$\citep{Baselmans2016,bueno2017full}) although they would need to be lowered further, by a factor of at least three, for the spectrometer to be background limited. The GEP would serve as a pathfinder for the Origins Space Telescope (\S\ref{origins}), which would have a greater reach in redshift by virtue of its larger telescope. SOFIA and balloons will also serve as technology demonstrators for the GEP and OST. 

The technology drivers for the GEP center on detector array size and readout technology. While KID arrays with $10^4 - 10^5$ pixels are within reach, investment must be made in development of low power-consumption readout technology (\S\ref{readschemes}). Large KID (or other direct detection technology) arrays with low-power readouts on SOFIA and balloons would raise their respective TRLs, enabling the GEP and OST.

\subsection{The Origins Space Telescope}\label{origins}
As part of preparations for the 2020 Decadal Survey, NASA is supporting four studies of flagship astrophysics missions. One of these studies is for a far-infrared observatory. A science and technology definition team (STDT) is pursuing this study with support from NASA GSFC. The STDT has settled on a single-dish telescope, and coined the name ``Origins Space Telescope'' (OST). The OST will trace the history of our origins, starting with the earliest epochs of dust and heavy element production, through to the search for extrasolar biomarkers in the local universe. It will answer textbook-altering questions, such as: ``How did the universe evolve in response to its changing ingredients?'' and ``How common are planets that support life?''

\begin{figure}
\begin{center}
\includegraphics[width=8cm,angle=0]{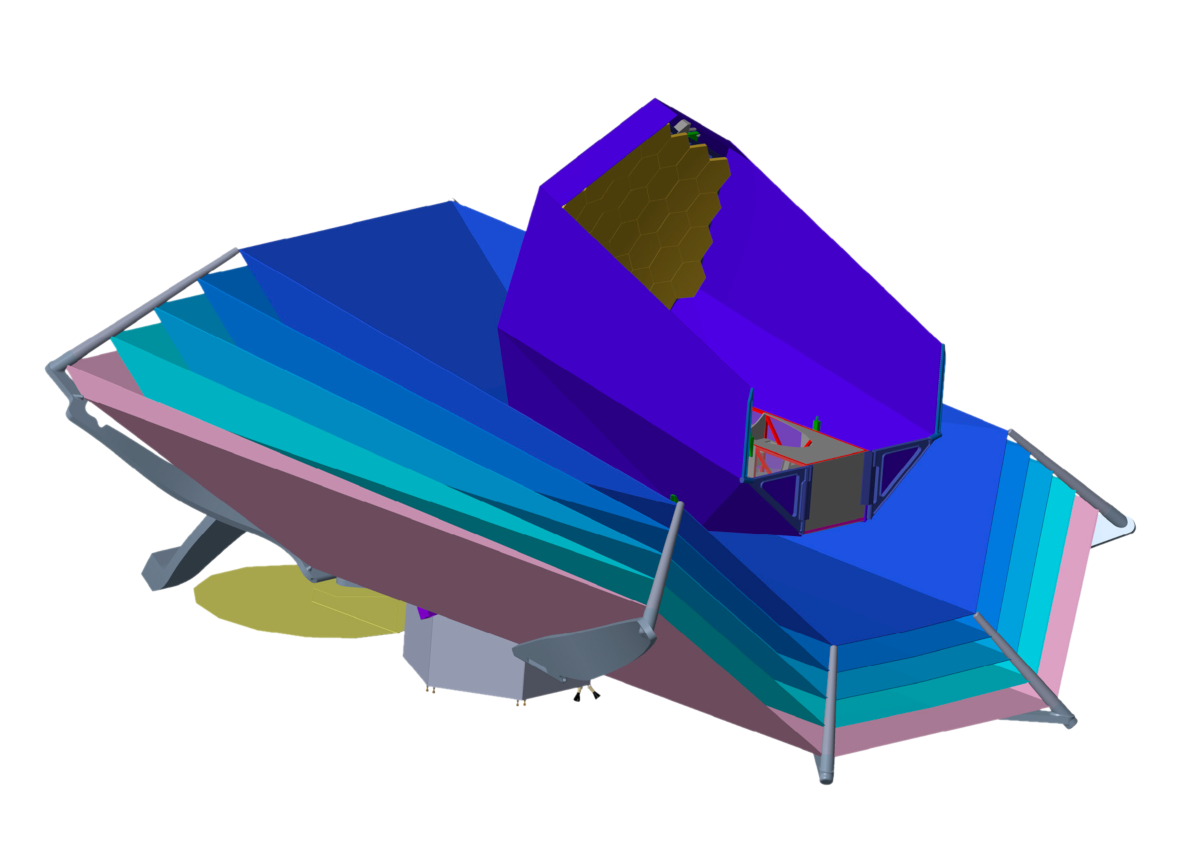}
\caption[Satellite concept - OST]{A concept image for the proposed Origins Space Telescope (OST, \S\ref{origins}). This image shows a design for the more ambitious ``Concept 1''. The design includes nested sunshields and a boom, in which the instrument suite is located. The color coding of the image gives a qualitative indication of telescope temperature.} 
\label{fig:satconceptost}
\end{center} 
\end{figure}

Two concepts for the OST are being investigated, based on an Earth-Sun L2 orbit, and a telescope and instrument module actively cooled with 4\,K-class cryocoolers. Concept 1 (Figure \ref{fig:satconceptost}) has an open architecture, like that of JWST. It has a deployable segmented 9-m telescope with five instruments covering the mid-infrared through the submillimeter. Concept 2 is smaller and simpler, and resembles the Spitzer Space Telescope architecturally. It has a 5.9-m diameter telescope (with the same light collecting area as JWST) with no deployable components. Concept 2 has four instruments, which span the same wavelength range and have comparable spectroscopic and imaging capabilities as the instruments in Concept 1.

Because OST would commence in the middle of the next decade, improvements in far-infrared detector arrays are anticipated, both in per-pixel sensitivity and array format, relative to what is currently mature for spaceflight (\S\ref{directdetect}). Laboratory demonstrations, combined with initial OST instrument studies which consider the system-level readout requirements, suggest that total pixel counts of 100,000 to 200,000 will be possible, with each pixel operating at the photon background limit. This is a huge step forward over the $3200$ pixels total on {\itshape Herschel} PACS and SPIRE, and the $\sim4000$ pixels anticipated for SPICA.

The OST is studying the impact of confusion on both wide and deep survey concepts. Their approach is as follows. First, a model of the far-infrared sky is used to generate a three-dimensional hyperspectral data cube. Each slice of the cube is then convolved with the telescope beam, and the resulting cube is used to conduct a search for galaxies with no information given on the input catalogs. Confusion noise is then estimated by comparing the input galaxy catalog to the recovered galaxy catalog. The results from this work are not yet available, but this approach is a significant step forward in robustness compared to prior methods\citep{kogut15}.

\subsection{CubeSats}\label{ssect:cube}
CubeSats are small satellites built in multiples of 1U (10\,cm $\times$ 10\,cm $\times$ 10\,cm, $<$1.33 kg). Because they are launched within containers, they are safe secondary payloads, reducing the cost of launch for the payload developer. In addition, a large ecosystem of CubeSat vendors and suppliers is available, which further reduces costs. CubeSats thus provide quick, affordable access to space, making them attractive technology pathfinders and risk mitigation missions towards larger observatories. Moreover, according to a 2016 National Academies report\citep{NAP23503}, CubeSats have demonstrated their ability to perform high-value science, especially via missions to make a specific measurement, and/or that complement a larger project. To date, well over 700 CubeSats have been launched, most of them 3U's.  

Within general astrophysics, CubeSats can produce competitive science, although the specific area needs to be chosen carefully\citep{ardila2017,shk18}. For example, long-duration pointed monitoring is a unique niche. So far the Astrophysics division within NASA's Science Mission Directorate has funded four CubeSat missions: in $\gamma$-rays (BurstCube, \cite{perkins2018}), X-rays (HaloSat, \cite{kaar16}), and in the ultraviolet (SPARCS, \cite{shkolnik2018}; CUTE, \cite{fleming2018}). 

For the far-infrared, the CubeSat technology requirements are daunting. Most far-infrared detectors require cooling to reduce the thermal background to acceptable levels, to 4\,K or even 0.1\,K, although CubeSats equipped with Schottky-based instruments that do not require active cooling may be sufficiently sensitive for certain astronomical and Solar System applications (see also e.g. \citep{jos18}). CubeSat platforms are thus constrained by the lack of low-power, high efficiency cryocoolers. Some applications are possible at 40\,K, and small Stirling coolers can provide 1\,W of heat lift at this temperature (see also \S\ref{ssect:cryoc}). However, this would require the majority of the volume and power budget of even a large CubeSat (which typically have total power budgets of a few tens of watts), leaving little for further cooling stages, electronics, detector systems, and telescope optics.  

CubeSats are also limited by the large beam size associated with small optics. A diffraction-limited $10$\,cm aperture operating at $100\,\mu$m would have a beam size of about $3.5\arcmin$. There are concepts for larger, deployable apertures \citep{agasid2013}, up to $\sim$20\,cm, but none have been launched. 

\begin{figure}
\includegraphics[width=8cm,angle=0]{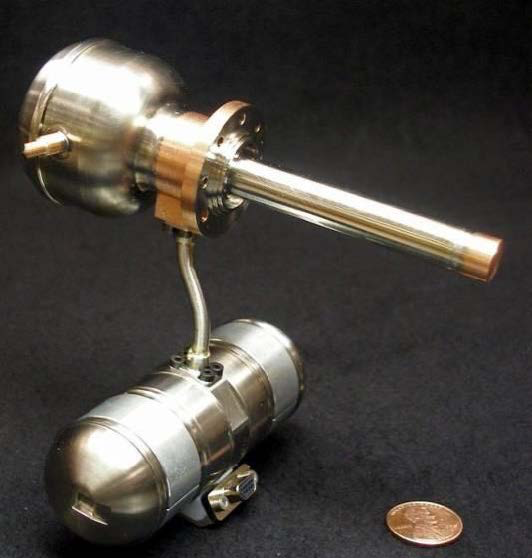}
\caption[A example of a 120\,K cryocooler for a CubeSat]{The Lockheed Martin Coaxial Micro Pulse Tube Cryocooler, which will provide cooling to 120\,K for the CubeSat Infrared Atmospheric Sounder (CIRAS), scheduled for launch in 2019 \citep{pag16}. This cooler weighs less than 0.4\,kg, and has reached TRL $\geq6$.} 
\label{fig:lm_cryo} 
\end{figure}

For these reasons, it is not currently feasible to perform competitive far-infrared science with CubeSats. However, CubeSats can serve to train the next generation of space astronomers, as platforms for technology demonstrations that may be useful to far-infrared astronomy, and as complements to larger observing systems. For example, the CubeSat Infrared Atmospheric Sounder (CIRAS) is an Earth Observation 6U mission with a $4.78 - 5.09\,\mu$m imaging spectrograph payload. The detector array will be cooled to 120\,K, using a Lockheed Martin Coaxial MPT Cryocooler, which provides a 1\,W heat lift (Figure \ref{fig:lm_cryo}). At longer wavelengths, the Aerospace Corporation's CUMULOS \citep{ardila2016} has demonstrated $8-15\,\mu$m Earth imaging with an uncooled bolometer from a CubeSat. CubeSats can also serve as support facilities. In the sub-millimeter range, {\itshape CalSat} uses a CubeSat as a calibration source for CMB polarization observatories \cite{johnson15}.

\subsection{The International Space Station}\label{ssect:iss}
The International Space Station (ISS) is a stable platform for both science and technology development. Access to the ISS is currently provided to the US astronomical community through Mission of Opportunity calls which occur approximately every two years and have $\sim\$60$M cost caps. Several payload sites are available for hosting US instruments, with typically $1 \,$m$^{3}$ of volume, at least 0.5 and up to 6\,kW of power, wired and wireless ethernet connectivity, and at least 20\,kbps serial data bus downlink capability\citep{ISSPropGuide}.  

In principle, the ISS is an attractive platform for astrophysics, as it offers a long-term platform at a mean altitude of 400\,km, with the possibility for regular instrument servicing. Infrared observatories have been proposed for space station deployment at least as far back as 1990\citep{brown90}. There are however formidable challenges in using the ISS for infrared astronomy. The ISS environment is, for infrared science, significantly unstable, with sixteen sunrises every 24 hour period, ``glints'' from equipment near the FoV, and vibrations and electromagnetic fields from equipment in the ISS. Furthermore, the  external instrument platforms are not actively controlled, and are subject to various thermal instabilities over an orbit, which would require active astrometric monitoring. 

Even with these challenges, there are two paths forward for productive infrared astronomy from the ISS:

\begin{itemize}

\item For hardware that can tolerate and mitigate the dynamic environment of the ISS, there is ample power and space for the deployment of instruments, potentially with mission lifetimes of a year or more. Example applications that may benefit from this platform include monitoring thermal emission from interplanetary dust, or time domain astronomy.

\item The long-term platform, freely available power, and opportunities for direct servicing by astronauts, make the ISS an excellent location to raise TRLs of technologies so that they can be deployed on other space-based platforms.

\end{itemize}

\noindent Efforts thus exist to enable infrared observing from the ISS. For example, the Terahertz Atmospheric/Astrophysics Radiation Detection in Space (TARDiS) is a proposed infrared experiment that will observe both in the upper atmosphere of Earth, and in the ISM of the Milky Way.

\section{New Instruments and Methods}\label{sect:newmod}
Continuing advances in telescope and detector technology will enable future-generation observatories to have much greater capabilities than their predecessors. Technological advancement also raises the possibility of new observing techniques in the far-infrared, with the  potential for transformational science. We discuss two such techniques in this section; interferometry, and time-domain astronomy.

\subsection{Interferometry}\label{ssect:firint}
\begin{figure*}
\includegraphics[width=16cm,angle=0]{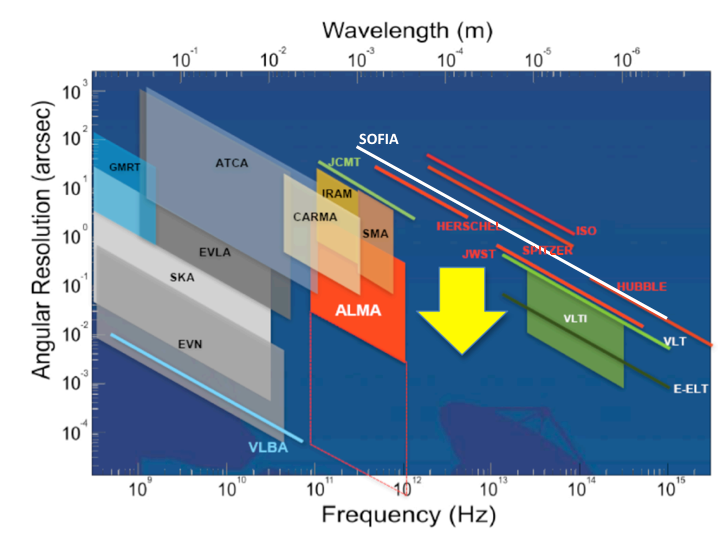}
\caption[The far-infrared resolution ``gap'', and the potential for far-infrared interferometry]{The angular resolutions of selected facilities as a function of wavelength. Very high spatial resolutions are achievable at millimeter to radio wavelengths using ground-based interferometers, while current and next-generation large-aperture telescopes can achieve high spatial resolutions in the optical and near-infrared. In the mid/far-infrared however the best achievable spatial resolutions still lag several orders of magnitude behind those achievable at other wavelengths. Far-infrared interferometry from space will remedy this, providing an increase in spatial resolution shown by the yellow arrow. A version of this figure originally appeared in the FISICA (Far-Infrared Space Interferometer Critical Assessment) report, courtesy of Thijs de Graauw.} 
\label{fig:firresgap} 
\end{figure*}

Most studies of future far-infrared observatories focus on single-aperture telescopes. There is however enormous potential for interferometry in the far-infrared (Figure \ref{fig:firresgap}). Far-infrared interferometry is now routine from the ground (as demonstrated by ALMA, NOEMA, and the SMA), but has been barely explored from space- and balloon-based platforms. However, the combination of access to the infrared without atmospheric absorption and angular resolutions that far exceed those of any single-aperture facility, enables entirely new areas of investigation\citep{sauvage2013sub,leis13,juanola2016far}.

In our solar system, far-infrared interferometry can directly measure the emission from icy bodies in the Kuiper belt and Oort cloud. Around other stars, far- infrared interferometry can probe planetary disks to map the spatial distribution of water, water ice, gas, and dust, and search for structure caused by planets. At the other end of the scale, far-infrared interferometry can measure the rest-frame near/mid-infrared emission from high-redshift galaxies without the information-compromising effects of spatial confusion. This was recognized within NASA's 2010 long-term roadmap for Astrophysics, {\it Enduring Quests/Daring Visions} \citep{endques14}, which stated that, within the next few decades, scientific goals will begin to outstrip the capabilties of single aperture telescopes. For example, imaging of exo-Earths, determining the distribution of molecular gas in protoplanetary disks, and directly observing the event horizon of a black hole all require single aperture telescopes with diameters of hundreds of meters, over an order of magnitude larger than is currently possible. Conversely, interferometry can provide the angular resolution needed for these goals with much less difficulty. 

Far-infrared interferometry is also an invaluable technology development platform. Because certain technologies for interferometry, such as ranging accuracy, are more straightforward for longer wavelengths, far-infrared interferometry can help enable interferometers operating in other parts of the electromagnetic spectrum\footnote{Interferometer technology has however been developed for projects outside the infrared; examples include the Keck Interferometer, CHARA, LISA Pathfinder, and the Terrestrial Planet Finder, as well as several decades of work on radio interferometry.}. This was also recognized within {\itshape Enduring Quests/Daring Visions}: ``...the technical requirements for interferometry in the far-infrared are not as demanding as for shorter wavelength bands, so far-infrared interferometry may again be a logical starting point that provides a useful training ground while delivering crucial science.'' Far-infrared interferometry thus has broad appeal, beyond the far-infrared community, as it holds the potential to catalyze development of space-based interferometry across multiple wavelength ranges.

\begin{figure*}
\centering
\includegraphics[width=12cm,angle=0]{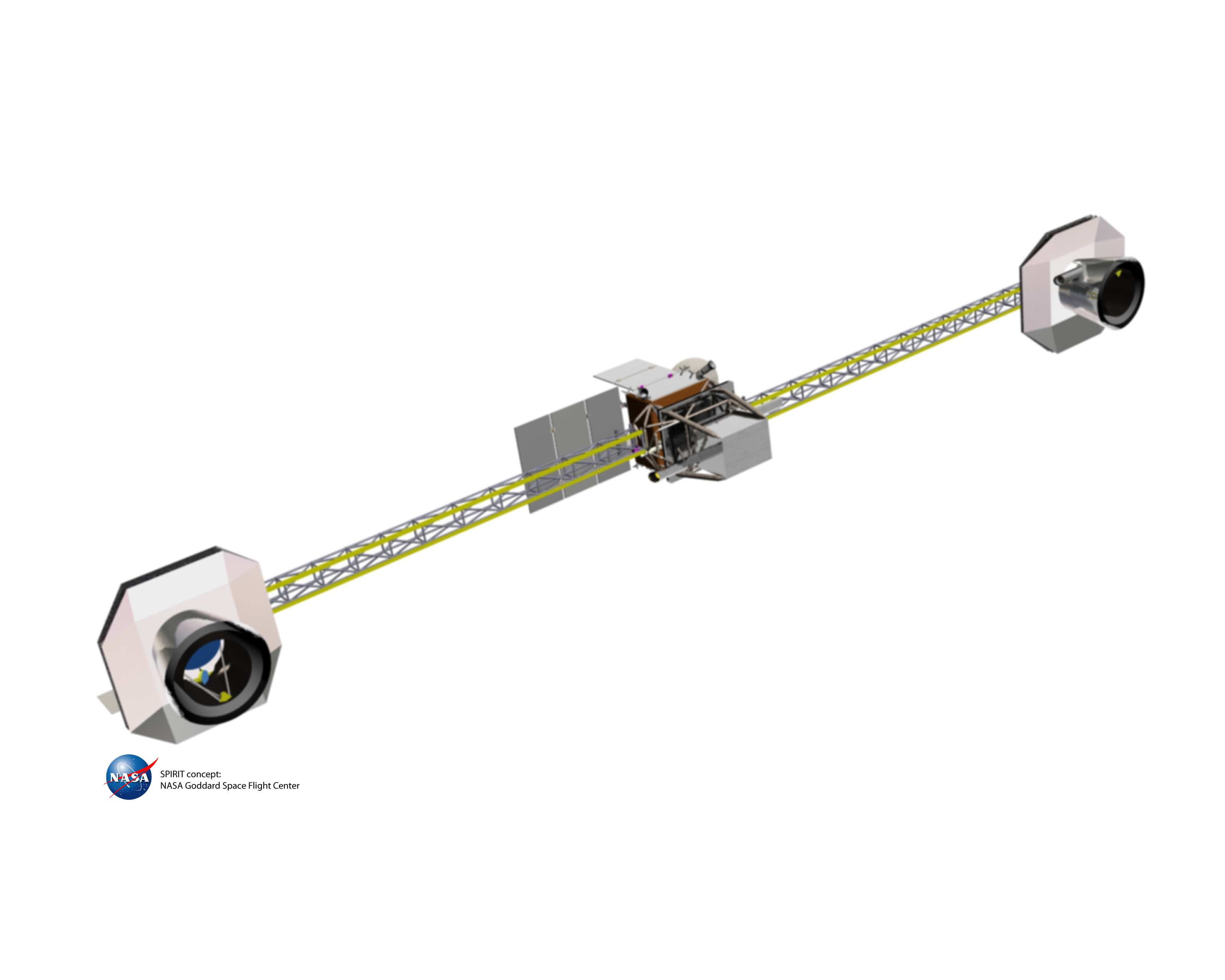}
\vspace{-1cm}
\caption[SPIRIT as an example of a structurally connected interferometer]{The SPIRIT structurally connected interferometer concept \citep{leis07}. SPIRIT is a spatio-spectral ``double Fourier'' interferometer that has been developed to Phase A level. SPIRIT has sub-arcsecond resolution at $100\,\mu$m, along with $R\sim4,000$ spectral resolution. The maximum interferometric baseline is 36\,m.} 
\label{fig:firinter} 
\end{figure*}

The 2000 Decadal Survey\citep{NAP9839} recommended development of a far-infrared interferometer, and the endorsed concept (the Submillimeter Probe of the Evolution of Cosmic Structure: SPECS) was subsequently studied as a ``Vision Mission'' \citep{har06}. Recognizing that SPECS was extremely ambitious, a smaller, structurally-connected interferometer was studied as a potential Origins Probe -- the Space Infrared Interferometric Telescope (SPIRIT\citep{leis07}, Figure \ref{fig:firinter}). At around the same time, several interferometric missions were studied in Europe, including FIRI\citep{helm09} and the heterodyne interferometer ESPRIT\citep{wild08}. Another proposed mission, TALC\citep{dur14,sauv16talc}, is a hybrid between a single-aperture telescope and an interferometer and thus demonstrates technologies for a structurally connected interferometer. There are also concepts using nanosats\citep{dohlena2014design}. Recently, the European community carried out the Far-Infrared Space Interferometer Critical Assessment (FP7-FISICA), resulting in a design concept for the Far-Infrared Interferometric Telescope (FIRIT). Finally, the ``double Fourier'' technique that would enable simultaneous high spatial and spectral observations over a wide FoV is maturing through laboratory experimentation, simulation, and algorithm development\citep{elias07,leis12,grain12,7460625,brack16}.

Two balloon payloads have been developed to provide scientific and technical demonstration of interferometry. They are the Far-Infrared Interferometric Telescope Experiment (FITE\citep{shib10}), and the Balloon Experimental Twin Telescope for Infrared Interferometry (BETTII\citep{rine14}), first launched in June 2017. The first BETTII launch resulted in a successful engineering flight, demonstrating nearly all of the key systems needed for future science flights. Unfortunately, an anomaly at the end of the flight resulted in complete loss of the payload. A rebuilt BETTII should fly before 2020. 

Together, BETTII and FITE will serve as an important development step towards future space-based interferometers, while also providing unique scientific return. Their successors, taking advantage of many of the same technologies as other balloon experiments (e.g. new cryocoolers, lightweight optics), will provide expanded scientific capability while continuing the path towards space-based interferometers.

Far-infrared interferometers have many of the same technical requirements as their single aperture cousins. In fact, an interferometer could be used in ``single aperture'' mode, with instruments similar to those on a single aperture telescope. However, in interferometric mode, the development requirements for space-based far-infrared interferometry are:

\begin{itemize}

\item Detailed simulations, coupled with laboratory validation, of the capabilities of interferometers. For example, imaging with an interferometer is sometimes assumed to require full coverage of the synthetic aperture; however, for many science cases, partial coverage (akin to coverage of ground-based radio interferometers) may be sufficient. 

\item High speed detector arrays are desirable for interferometry missions, to take advantage of fast-scanning techniques. 

\item Free-flying interferometers can benefit from advances in sub-newton thruster technology, as well as techniques for efficient formation flying. 

\item Structurally connected interferometers can benefit from studying deployment of connected structures and boom development.

\item Demonstration of the system-level integration of interferometers. Balloon-borne pathfinders provide an ideal platform for doing this. 

\end{itemize}

Finally, we comment on temporal performance requirements. The temporal performance requirements of different parts of an interferometer depend on several factors, including the FoV, sky and telescope backgrounds, rate of baseline change, and desired spectral resolution. We do not discuss these issues in detail here, as they are beyond the scope of a review paper. We do however give an illustrative example; a $1\arcmin$ FoV, with a baseline of 10\,m, spectral resolution of $R=100$, and 16 points per fringe, results in a readout speed requirement of 35\,Hz. However, increasing the spectral resolution to $R=1000$ (at the same scan speed) raises the readout speed requirement to 270\,Hz. These correspond to detector time constants of 17\,ms and 3\,ms. A baseline requirement for a relatively modest interferometer (e.g. SHARP-IR \citep{rine16}) is thus a detector time constant of a few milliseconds. The exact value is however tied tightly to the overall mission architecture and operations scheme.

\subsection{Time-domain \& Rapid Response Astronomy}\label{ssect:timed}
Time domain astronomy is an established field at X-ray through optical wavelengths, with notable observations including {\it Swift}'s studies of transient high-energy events, and the {\it Kepler} mission using optical photometry to detect exoplanets. Time domain astronomy in the far-infrared holds the potential for similarly important studies of phenomena on timescales of days to years; (1) searches for infrared signatures of (dust-obscured) $\gamma$-ray bursts, (2) monitoring the temporal evolution of waves in debris disks to study the earliest stages of planet formation, and (3) monitoring supernovae light curves to study the first formation stages of interstellar dust. To date however such capabilities in the far-infrared have been limited. For example, {\itshape Spitzer} was used to measure secondary transits of exoplanets\citep{dem07}, but only when the ephemeris of the target was known.

The limitations of far-infrared telescopes for time-domain astronomy are twofold. First, to achieve high photometric precision in the time domain, comparable to that provided by {\it Kepler}, the spacecraft must be extremely stable, to requirements beyond those typically needed for cameras and spectrographs. This is not a fundamental technological challenge, but the stability requirements must be taken into consideration from the earliest design phase of the observatory. Second, if the intent is to {\itshape discover} transient events in the far-infrared (rather than monitor known ones) then the FoV of the telescope must be wide, since most transient events cannot be predicted and thus must be found via observations of a large number of targets.

\section{Technology Priorities}\label{sect:firprior}
The anticipated improvements in existing far-infrared observatories, as well as the realization of next-generation space-based far-infrared telescopes, all require sustained, active development in key technology areas. We here review the following areas; direct detector arrays (\S\ref{directdetect}), medium-resolution spectroscopy (\S\ref{ssect:medresspec}), heterodyne spectroscopy (\S\ref{ssect:het}), Fabry-Perot interferometry (\S\ref{ssect:fabry}), cooling systems (\S\ref{ssect:cryoc}), and mirrors (\S\ref{ssect:mirr}). We briefly discuss a selection of other topics in \S\ref{ssect:general}.

\subsection{Direct Detector Arrays}\label{directdetect}
A key technical challenge for essentially any future far-infrared space observatory (whether single-aperture or interferometer) is the development of combined direct detector + multiplexer readout systems. These systems are not typically developed by the same industrial teams that build near- and mid-infrared devices. Instead, they are usually developed by dedicated groups at universities or national labs. These systems have two core drivers:

\begin{enumerate}
\item {\bfseries Sensitivity:} The per-pixel sensitivity should meet or exceed the photon background noise set by the unavoidable backgrounds: zodiacal light, galactic cirrus, and the microwave background (Figure \ref{fig:backgr}). An especially important target is that for moderate-resolution (R$\sim$1000) spectroscopy, for which the per-pixel NEP is 3$\times$10$^{-20}\rm\,W Hz^{-1/2}$. For the high-resolution direct detection spectrometers considered for the OST, the target NEP is $\sim10^{-21}\rm\,W Hz^{-1/2}$. A representative set of direct detector sensitivities and requirements is given in Table \ref{tbl:requirements}.
\vspace{0.1cm}

\item {\bfseries High pixel counts:} Optimal science return from a mission like the OST demands total pixel counts (in all instruments) in the range $10^{5-6}$.  This is still a small number compared with arrays for the optical and near-infrared, for which millions of pixels can be fielded in a single chip, but $\sim$100$\times$ larger than the total number of pixels on {\itshape Herschel}. Moreover, mapping speed is also influenced by the per-pixel aperture efficiency. Large, high-efficiency feedhorn systems (such as that used on {\itshape Herschel} SPIRE), can offer up to twice the mapping speed {\itshape per detector}, though such systems are slower per unit focal plane area than more closely packed horns or filled arrays \citep{griffin02}.
\end{enumerate}

\noindent There are also the challenges of interference and dynamic range (\S\ref{spaceobs}).

The world leaders in far-infrared detector technology include SRON in the Netherlands, Cambridge and Cardiff in the UK, and NASA in the USA, with at least three approaches under development. In order of technical readiness they are: 

\begin{itemize}

\item {\bfseries Superconducting transition-edge-sensed (TES) bolometers}, which have been used in space-based instruments, as well as many atmosphere-based platforms. 

\item {\bfseries Kinetic inductance detectors (KIDs)}, which have progressed rapidly, and have been used on several ground-and atmosphere-based instruments. The best KID sensitivities are comparable to TES detectors and have been demonstrated at larger (kilopixel) scales, though the sensitivities needed for spectroscopy with future large space missions remain to be demonstrated. While KIDs lead in some areas (e.g., pixel count), overall they are a few years behind TES-based systems in technological maturity.

\item {\bfseries Quantum capacitance detectors (QCDs)}, which have demonstrated excellent low-background sensitivity but at present have modest yield, and are substantially behind both TES and KID-based systems in terms of technological maturity. 

\end{itemize}

\noindent All are potentially viable for future far-infrared missions. We consider each one in turn, along with a short discussion of multiplexing.

\begin{table*}
\begin{threeparttable}[b]
\caption[Far-infrared to mm-wave detector arrays: examples and requirements]{Selected examples of sensitivities achieved by far-infrared to mm-wave detector arrays, along with some required for future missions} \label{tbl:requirements}
{\scriptsize 
\begin{tabular}{lcccccccl}
\hline
\hline
 Observatory \&                  & Waveband   & Aperture & $T_{aper}$ & $T_{det}$ & NEP                             & Detector        & Detector     & Notes        \\
 instrument                      & $\mu$m     & meters   &  K         &  K        & $\mathrm{W}/\sqrt{\mathrm{Hz}}$ & Technology      & Count        &              \\
\hline 
JCMT - SCUBA                     & 450/850    & 15       & 275        & 0.1       & $2\times 10^{-16}$              & Bolometers      & 91/36        &              \\  
JCMT - SCUBA2                    & 450/850    & 15       & 275        & 0.1       & $2\times 10^{-16}$              & TES             & 5000/5000    &              \\  
APEX - ArTeMis                   & 200-450    & 12       & 275        & 0.3       & $4.5\times 10^{-16}$            & Bolometers      & 5760         &              \\
APEX - A-MKID                    & 350/850    & 12       & 275        & 0.3       & $1\times 10^{-15}$              & KIDS            & 25,000       &              \\  
APEX - ZEUS-2                    & 200-600    & 12       & 275        & 0.1       & $4\times 10^{-17}$              & TES             & 555          & $R\sim1000$  \\  
CSO - MAKO                       & 350        & 10.4     & 275        & 0.2       & $7\times 10^{-16}$              & KIDS            & 500          & Low-\$/pix   \\
CSO - Z-Spec                     & 960-1500   & 10.4     & 275        & 0.06      & $3\times 10^{-18}$              & Bolometers      & 160          &              \\
IRAM - NIKA2                     & 1250/2000  & 30       & 275        & 0.1       & $1.7\times 10^{-17}$            & KIDS            & 4000/1000    &              \\
LMT - TolTEC                     & 1100       & 50       & 275        & 0.1       & $7.4\times 10^{-17}$            & KIDS            & 3600         & Also at 1.4\,mm, 2.1\,mm             \\
SOFIA - HAWC+                    & 40-250     & 2.5      & 240        & 0.1       & $6.6\times 10^{-17}$            & TES             & 2560         &               \\
SOFIA - HIRMES                   & 25-122     & 2.5      & 240        & 0.1       & $2.2\times 10^{-17}$            & TES             & 1024         & Low-res channel \\
BLAST-TNG                        & 200-600    & 2.5      & 240        & 0.3       & $3\times 10^{-17}$              & KIDS            & 2344         &               \\  
{\itshape Herschel} - SPIRE      & 200-600    & 3.5      & 80         & 0.3       & $4\times 10^{-17}$              & Bolometers      & 326          &               \\  
{\itshape Herschel} - PACS bol.  & 60-210     & 3.5      & 80         & 0.3       & $2\times 10^{-16}$              & Bolometers      & 2560         &              \\
{\itshape Herschel} - PACS phot. & 50-220     & 3.5      & 80         & 1.7       & $5\times 10^{-18}$              & Photoconductors & 800          & $R\sim2000$  \\  
{\itshape Planck} - HFI          & 300-3600   & 1.5      & 40         & 0.1       & $1.8\times 10^{-17}$            & Bolometers      & 54           &              \\  
SuperSpec                        & 850-1600   & --       & N/A        & 0.1       & $1.0\times 10^{-18}$            & KIDS            & $\sim10^{2}$ & $R\lesssim700$ \\  
SPACEKIDS                        & --         & --       & N/A        & 0.1       & $3\times 10^{-19}$              & KIDS            & 1000         &                \\  
\hline
SPICA - SAFARI                   & 34-210     & 3.2      & $<6$       & 0.05      & $2\times 10^{-19}$              &                 & 4000         &               \\
SPIRIT                           & 25-400     & 1.4      & $4$        & 0.05      & $1\times 10^{-19}$              &                 & $\sim10^{2}$ &       \\  
OST - imaging                    & 100-300    & 5.9-9.1  & $4$        & 0.05      & $2\times 10^{-19}$              &                 & $\sim10^{5}$ &                \\  
OST - spectroscopy               & 100-300    & 5.9-9.1  & $4$        & 0.05      & $2\times 10^{-20}$              &                 & $\sim10^{5}$ & $R\sim500$    \\  
\hline
\hline
\end{tabular}
}
 \begin{tablenotes}
 {\footnotesize 
 \item Requirements for the SPICA/SAFARI instrument are taken from \citep{jackson2012spica}. Requirements for the SPIRIT interferometer (whose aperture is the effective aperture diameter for an interferometer with two 1-m diameter telescopes) are taken from  \citep{benford2007cryogenic}.
 }

\end{tablenotes}
 \end{threeparttable}
\end{table*}

\subsubsection{Transition Edge Sensors}\label{ssect:tes}
\begin{figure*}
\includegraphics[width=16cm,angle=0]{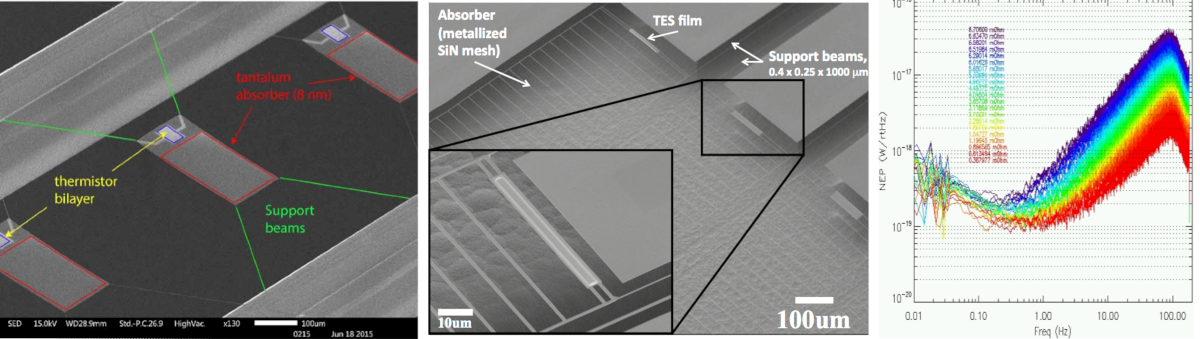}
\caption[Transition-edge sensed (TES) bolometers]{Transition-edge sensed (TES) bolometers developed at SRON (left) and JPL (center), targeting high sensitivity for far-infrared spectroscopy from cold telescopes. These are silicon-nitride suspensions, similar to the {\itshape Herschel} and {\itshape Planck} bolometers, but they feature long ($\sim1\,$mm), narrow ($\sim0.4\,\mu$m) suspension legs, and are cooled to below 100\,mK. Both programs have demonstrated NEPs of $1-3\times10^{-19}\,$W\,Hz$^{-1/2}$\citep{suz16dev}. An example NEP measurement of the JPL system is shown at right.} 
\label{fig:dettes} 
\end{figure*}

A transition edge sensor (TES, Figure \ref{fig:dettes}) consists of a superconducting film operated near its superconducting transition temperature. This means that the functional form of the temperature dependence of resistance, $R(T)$, is very sharp. The sharpness of the $R(T)$ function allows for substantially better sensitivity than semi-conducting thermistors (though there are other factors to consider, such as readout schemes, see \S\ref{readschemes}). Arrays of transition-edged-sensed (TES) bolometers have been used in CMB experiments \citep{bicep15,hend16,hubm16,thorn16,denis16}, as well as in calorimeters in the $\gamma$-ray\citep{noro13}, X-ray\citep{irwin1996x,woll00}, ultraviolet, and optical. They are also anticipated for future X-ray missions, such as Athena\citep{smith16,gott16}. 

In the infrared, TES bolometers are widely used. A notable ground-based example is SCUBA2 on the JCMT\citep{holl13} (Table \ref{tbl:requirements}). Other sub-orbital examples include HAWC+ and the upcoming HIRMES instrument, both on SOFIA. TES bolometers are also planned for use in the SAFARI instrument for SPICA\citep{tessfara,gold16,khos16,hijm16}. In terms of sensitivity, groups at SRON and JPL have demonstrated TES sensitivities of 1$\times$10$^{-19}\,\rm W\,Hz^{-1/2}$\citep{beyer12,karas14,khos16}.  

The advantages of TES arrays over KIDs and QCD arrays are technological maturity and versatility in readout schemes (see \S\ref{readschemes}). However, TES detector arrays do face challenges. The signal in TES bolometers is a current through a (sub-$\Omega$) resistive film at sub-kelvin temperatures, so conventional amplifiers are not readily impedance matched to conventional low-noise amplifiers with high input impedance. Instead, superconducting quantum interference devices (SQUIDs) are used as first-stage amplifiers and SQUID-based circuits have been fashioned into a switching time-domain multiplexers (the TDM, from NIST and UBC\citep{irwin2005transition}), which has led to array formats of up to $\sim$10$^4$ pixels. While this time-domain multiplexing system is mature and field tested in demanding scientific settings, it is not an approach that can readily scale above $\sim10^4$ pixels, due primarily to wire count considerations. Other issues with TES arrays include; (1) challenging array fabrication, (2) relatively complex SQUID-based readout systems and no on-chip multiplexing (yet).

\subsubsection{Kinetic Inductance Detectors}\label{ssect:kids}
\begin{figure*}
\includegraphics[width=16cm,angle=0]{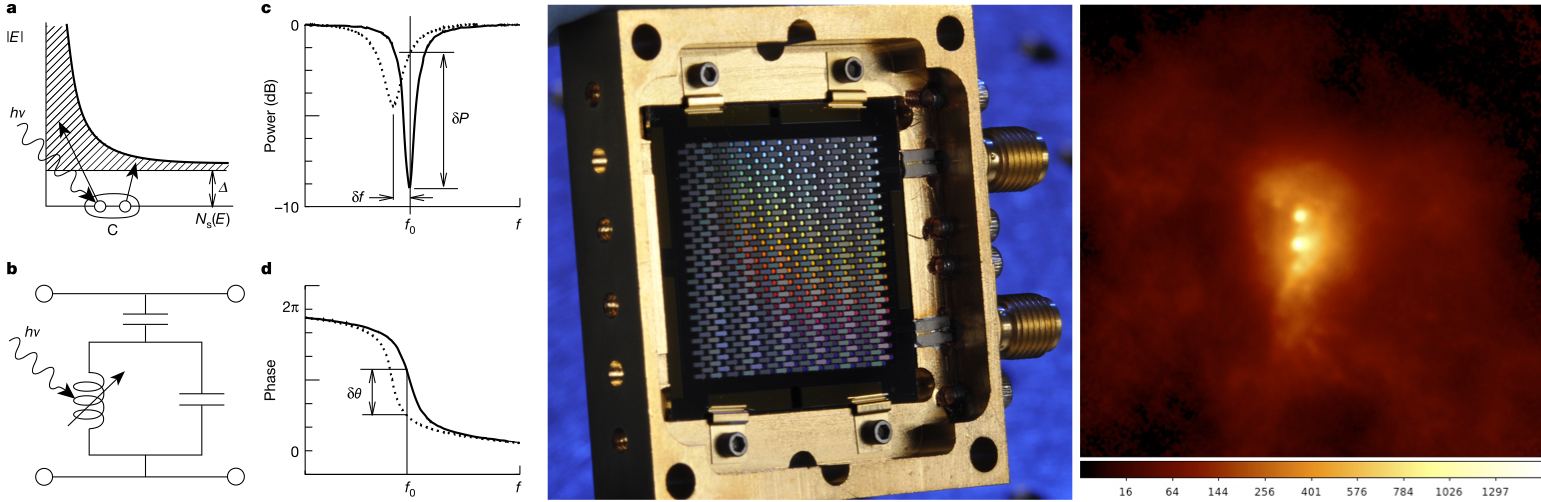}
\caption[Kinetic Inductance Detectors (KIDs).]{Kinetic Inductance Detectors (KIDs). The schematic at left is reprinted from\citep{day03}; (a) Photons are absorbed in a superconducting film operated below its transition temperature, breaking Cooper pairs to create free electrons; (b) The increase in free electron density increases the inductance (via the kinetic inductance effect) of an RF or microwave resonator, depicted schematically here as a parallel LC circuit which is capacitively coupled to a through line. (c) On resonance, the LC circuit loads the through line, producing a dip in its transmission. The increase in inductance moves the resonance to lower frequency ($f\sim 1/\sqrt{L}$), which produces a phase shift (d) of a RF or microwave probe signal transmitted though the circuit. Because the resonators can have high quality factor, many hundreds to thousands can be accessed on a single transmission line. Center shows the 432-pixel KID array in the Caltech / JPL MAKO camera, and right shows an image of SGR B2 obtained with MAKO at the CSO.} 
\label{fig:detkid} 
\end{figure*}

The simplest approach to high-multiplex-factor frequency domain multiplexing (FDM, see also \S\ref{readschemes}) thus far is the kinetic inductance detector (KID\citep{day03,zmu12}, Figure \ref{fig:detkid}). In a KID, photons incident on a superconducting film break Cooper pairs, which results in an increase in the inductance of the material. When embedded in a resonant circuit, the inductance shift creates a measureable frequency shift, which is encoded as a phase shift of the probe tone. KIDs originated as far-infrared detector arrays, with on-telescope examples including MAKO \citep{mako12} and MUSIC\citep{maloney2010music} at the CSO, A-MKID\citep{heym10} at APEX, NIKA/NIKA2\citep{monf10,monf11,adam18} at IRAM, the extremely compact $\mu$-Spec \citep{cataldo2014micro,barrentine2016design}, SuperSpec\citep{shirokoff2014design}, and the submillimeter wave imaging spectrograph DESHIMA\citep{endo12}. KIDs were later adapted for the optical / near-infrared\citep{mazin12}, where they provide advances in time resolution and energy sensitivity. Examples include ARCONS\citep{arcons13},  DARKNESS \& MEC\citep{mee15,maz15}, the KRAKENS IFU\citep{mazkr15}, and PICTURE-C\citep{picc}. KIDs are also usable for millimeter-wave/CMB studies \citep{calvo10,kara12,mcc14,low14,oguri2016}, although there are challenges in finding materials with suitably low $T_c$'s when operating below 100\,GHz. KIDs are now being built in large arrays for several ground-based and sub-orbital infrared observatories, including the BLAST-Pol2 balloon experiment. 

There exist three primary challenges in using KIDS in space-based infrared observatories:
\vspace{0.1cm}

\noindent {\bfseries Sensitivity}: Sub-orbital far-infrared observatories have relatively high-backgrounds, and thus sensitivities that are $2-3$ orders of magnitude above those needed for background-limited observations from space. For space-based KIDs instruments, better sensitivities are needed. The state of the art is from SPACEKIDs, for which NEPs of 3$\times$10$^{-19}\,\rm W\,Hz^{-1/2}$ have been demonstrated in aluminum devices coupled via an antenna \citep{devis14,griff16,Baselmans2016}. This program has also demonstrated 83\% yield in a 961-pixel array cooled to $120\,$mK. A further, important outcome of the SPACEKIDs program was the demonstration that the effects of cosmic ray impacts can be effectively minimised\citep{Baselmans2016,monfabas16}. In the US, the Caltech / JPL group and the SuperSpec collaboration have demonstrated sensitivities below 1$\times$10$^{-18}\,\rm W\,Hz^{-1/2}$ in a small-volume titanium nitride devices at $100\,$mK, also with radiation coupled via an antenna. 
\vspace{0.1cm}

\noindent {\bfseries Structural considerations}: KIDs must have both small active volume (to increase response to optical power) and a method of absorbing photons directly without using superconducting transmission lines. Options under development include: 

\begin{itemize}

\item Devices with small-volume meandered absorbers / inductors, potentially formed via electron-beam lithography for small feature widths.

\item Thinned-substrate devices, in which the KID inductor is patterned on a very thin (micron or sub-micron) membrane which may help increase the effective lifetime of the photo-produced quasiparticles, thereby increasing the response of the device. 

\end{itemize}
\vspace{0.1cm}

\noindent {\bfseries Antenna coupling at high frequencies}: While straightforward for the submillimeter band, the antenna coupling becomes non-trivial for frequencies above the superconducting cutoff of the antenna material (e.g., $\sim714\,$GHz for Nb and $1.2\,$THz for NbTiN). To mitigate this, one possible strategy is to integrate the antenna directly into the KID, using only aluminium for the parts of the detector that interact with the THz signal. This approach has been demonstrated at $1.55\,$THz, using a thick aluminium ground plane and a thin aluminium central line to limit ground plane losses to 10\% \citep{bueno2017full,Baselmans2016}. This approach does not rely on superconducting stripline technology and could be extended to frequencies up to $\sim10\,$THz.
\vspace{0.1cm}

A final area of research for KIDs, primarily for CMB experiments, is the KID-sensed bolometer, in which the thermal response of the KID is used to sense the temperature of a bolometer island. These devices will be limited by the fundamental phonon transport sensitivity of the bolometer, so are likely to have sensitivity limits comparable to TES bolometers, but may offer advantages including simplified readout, on-array multiplexing, lower sensitivity to magnetic fields, and larger dynamic range.

\subsubsection{Quantum Capacitance Detectors}\label{ssect:qcd}
The Quantum Capacitance Detector (QCD\citep{Shaw09,Bueno10,Bueno11,Stone12,Echternach13}) is based on the Single Cooper Pair Box (SCB), a superconducting device initially developed as a qubit for quantum computing applications. The SCB consists of a small island of superconducting material connected to a ground electrode via a small (100\,nm $\times$ 100\,nm) tunnel junction. The island is biased with respect to ground through a gate capacitor, and because it is sufficiently small to exhibit quantum behaviour, its capacitance becomes a strong function of the presence or absence of a single free electron. By embedding this system capacitively in a resonator (similar to that used for a KID), a single electron entering or exiting the island (via tunneling through the junction) produces a detectable frequency shift.

To make use of this single-electron sensitivity, the QCD is formed by replacing the ground electrode with a superconducting photon absorber. As with the KIDs, photons with energy larger than the superconducting gap breaks Cooper pairs, establishing a density of free electrons in the absorber that then tunnel onto (and rapidly back out of) the island through the tunnel junction. The rate of tunneling into the island, and thus the average electron occupation in the island, is determined by the free-electron density in the absorber, set by the photon flux. Because each photo-produced electron tunnels back and forth many times before it recombines, and because these tunneling events can be detected individually, the system has the potential to be limited by the photon statistics with no additional noise.

\begin{figure*}
\includegraphics[width=16cm,angle=0]{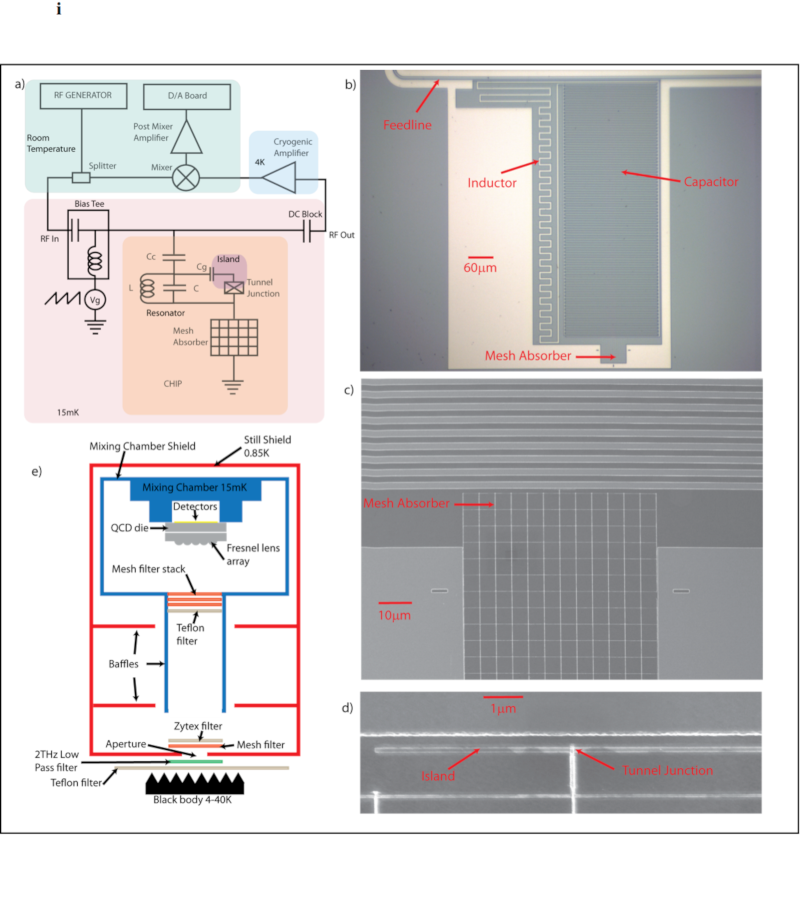}
\vspace{-1cm}
\caption[Quantum capacitance detector]{The Quantum Capacitance Detector (QCD). (a) Schematic representation showing the mesh absorber QCD with its LC resonator coupling it to the readout circuit. The single Cooper-pair box (SCB) island is formed between the tunnel junction and the gate capacitor. The tunnel junction is connected to the mesh absorber which in turn is connected to ground plane. The SCB presents a variable capacitance in parallel with an LC resonator. (b) Optical microscope picture of a device, showing the feedline, the inductor, the interdigitated capacitor, all fabricated in Nb and the Al mesh absorber. (c) SEM picture of mesh absorber consisting of 50nm wide aluminum lines on a $5\,\mu$m pitch grid. (d) Detail of the SCB showing the aluminum island (horizontal line) in close proximity to the lowest finger of the interdigitated capacitor and the tunnel junction (overlap between the island and vertical line connecting to the mesh absorber below). (e) Optical setup schematic showing temperature-tunable blackbody, aperture, and filters which define the spectral band.   This device has demonstrated an optical NEP of 2$\times$10$^{-20}\,\rm W\,Hz^{-1/2}$ at $200\,\mu$m, as well as the ability to count individual photons\citep{Echternach13,echter18}.} 
\label{fig:detqcd} 
\end{figure*}

This has indeed been demonstrated. QCDs have been developed to the point where a 25-pixel array yields a few devices which are photon noise limited for $200\,\mu$m radiation under a load of 10$^{-19}\,\rm W$, corresponding to a NEP of $2\times10^{-20}\,\rm W Hz^{-1/2}$. The system seems to have good efficiency as well, with inferred detection of 86\% of the expected photon flux for the test setup. As an additional demonstration, a fast-readout mode has been developed which can identify individual photon arrival events based on the subsequent increase in tunneling activity for a timescale on order the electron recombination time (Figure \ref{fig:detqcd}).

With its demonstrated sensitivity and natural frequency-domain multiplexing, the QCD is promising for future far-infrared space systems. Optical NEPs of below $10^{-20}\,\rm W Hz^{-1/2}$ at $200\,\mu$m have been demonstrated, with the potential for photon counting at far-infrared wavelengths \citep{echter18}. However, QCDs are some way behind both TES and KIDs arrays in terms of technological maturity. To be viable for infrared instruments, challenges in (1) yield and array-level uniformity, (2) dark currents, and (3) dynamic range must all be overcome. The small tunnel junctions are challenging, but it is hoped that advances in lithography and processing will result in improvements.

\subsubsection{System Considerations for Direct Detector Readouts}\label{readschemes}
There exist three commonly used multiplexing (muxing) schemes\citep{ramas09} for readout of arrays; Frequency Domain Muxing (FDM), Time Domain Muxing (TDM), and Code Division Muxing (CDM). In this section we briefly review their applicability and advantages. 

FDM is a promising path to reading out the large arrays anticipated in future infrared observatories. In FDM, a single readout circuit services up to $\sim1000$ pixels, each coupled through a micro-resonator tuned to a distinct frequency. Each pixel is then probed individually with an RF or microwave tone at its particular frequency. The warm electronics must create the suite of tones which is transmitted to the array for each circuit, then digitize, Fourier-transform, and channel the output data stream to measure the phase and amplitude shifts of each tone independently. The number of resonators (and thus pixels) that can be arrayed onto a single readout circuit depends on the quality factor (Q) of the resonators and the bandwidth available in the circuit. For micro-resonators patterned in superconducting films, resonator Q's exceeding $10^7$ are possible but more typical values are around $10^5$, which permits approximately $10^3$ pixels per octave of readout bandwidth to be operated with sufficiently low cross-talk. 

In these systems, all of the challenging electronics are on the warm side, and the detector array is accessed via low-loss RF / microwave lines (one from the warm side down through the crysotat stages, another for the return signal). Moreover, FDM readout schemes can be applied to both TES and KIDs arrays, while other multiplexing schemes are TES-only. An example of recent progress is the development of an FDM scheme that can read out 132 TES pixels simultaneously, using a single SQUID, without loss of sensitivity \citep{hijm16}. This is very close to the 160 detectors per SQUID targeted for SPICA/SAFARI. 

There are, however, limitations to FDM schemes:

\begin{enumerate} 

\item {\bfseries Thermal constraints:} While the detector arrays themselves are essentially passive, the conductors, whether coaxial or twisted pair, will have thermal conduction from the warm stages, impacting the overall thermal design. Additionally these systems require a single low-noise amplifier (LNA) on each circuit, likely deployed somewhere between 4\,K and 20\,K, and the LNAs will have some dissipation.

\item {\bfseries Signal processing:} FDM schemes pose significant challenges for backend electronics processing capability: they must digitize the returning waveforms, then Fourier transform in real time at the science sampling rate and extract the full array of tone phases which encode the pixel signal levels. These hurdles become non-trivial for the large arrays envisaged for future missions. 

\end{enumerate}

A further challenge, that applies to readout schemes for any far-infrared resonant detector array (including TES, KID, and QCD systems), is the power required to read out $10^{4-5}$ detector arrays, due in part to the signal processing requirements. The power requirements are such that they may pose a significant obstacle to reading out $\sim10^5$ detector arrays on {\it any} balloon- or space-based platform.

For the OST, power dissipation in the warm electronics will be a particular challenge. An example is the medium-resolution survey spectrometer (MRSS), which targets 200,000 pixels among all six spectrometer bands. The concept assumes resonator frequencies between 75\,MHz and 1\,GHz, and that $1500$ pixels can be arrayed in this bandwidth (a relatively comfortable multiplexing density assuming $400$ per readout octave). This requires 130 readout circuits, each with two coaxial lines all the way to the cold stage, and a cold amplifier on the output. The conducted loads through the coaxial lines, as well as reasonable assumptions about the LNA dissipation (1\,mW at 4\,K plus 3\,mW at 20\,K for each circuit) do not stress the observatory thermal design. However, the electronics for each circuit requires a 2 giga-sample per second analog to digital converter (ADC) working at $\sim$12 bits depth, followed by FFTs of this digital signal stream in real time - 1024 point FFTs every $0.5\,\mu$s. Systems such as these implemented in FPGAs used in the laboratory dissipate $\sim$100\,W for each readout circuit, which is not compatible with having 130 such systems on a space mission.

For these reasons, development of muxing schemes is a high priority for large-format arrays, irrespective of the detector technology used. A promising path for such development is to employ a dedicated application specific integrated circuit (ASIC), designed to combine the digitization, FFT, and tone extraction in a single chip. Power dissipation estimates obtained for the MRSS study based on custom spectrometer chips developed for flight systems, and extrapolating to small-gate CMOS technology, suggest that such a custom chip could have a power dissipation of $\sim$14\,W per circuit, including all aspects. At this level, the total scales to $\sim1.8$\,kW. This power dissipation is well within the range of that of other sub-systems on future missions - for example, such missions will require several kW to operate the cryocoolers - and thus does not pose a unique problem.
\vspace{0.1cm}

\noindent Finally, we make four observations:
\vspace{0.2cm}

\noindent (1) While the power scaling calculations are straightforward, the development of this silicon ASIC is a substantial design effort, in large part because of the 12-bit depth; most fast digital spectrometers implemented in CMOS operate at 3 or 4 bits depth.
\vspace{0.1cm}

\noindent (2) The power dissipation scales as the total bandwidth, so the per-pixel electronics power dissipation could be reduced if lower resonant frequencies were used. The downside of this though is that the physical size of the resonators scale approximately as $1/\sqrt{f}$, and (with current designs) becomes several square millimeters per resonator for frequencies below $\sim50\,$MHz.
\vspace{0.2cm}

\noindent (3) Hybrid schemes, such as combining CDM with frequency domain readout, are attractive for their power efficiency, both at 4\,K due to lower number of high electron mobility transistors (HEMTs) or Parametric Amps, and for the warm electronics due to lower bandwidths and lower wire counts. These schemes however are only applicable to TES based systems.
\vspace{0.2cm}

\noindent (4) With $Q = 10^5$ and 1000 resonators per octave, the FDM scheme utilizes only a few percent of readout bandwidth. Factors of 10 or more improvement in multiplexing density and reduction in readout power are possible if the resonator frequency placement could be improved to avoid collisions, e.g. through post-fabrication trimming\footnote{Post-fabrication trimming (PFT) is a family of techniques that permanently alter the refractive index of a material to change the optical path length\citep{spara05,schrau08,atab13}. The advantage of PFT is that it does not require complex control electronics, but concerns have been raised over the long-term stability of some of the trimming mechanisms.}.

\subsection{Medium-resolution spectroscopy}\label{ssect:medresspec}
A variety of spectrometer architectures can be used to disperse light at far-infrared wavelengths. Architectures that have been successfully used on air-borne and space instruments include grating dispersion like FIFI-LS on SOFIA \citep{Klein2006} and PACS on \textit{Herschel} \cite{pog10}, Fourier Transform spectrometers like the \textit{Herschel}/SPIRE-FTS \citep{gri10}, and Fabry-Perot etalons like FIFI on the KAO telescope \citep{Poglitsch1990}.  These technologies are well understood and can achieve spectral resolutions of $R = 10^2 - 10^4$. However, future spectrometers will need to couple large FoVs to many thousands of imaging detectors, a task for which all three of these technologies have drawbacks. Grating spectrometers are mechnically simple devices that can achieve $R \sim 1000$, but are challenging to couple to wide FoVs since the spectrum is dispersed along one spatial direction on the detector array. FTS systems require moving parts and suffer from noise penalties associated with the need for spectral scanning. They are also not well-suited to studies of faint objects because of systematics associated with long-term stability of the interferometer and detectors \citep{zmuid03}. Fabry-Perot systems are also mechanically demanding, requiring tight parallelism tolerances of mirror surfaces, and typically have restricted free spectral range due to the difficulty of manufacturing sufficiently precise actuation mechanisms \citep{Parshley2014}. A new technology that can couple the large FoVs anticipated in next-generation far-infrared telescopes to kilo-pixel or larger detector arrays would be transformative for far-infrared spectroscopy.

A promising approach to this problem is far-infrared filter bank technology\citep{Kovacs2012,Wheeler2016}. This technology has been developed as a compact solution to the spectral dispersion problem, and has potential for use in space. These devices require the radiation to be dispersed to propagate down a transmission line or waveguide. The radiation encounters a series of tuned resonant filters, each of which consists of a section of transmission line of length $\lambda_{i}/2$, where $\lambda_{i}$ is the resonant wavelength of channel $i$. These half-wave resonators are evanescently coupled to the feedline with designable coupling strengths described by the quality factors $Q_{\rm feed}$ and $Q_{\rm det}$ for the feedline and detector, respectively. The filter bank is formed by arranging a series of channels monotonically increasing in frequency, with a spacing between channels equal to an odd multiple of $\lambda_{i}/4$.  The ultimate spectral resolution $R=\lambda / \Delta \lambda$ is given by:

\begin{equation}
\frac{1}{R} = \frac{1}{Q_{\rm filt}} = \frac{1}{Q_{\rm feed}} +
\frac{1}{Q_{\rm det}} + \frac{1}{Q_{\rm loss}}.
\end{equation}

\noindent where $Q_{\rm loss}$ accounts for any additional sources of dissipation in the circuit and $Q_{\rm filt}$ is the net quality factor. This arrangement has several advantages in low and medium-resolution spectroscopy from space, including: (1) compactness (fitting on a single chip with area of tens of square cm), (2) integrated on-chip dispersion and detection, (3) high end-to-end efficiency equal to or exceeding existing technologies, and (4) a mechanically stable architecture. A further advantage of this architecture is the low intrinsic background in each spectrometer, which only couples to wavelengths near its resonance. This means that very low backgrounds can be achieved, requiring detector NEPs below $10^{-20}$ W Hz$^{-1/2}$. Filter banks do however have drawbacks\citep{Kovacs2012}. For example, while filter banks are used in instruments operating from millimeter to radio wavelengths, they are currently difficult to manufacture for use at wavelengths shortward of about 500\,$\mu$m.

Two ground-based instruments are being developed that make use of filter banks. A prototype transmission-line system has been fabricated for use in SuperSpec \citep{Shirokoff2012,SHD2014} for the LMT. SuperSpec will have $R \sim 300$ near 250\,GHz and will allow photon-background limited performance. A similar system is WSPEC, a 90\,GHz filter bank spectrometer that uses machined waveguide to propagate the radiation \cite{Che2015}. This prototype instrument has 5 channels covering the $130 {-} 250$\,GHz band. Though neither instrument is optimized for space applications, this technology can be adapted to space, and efforts are underway to deploy it on sub-orbital rockets.

\subsection{High-resolution spectroscopy}\label{ssect:het}
Several areas of investigation in mid/far-infrared astronomy call for spectral resolution of $R\geq10^{5}$, higher than can be achieved with direct detection approaches. At this very high spectral resolution, heterodyne spectroscopy is routinely used \citep{Schieder2008,golds17}, with achievable spectral resolution of up to $R\simeq10^{7}$. In heterodyne spectroscopy, the signal from the ``sky'' source is mixed with a spectrally-pure, large-amplitude, locally-generated signal, called the ``Local Oscillator (LO)'', in a nonlinear device. The nonlinearity generates the sum and difference of the sky and LO frequencies. The latter, the ``Intermediate Frequency (IF)'', is typically in the $1-10$\,GHz range, and can be amplified by low-noise amplifiers and subsequently sent to a spectrometer, which now is generally implemented as a digital signal processor. A heterodyne receiver is a coherent system, preserving the phase and amplitude of the input signal. While the phase information is not used for spectroscopy, it is available and can be used for e.g. interferometry. 

The general requirements for LOs are as follows; narrow linewidth, high stability, low noise, tunability over the required frequency range, and sufficient output power to couple effectively to the mixer. For far-infrared applications, LO technologies are usually one of two types: multiplier chain, and Quantum Cascade Laser (QCL). Multiplier chains offer relatively broad tuning, high spectral purity, and known output frequency. The main limitation is reaching higher frequencies ($>3\,$THz). QCL’s are attractive at higher frequencies, as their operating frequency range extends to 5\,THz and above, opening up the entire far-infrared range for high resolution spectroscopy.

For mixers, most astronomical applications use one or more of three technologies: Schottky diodes, Superconductor-Insulator-Superconductor (SIS) mixers, and Hot Electron Bolometer (HEB) mixers\citep{klapw17}. Schottky diodes function at temperatures of $>70\,$K, can operate at frequencies as high as $\sim3\,$THz ($100\,\mu$m), and provide large IF bandwiths of $>8\,$GHz, but offer sensitivities that can be an order of magnitude or more poorer than either SIS or HEB mixers. They also require relatively high LO power, of order 1\,mW. SIS and HEB mixers, in contrast, have operating temperatures of $\sim4\,$K and require LO powers of only $\sim1\mu$W. SIS mixers are most commonly used at frequencies up to about 1\,THz, while HEB mixers are used over the 1-6\,THz range. At present, SIS mixers offer IF bandwidths and sensitivities both a factor of 2-3 better than HEB mixers. All three mixer types have been used on space-flown hardware; SIS and HEB mixers in the {\itshape Herschel} HIFI instrument\citep{deg10,roelf12}, and Schottky diodes on instruments in SWAS and Odin. 

Heterodyne spectroscopy can currently achieve spectral resolutions of $R\simeq10^{7}$, and in principle the achievable spectral resolution is limited only by the purity of the signal from the LO. Moreover, heterodyne spectroscopy preserves the phase of the sky signal as well as its frequency, lending itself naturally to interferometric applications. Heterodyne arrays are used on SOFIA, as well as many ground-based platforms. They are also planned for use in several upcoming observatories, including GUSTO. A further example is FIRSPEX, a concept study for a small-aperture telescope with heterodyne instruments to perform several large-area surveys targeting bright far-infrared fine-structure lines, using a scanning strategy similar to that used by {\itshape Planck}\citep{rig16}.  

There are however challenges for the heterodyne approach. We highlight five here:

\begin{itemize}

\item {\bfseries The antenna theorem:} Coherent systems are subject to the antenna theorem that allows them to couple to only a single spatial mode of the electromagnetic field. The result is that the product of the solid angle subtended by the beam of a heterodyne receiver system ($\Omega$) and its collecting area for a normally incident plane wave ($A_e$) is determined; $A_e\Omega = \lambda^2$ \citep{gold02}. 

\item {\bfseries The quantum noise limit:} A heterodyne receiver, being a coherent system, is subject to the quantum noise limit on its input noise temperature, $T \ge hf/k$ (e.g.\citep{zmuid03}. While SIS mixers have noise temperatures only a few times greater than the quantum noise limit, HEB mixer receivers typically have noise temperatures $\sim10$ times the quantum noise limit, e.g. $10\times91$\,K at $f = 1900$\,GHz. Improved sensitivity for HEB mixers, and SIS mixers operating at higher frequencies will offer significant gains in astronomical productivity.

\item {\bfseries Limited bandwith:} There is a pressing need to increase the IF bandwidth of HEB mixers, with a minimum  of 8\,GHz bandwidth required at frequencies of $\sim3$\,THz. This will allow for complete coverage of Galactic spectral lines with a single LO setting, as well as the lines of nearby galaxies. Simultaneous observation of multiple lines also becomes possible, improving both efficiency and relative calibration accuracy.

\item {\bfseries Array size:} The largest arrays currently deployed (such as in upGREAT on SOFIA) contain fewer than 20 pixels although a 64-pixel ground-based array operating at 850\,$\mu$m has been constructed \citep{groppi10}. Increasing array sizes to hundreds or even thousands of pixels will require SIS and HEB mixers that can be reliably integrated into these new large-format arrays, low power IF amplifiers, and efficient distribution of LO power.

\item {\bfseries Power requirements:} Existing technology typically demands significantly more power per pixel than is available for large-format arrays on satellite-based platforms.

\end{itemize}

On a final note: for the higher frequency ( $>3$\,THz) arrays, high-power ($5-10$\,mW) QCL LO's are a priority for development, along with power division schemes (e.g., Fourier phase gratings) to utilize QCLs effectively \citep{hayton20144,richter20154,richter2015performance}. At $<3\,$THz, frequency-multiplied sources remain the system of choice, and have been successfully used in missions including SWAS, {\itshape Herschel}-HIFI, STO2, and in GREAT and upGREAT on SOFIA. However, to support large-format heterodyne arrays, and to allow operation with reduced total power consumption for space missions, further development of this technology is necessary. Further valuable developments include SIS and HEB mixers that can operate at temperatures of $>20$\,K, and integrated focal planes of mixers and low-noise IF amplifiers.

\subsection{Fabry-Perot Interferometry}\label{ssect:fabry}
Fabry-Perot Interferometers (FPIs) have been used for astronomical spectroscopy for decades, with examples such as FIFI\citep{pogli91}, KWIC\citep{stacey93}, ISO-SWS/LWS\citep{degraa96,clegg96}, and SPIFI\citep{bradford02}. FPIs similar to the one used in ISO have also been developed for balloon-borne telescopes\citep{pepe94}. 

FPIs consist of two parallel, highly reflective (typically with reflectivities of $\sim96\%$), very flat mirror surfaces. These two mirrors create a resonator cavity. Any radiation whose wavelength is an integral multiple of twice the mirror separation meets the condition for constructive interference and passes the FPI with high transmission. Since the radiation bounces many times between the mirrors before passing, FPIs can be fabricated very compactly, even for high spectral resolution, making them attractive for many applications. In addition, FPIs allow for large FoVs, making them an excellent choice as devices for spectroscopic survey instruments.

Observations with FPI are most suitable for extended objects and surveys of large fields, where moderate to high spectral resolution ($R\sim10^2 - 10^5$) is required. for example:

\begin{itemize}

\item Mapping mearby galaxies in multiple molecular transitions and atomic or ionic fine-structure lines in the far-infrared. This traces the properties of the interstellar medium, and relates small-scale effects like star forming regions to the larger-scale environment of their host galaxies. 

\item For high-redshift observations, FPI is suited to survey large fields and obtain a 3D data cube by stepping an emission line over a sequence of redshift bins. This results in line detections from objects located at the corresponding redshift bins and allows e.g. probing ionization conditions or metallicities for large samples simultaneously.

\end{itemize}

FPIs do however face challenges. We highlight four examples:
\vspace{0.2cm}

\noindent (1) To cover a certain bandwidth, the FPI mirror separation has to be continuously or discretely changed, i.e. the FPI has to be scanned, which requires time, and may result in poor channel-to-channel calibration in the spectral direction if the detector system is not sufficiently stable.  
\vspace{0.2cm}

\noindent (2) Unwanted wavelengths that fulfill the resonance criteria also pass through the FPI and need to be filtered out. Usually, additional FPIs operated in lower order combined with band-pass or blocking/edge filters are used for order sorting. However, since most other spectrometers need additional filters to remove undesired bands, the filtration of unwanted orders in FPIs is not a profound disadvantage.
\vspace{0.2cm}

\noindent (3) In current far-infrared FPIs, the reflective components used for the mirrors are free-standing metal meshes. The finesse\footnote{The spectral range divided by the FWHMs of individual resonances, see e.g.\citep{Ismail16}.} of the meshes changes with wavelength and therefore a FPI is only suitable over a limited wavelength range. Also, the meshes can vibrate, which requires special attention especially for high spectral resolution, where the diameters can be large. Replacing the free-standing metal meshes with a different technology is therefore enabling for broader applications of FPI. For example, flat silicon wafers with an anti-reflection structure etched on one side and the other side coated with a specific thin metal pattern, optimized for a broader wavelength range, can substitute for a mirror. This silicon wafer mirror is also less susceptible to vibrations and could be fabricated with large enough diameters. 
\vspace{0.2cm}

\noindent (4) Improving cryogenic scanning devices. Currently, FPIs usually use piezoelectric elements (PZTs) for scanning. However, PZTs have limited travel range, especially at 4\,K. Moreover, mechanical devices or PZT-driven motors are still not reliable enough at cryogenic temperatures, or too large to be used in the spaces available inside the instruments. It is thus important to develop either smaller PZT-driven devices which can travel millimeters with resolutions of nanometers at a temperature of 4\,K, or an alternative scanning technology that overcomes the limitations of PZT devices and satisfies the requirements of FPIs.

\begin{table*}
\begin{threeparttable}[b]
\caption[On-orbit Mechanical cryocooler lifetimes]{Long-life space cryocooler operating experiences as of May 2016.} \label{tab:coolers}
{\scriptsize      
 \begin{tabular}{cr|ccl}
 \hline
  \hline           
 \multicolumn{2}{l}{Cooler, Mission, \& Manufacturer}                               & T (K) & Hours/unit    & Notes \\
 \hline            
 \hline      
\multicolumn{2}{|l|}{\cellcolor{gray!30}{\bfseries Turbo Brayton}}                  & & &      \\
 \multicolumn{2}{r} {International Space Station - MELFI (Air Liquide)}             & 190   & 85,600   &  Turn-on 7/06, ongoing, no degradation      \\
 \multicolumn{2}{r} {HST - NICMOS (Creare)}                                          & 77    & 57,000   & 3/02 thru 10/09, off, coupling to load failed      \\
\multicolumn{2}{|l|}{\cellcolor{gray!30}{\bfseries Stirling}}                       & & &      \\
 \multicolumn{2}{r} {HIRDLS: 1-stage (Ball Aerospace)}                              & 60    & 83,800   & 8/04 thru 3/14, instrument failed 03/08, data turned off 3/14      \\  %% High Resolution Dynamics Limb Sounder (HIRDLS)
 \multicolumn{2}{r} {TIRS: 2-stage (Ball Aerospace)}                                & 35    & 27,900   & Turn-on 6/13, ongoing, no degradation      \\  %% Thermal Infrared Sensor (TIRS)
 \multicolumn{2}{r} {ASTER-TIR (Fujitsu)}                                           & 80    & 141,7000 & Turn-on 3/00, ongoing, no degradation       \\  %% ADVANCED SPACEBORNE THERMAL EMISSION AND REFLECTION RADOIMETER
 \multicolumn{2}{r} {ATSR-1 on ERS-1 (RAL)}                                         & 80    & 75,300   & 7/91 thru 3/00, satellite failed        \\
 \multicolumn{2}{r} {ATSR-2 on ERS-2 (RAL)}                                         & 80    & 112,000  & 4/95 thru 2/08, instrument failed     \\
 \multicolumn{2}{r} {Suzaku: one stage (Sumitomo)}                                  & 100   & 59,300   & 7/05 thru 4/12, mission end, no degradation        \\
 \multicolumn{2}{r} {SELENE/Kaguya GRS: one stage (Sumitomo)}                       & 70    & 14,600   & 10/07 thru 6/09, mission end, no degradation       \\
 \multicolumn{2}{r} {\cellcolor{green!15}AKARI: two stage (Sumitomo)}               &\cellcolor{green!15} 20    &\cellcolor{green!15} 39,000   &\cellcolor{green!15} 2/06 thru 11/11, mission end  \\
 \multicolumn{2}{r} {RHESSI (Sunpower)}                                             & 80    & 124,600  & Turn-on 2/02, ongoing, modest degradation          \\
 \multicolumn{2}{r} {CHIRP (Sunpower)}                                              & 80    & 19,700   & 9/11 thru 12/13, mission end, no degradation   \\
 \multicolumn{2}{r} {ASTER-SWIR (Mitsubishi)}                                       & 77    & 137,500  & Turn-on 3/00, ongoing, load off at 71,000 hours      \\
 \multicolumn{2}{r} {ISAMS (Oxford/RAL)}                                            & 80    & 15,800   & 10/91 thru 7/92, instrument failed       \\
 \multicolumn{2}{r} {HTSSE-2 (Northrop Grumman)}                                    & 80    &  24,000  & 3/99 thru 3/02, mission end, no degradation       \\
 \multicolumn{2}{r} {HTSSE-2 (BAe)}                                                 & 80    & 24,000   & 3/99 thru 3/02, mission end, no degradation    \\
 \multicolumn{2}{r} {MOPITT (BAe)}                                                  & 50-80 & 138,600  & Turn on 3/00, lost one disp. at 10.300 hours      \\
 \multicolumn{2}{r} {\cellcolor{green!15}Odin (Astrium)}                            &\cellcolor{green!15} 50-80 &\cellcolor{green!15} 132,600  &\cellcolor{green!15} Turn-on 3/01, ongoing, no degradation       \\
 \multicolumn{2}{r} {ERS-1: AATSR \& MIPAS (Astrium)}                               & 50-80 & 88,200   & 3/02 thru 4/12, no degradation, satellite failed  \\
 \multicolumn{2}{r} {INTEGRAL (Astrium)}                                            & 50-80 & 118,700  & Turn-on 10/02, ongoing, no degradation            \\
 \multicolumn{2}{r} {Helios 2A (Astrium)}                                           & 50-80 & 96,600   & Turn-on 4/05, ongoing, no degradation             \\
 \multicolumn{2}{r} {Helios 2B (Astrium)}                                           & 50-80 & 58,800   & Turn-on 4/10, ongoing, no degradation             \\
 \multicolumn{2}{r} {SLSTR (Airbus)}                                                & 50-80 & 1,4000   & Turn-on 3/16, ongoing, no degradation             \\
\multicolumn{2}{|l|}{\cellcolor{gray!30}{\bfseries Pulse-Tube}}                     & & &      \\
 \multicolumn{2}{r} {CX (Northrop Grumman)}                                         & 150   & 161,600  & Turn-on 2/98, ongoing, no degradation   \\
 \multicolumn{2}{r} {MTI (Northrop Grumman)}                                        & 60    & 141,600  & Turn-on 3/00, ongoing, no degradation      \\
 \multicolumn{2}{r} {Hyperion (Northrop Grumman)}                                   & 110   & 133,600  & Turn-on 12/00, ongoing, no degradation         \\
 \multicolumn{2}{r} {SABER on TIMED (Northrop Grumman)}                             & 75    & 129,600  & Turn-on 1/02, ongoing, no degradation         \\
 \multicolumn{2}{r} {AIRS (Northrop Grumman)}                                       & 55    & 121,600  & Turn-on 6/02, ongoing, no degradation         \\
 \multicolumn{2}{r} {TES (Northrop Grumman)}                                        & 60    & 102,600  & Turn-on 8/04, ongoing, no degradation         \\
 \multicolumn{2}{r} {JAMI (Northrop Grumman)}                                       & 65    &  91,000  & 4/05 thru 12/15, mission end, no degradation       \\
 \multicolumn{2}{r} {IBUKI/GOSAT (Northrop Grumman)}                                & 65    &  63,300  & Turn-on 2/09, ongoing, no degradation         \\
 \multicolumn{2}{r} {OCO-2 (Northrop Grumman)}                                      & 110   &  14,900  & Turn-on 8/14, ongoing, no degradation         \\
 \multicolumn{2}{r} {Himawari-8 (Northrop Grumman)}                                 & 65    &  12,800  & Turn-on 12/14, ongoing, no degradation         \\
\multicolumn{2}{|l|}{\cellcolor{gray!30}{\bfseries Joule-Thompson}}                 & & &      \\
 \multicolumn{2}{r} {International Space Station - SMILES (Sumitomo)}               & 4.5   &  4,500   & 10/09 thru 04/10, instrument failed      \\
 \multicolumn{2}{r} {\cellcolor{green!15}{\itshape Planck} (RAL/ESA)}               &\cellcolor{green!15} 4     &\cellcolor{green!15} 38,500   &\cellcolor{green!15} 5/09 thru 10/13, mission end, no degradation      \\                                                           
 \multicolumn{2}{r} {\cellcolor{green!15}{\itshape Planck} (JPL)}                   &\cellcolor{green!15} 18    &\cellcolor{green!15} 27,500   &\cellcolor{green!15} FM1: 8/10-10/13 (EOM), FM2: failed at 10,500 hours      \\
\hline
\hline
 \end{tabular}
 }
 \begin{tablenotes}
 {\footnotesize 
 \item Almost all cryocoolers have continued to operate normally until turned off at end of instrument life. Mid/far-infrared \& CMB astrophysics observatories are highlighted in green. The data in this table are courtesy of Ron Ross, Jr. }
\end{tablenotes}
 \end{threeparttable}
\end{table*}

\subsection{Small and Low-Power Coolers}\label{ssect:cryoc}
For any spaceborne observatory operating at mid/far-infrared wavelengths, achieving high sensitivity requires that the telescope, instrument, and detectors be cooled, with the level of cooling dependent on the detector technology, the observation wavelength, and the goals of the observations. Cooling technology is thus fundamentally enabling for all aspects of mid/far-infrared astronomy.   

The cooling required for the telescope depends on the wavelengths being observed (Figure \ref{fig:backgr}). For some situations, cooling the telescope to $30-40$\,K is sufficient. At these temperatures it is feasible to use radiative (passive) cooling solutions if the telescope is space-based, {\itshape and} if the spacecraft orbit and attitude allow for a continuous view of deep space \citep{haw92}. Radiative coolers typically resemble a set of thermal/solar shields in front of a black radiator to deep space (Figure \ref{fig:launchir}). This is a mature technology, having been used on {\itshape Spitzer}, {\itshape Planck}, and JWST (for an earlier proposed example, see\citep{thron95}). 

For many applications however, cooling the telescope to a few tens of kelvins is sub-optimal. Instead, cooling to of order 4\,K is required, for e.g., zodiacal background limited observations (see also \S\ref{spaceobs}). Moreover, detector arrays require cooling to at least this level. For example, SIS and HEB mixers need cooling to 4\,K, while TES, KID, and QCD arrays need cooling to 0.1\,K or below. Achieving cooling at these temperatures requires a cooling chain - a staged series of cooling technologies selected to maximize the cooling per mass and per input power.  

\begin{figure*}
\includegraphics[width=16cm,angle=0]{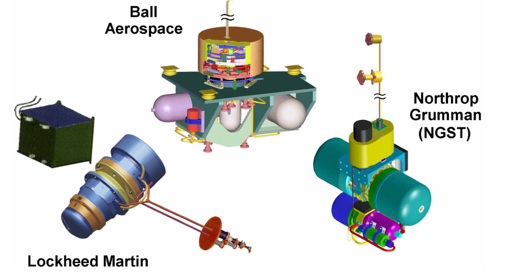}
\caption[Three cryocoolers for 6\,K cooling]{Three cryocoolers for 6\,K cooling developed through the Advanced Cryocooler Technology Program (ACTDP).} 
\label{fig:cryod} 
\end{figure*}

To achieve temperatures below $\sim40$\,K, or where a continuous view of deep space is not available, cryocoolers are necessary. In this context, the Advanced Cryocooler Technology Development Program (ACTDP\citep{ross06}), initiated in 2001, has made excellent progress in developing cryogen-free multi-year cooling for low-noise detector arrays at temperatures of 6\,K and below (Figure \ref{fig:cryod}). The state-of-the-art for these coolers include those on-board {\itshape Planck}, JWST, and {\itshape Hitomi}\citep{shirron16hit}. Similar coolers that could achieve 4\,K are at TRL 4-5, having been demonstrated as a system in a laboratory environment\citep{ross04}, or as a variant of a cooler that has a high TRL (JWST/MIRI). Mechanical cryocoolers for higher temperatures have already demonstrated impressive on-orbit reliability (Table \ref{tab:coolers}). The moving components of a 4\,K cooler are similar (expanders) or the same (compressors) as those that have flown. Further development of these coolers to maximize cooling per input power for small cooling loads ($<100$\,mW at 4\,K) and lower mass is however needed. There is also a need to minimize the vibration from the cooler system. The miniature reverse Brayton cryocoolers in development by Creare are examples of reliable coolers with negligible exported vibration. These coolers are at TRL 6 for 80\,K and TRL 4 for 10\,K operation.

\begin{figure}
\includegraphics[width=8cm,angle=0]{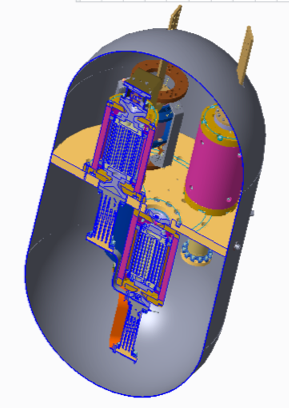}
\caption[The Continuous Adiabatic Demagnetization Refrigerator (CADR) under development at GSFC]{The Continuous Adiabatic Demagnetization Refrigerator (CADR) under development at NASA GSFC. This will provide $6\,\mu$W of cooling at 50\,mK. It also has a precooling stage that can be operated from 0.3 to 1.5\,K. The picture also shows a notional enclosing magnetic shield for a $<1\mu$\,T fringing field.} 
\label{fig:cryoe} 
\end{figure}

For cooling to below 0.1\,K, adiabatic demagnetization refrigerators (ADRs) are currently the only proven technology, although work has been funded by ESA to develop a continuously recirculating dilution refrigerator (CADR). A single shot DR was flown on {\itshape Planck} producing $0.1\,\mu$W of cooling at $100$\,mK for about 1.5 years, while a three-stage ADR was used on {\itshape Hitomi} producing $0.4\,\mu$W of cooling at 50\,mK with an indefinite lifetime. In contrast, a TRL 4 CADR has demonstrated $6\,\mu$W  of cooling at $50$\,mK with no life-limiting parts\citep{shirron01} (Figure \ref{fig:cryoe}). This technology is being advanced toward TRL 6 by 2020 via funding from the NASA SAT/TPCOS program\citep{tuttle17}. Demonstration of a 10\,K upper stage for this machine, as is planned, would enable coupling to a higher temperature cryocooler, such as that of Creare, that has near-zero vibration. The flight control electronics for this ADR are based on the flight-proven {\itshape Hitomi} ADR control, and has already achieved TRL 6. ADR coolers are the current reference design for the {\itshape Athena} X-ray observatory. For the OST, all three of the above technologies are required to maintain the telescope near 4\,K and the detector arrays near 50\,mK.

Continued development of 0.1\,K and $4\,$k coolers with cooling powers of tens of mW, high reliability, and lifetimes of 10+ years is of great importance for future far-infrared observatories. Moreover, the development of smaller, lighter, vibration resistant, power efficient cryo-coolers enables expansion of infrared astronomy to new observing platforms. An extremely challenging goal would be the development of a 0.1\,K cooler with power, space, and vibration envelopes that enable its use inside a 6U CubeSat, while leaving adequate resources for detector arrays, optics, and downlink systems (see also \S\ref{ssect:cube}). More generally, the ubiquity of cooling in infrared astronomy means that development of low-mass, low-power, and low cost coolers will reduce mission costs and development time across all observational domains.

\subsection{High Surface Accuracy Lightweight Mirrors}\label{ssect:mirr}
As far-infrared observing platforms mature and develop, there emerge new opportunities to use large aperture mirrors for which the only limitations are (1) mirror mass, and (2) approaches to active control and correction of the mirror surface. This raises the possibility of a high altitude, long duration far-infrared observing platform with a mirror factors of 2-5 larger than on facilities such as SOFIA or {\itshape Herschel}. 

The key enabling technology for such an observing platform is the manufacture of lightweight, high surface accuracy mirrors, and their integration into observing platforms. This is especially relevant for ULDBs, which are well-suited to this activity. Lightweight mirrors with apertures of three meters to several tens of meters are ideal for observations from balloon-borne platforms. Carbon-fiber mirrors are an attractive option; they are low mass and can offer high sensitivity in the far-infrared, at low cost of manufacture. Apertures of 2.5\,m are used on projects such as BLAST-TNG\citep{galitzki2014next}. Apertures of up to $\sim$10-m are undergoing ground-based tests, including the phase 2 NIAC study for the Large Balloon Reflector \citep{walk14,less15,cort16}.

A conceptually related topic is the physical size and mass of optical components. The physical scale of high resolution spectrometers in the far-infrared is determined by the optical path difference required for the resolution. For resolutions of $R\gtrsim10^5$, this implies scales of several meters for a grating spectrometer. This scale can be reduced by folding, but mass remains a potentially limiting problem. Moreover, larger physical sizes are needed for optical components to accommodate future large format arrays, posing challenges for uniformity, thermal control, and antireflection coatings. The development of low-mass optical elements suitable for diffraction limited operation at $\lambda \geq 25\,\mu$m would open the range of technical solutions available for the highest performance instruments.

\subsection{Other Needs}\label{ssect:general}
There exist several further areas for which technology development would be beneficial. We briefly summarize them below:
\vspace{0.2cm}

\noindent {\bfseries Lower-loss THz optics:} lenses, polarizers, filters, and duplexers.
\vspace{0.1cm}

\noindent {\bfseries Digital backends:} Low-power (of order a few watts or less) digital backends with $>1000$ channels covering up to several tens of GHz of bandwidth.
\vspace{0.1cm}

\noindent {\bfseries Wide-field imaging fourier transform spectrometers:} Expanding on the capabilities of e.g. SPIRE on {\itshape Herschel}, balloon or space-based IFTS with FoVs of tens of square arcminutes\citep{mail13}. Examples include the concept H2EX\citep{boul09}.
\vspace{0.1cm}

\noindent {\bfseries Deployable optics:} Development of deployable optics schemes across a range of aperture sizes would be enabling for a range of platforms. Examples range from 20-50\,cm systems for CubeSats to 5-10\,m systems for JWST. 
\vspace{0.1cm}

\noindent {\bfseries Data downlinking and archiving:} The advent of infrared observatories with large-format detector arrays presents challenges in downlinking and archiving. Infrared observatories have, to date, not unduly stressed downlinking systems, but this could change in the future with multiple instruments each with $10^4 - 10^5$ pixels on a single observatory. Moreover, the increasing number and diversity of PI and facility-class infrared observatories poses challenges to data archiving, in particular for enabling investigators to efficiently use data from multiple observatories in a single study. One way to mitigate this challenge is increased use of on-board data processing and compression, as is already done for missions operating at shorter wavelengths. 
\vspace{0.1cm}

\noindent {\bfseries Commonality and community in instrument software:} Many tasks are similar across a single platform, and even between platforms (e.g., pointing algorithms, focus, data download). Continued adherence to software development best practices, code sharing via repositories via GitHub, and fully open-sourcing software, will continue to drive down associated operating costs, speed up development, and facilitate ease of access.

\section{Conclusions: The Instrument Development Landscape for Infrared Astronomy}\label{sect:conc}
The picture that coalesces from this review is that far-infrared astronomy is still an emerging field, even after over forty years of development. Optical and near-infrared astronomy has a mature and well-understood landscape in terms of technology development for different platforms. In contrast, far-infrared astronomy has more of the ``Wild West'' about it; there are several observing platforms that range widely in maturity, all with overlapping but complementary domains of excellence. Moreover, considering the state of technology, {\itshape all} areas have development paths where huge leaps forward in infrared observing capability can be obtained. In some cases, entirely new platforms can be made possible.

To conclude this review, we bring together and synthesize this information in order to lay out how the capabilities of each platform can be advanced. To do so, we use the following definitions:

\begin{itemize}

\item {\bfseries Enabling:} Enabling technologies satisfy a capability need for a platform, allowing that platform to perform science observations in a domain that was hitherto impossible with that platform.

\item {\bfseries Enhancing:} Enhancing technologies provide significant benefits to a platform over the current state of the art, in terms of e.g., observing efficiency or cost effectiveness, but do not allow that platform to undertake observations in new science domains. 

\end{itemize}

\noindent These definitions correspond closely to the definitions of Enabling (a pull technology) and Enhancing (a push technology) as used in the 2015 NASA Technology Roadmap.

Since different technology fields vary in relevance for different platforms, technologies can be enabling for some platforms and enhancing for others. In Table \ref{tab:summary} we assess the status of selected technology areas as enabling or enhancing, as a function of observing platform. This table is solely the view of the authors, and not obtained via a community consultation. 

With this caveat in mind, based on Table \ref{tab:summary}, we present a non-exhaustive list of important technology development areas for far-infrared astronomy:
\vspace{0.3cm}

\noindent {\bfseries Large format detectors:} Existing and near-future infrared observatories include facilities with large FoVs, or those designed to perform extremely high resolution spectroscopy. These facilities motivate the development of large-format arrays that can fill telescope FoVs, allowing for efficient mapping and high spatial resolutions. A reference goal is to increase the number of pixels in arrays to $10^5$ for direct detectors, and $10^2$ for heterodyne detectors. This is a small number compared with arrays for optical and near-infrared astronomy, for which millions of pixels can be fielded in a single chip, but is still $1-2$ orders of magnitude larger than any array currently used in the far-infrared.
\vspace{0.2cm}

\noindent {\bfseries Detector readout electronics:} Increases in detector array sizes are inevitably accompanied by increases in complexity and power required for the readout electronics, and power dissipation of the cold amplifiers for these arrays. At present, the power requirements for $\gtrsim10^4$ detector array readout systems are a key limitation for their use in any space-based or sub-orbital platform, restricting them to use in ground-based facilities. For these reasons, development of multiplexing schemes is a high priority for large-format arrays, irrespective of the technology used. 

The main driver for power dissipation is the bandwidth of the multiplexers. Low-power cryogenic amplifiers, in particular parametric amplifiers, can mitigate this problem at 4\,K. Application Specific Integrated Circuits (ASICs), which combine digitization, FFT, and tone extraction in a single chip, can greatly reduce the power required for the warm readout system. A reference goal for the use of $\gtrsim10^4$ pixel arrays on space-based observatories such as the OST is a total power dissipation in the readout system of below $2\,$kW. This requires a denser spacing of individual channels in frequency domain multiplexers. For balloon-based facilities, sub-kW power dissipation is desirable. 
\vspace{0.2cm}

\noindent {\bfseries Direct detector sensitivity \& dynamic range:} The performance of $4\,$K-cooled space-based and high-altitude sub-orbital telescopes will be limited by astrophysical backgrounds such as zodiacal light, galactic cirrus, and the microwave background, rather than telescope optics or the atmosphere. Increasing pixel sensitivity to take advantage of this performance is of paramount importance to realize the potential of future infrared observatories. A reference goal is large-format detector arrays with per-pixel NEP of $2\times 10^{-20}\,\mathrm{W}/\sqrt{\mathrm{Hz}}$. This sensitivity is enabling for all imaging and medium resolution spectroscopy applications. It meets the requirement for R$\sim$1000 spectroscopy for the OST, and exceeds the medium resolution spectroscopy requirement for SPICA by a factor of five. However, for high spectral resolutions ($R>10^5$, e.g. the proposed HRS on the OST), even greater sensitivities are required, of $\sim 10^{-21}\,\mathrm{W}/\sqrt{\mathrm{Hz}}$, and ideally photon-counting. 

Turning to dynamic range; the dynamic range of detector arrays for high-background applications, such as ground-based observatories, is sufficient. However, the situation is problematic for the low background of cold space-based observatories. This is particularly true of observatories with $\gtrsim5\,$m apertures, since the saturation powers of currently proposed high-resolution detector arrays are within $\sim2$ orders of magnitude of their NEPs. It would be advantageous to increase the dynamic range of detector arrays to five or more orders of magnitude of their NEPs, as this would mitigate the need to populate the focal plane with multiple detector arrays, each with different NEPs. 
\vspace{0.2cm}

\noindent {\bfseries Local Oscillators for heterodyne spectroscopy:} The extremely high spectral resolutions achievable by heterodyne spectroscopy at mid/far-infrared wavelengths are of great value, both for scientific investigations in their own right, and for complementarity with the moderate spectral resolutions of facilities like JWST. This motivates continued development of high quality Local Oscillator sources to increase the sensitivity and bandwidth of heterodyne receivers. An important development area is high spectral purity, narrow-line, phase-locked, high-power ($5-10$\,mW) Quantum Cascade Laser (QCL) LOs, since the QCL LOs operate effectively for the higher frequency ( $>3$\,THz) arrays. A complementary development area is power division schemes (e.g., Fourier phase gratings) to utilize QCLs effectively
\vspace{0.2cm}

\noindent {\bfseries High bandwith heterodyne mixers:} The current bandwidth of heterodyne receivers means that only very small spectral ranges can be observed at any one time, meaning that some classes of observation, such as multiple line scans of single objects, are often prohibitively inefficient. There is thus a need to increase the IF bandwidth of $1-5\,$THz heterodyne mixers. A reference goal is a minimum of $8\,$GHz bandwidth required at frequencies of $\sim3$\,THz. This will allow for simultaneous observation of multiple lines, improving both efficiency and calibration accuracy. A related development priority is low-noise $1-5\,$THz mixers that can operate at temperatures of $>20$\,K. At present, the most promising paths towards such mixers align with the HEB and SIS technologies. 
\vspace{0.2cm}

\noindent {\bfseries Interferometry:} Ground-based observations have conclusively demonstrated the extraordinary power of interferometry in the centimeter to sub-millimeter, with facilities such as the VLA and ALMA providing orders of magnitude increases in spatial resolution and sensitivity over any existing single-dish telescope. As Table \ref{tab:summary} illustrates, the technology needs for space-based far-infrared interferometry are relatively modest, and center on direct detector developments. For interferometry, high-speed readout is more important than a large pixel count or extremely low NEP. For example, SPIRIT requires $14\times14$ pixel arrays of detectors with a NEP of $\sim 10^{-19}\,\mathrm{W}/\sqrt{\mathrm{Hz}}$ and a detector time constant of $\sim185\,\mu$s \citep{benford2007cryogenic}. Detailed simulations, coupled with rigorous laboratory experimentation and algorithm development, are the greatest priorities for interferometry.
\vspace{0.2cm}

\noindent {\bfseries Cryocoolers:} Since cooling to $4\,$K and $0.1\,$K temperatures is required for all far-infrared observations, improvements in the efficiency, power requirements, size, and vibration of cryocoolers are valuable for all far-infrared space- and sub-orbital-based platforms. For $<0.1\,$K coolers, there is a need for further development of both CADRs and DRs that enable cooling of up to tens of $\mu$W at $<0.1\,$K, to enable cooling of larger arrays. For $4\,$K coolers, further development to maximize cooling power per input power for small cooling loads ($<100$\,mW at $4\,$K) and lower mass is desirable, along with minimizing the exported vibration from the cooler system. For $\sim30\,$K coolers, development of a cooling solution with power, space, and vibration envelopes that enable its use inside a 6U CubeSat, while leaving adequate resources for detector arrays, optics, and downlink systems, would enable far-infrared observations from CubeSat platforms, as well as enhancing larger observatories.
\vspace{0.2cm}

\noindent {\bfseries Deployable and/or Light-weight telescope mirrors:} The advent of long-duration high-altitude observing platforms, and the expanded capabilities of future launch vehicles, enable the consideration of mirrors for far-infrared observatories with diameters 2-5 times larger than on facilities such as SOFIA and {\itshape Herschel}. The most important limitations on mirror size are then (a) mass, and (b) approaches to active control of the mirror surface. The development of large-aperture, lightweight, high surface accuracy mirrors is thus an important consideration, including those in a deployable configuration. A related area is the development of optical components that accomodate large-format arrays, or very high resolution spectroscopy. 
\vspace{0.2cm}

\noindent {\bfseries Technology maturation platforms:} Sub-orbital far-infrared platforms including ground-based facilities, SOFIA, and balloon-borne observatories, continue to make profound advances in all areas of astrophysics. However, they also serve as a tiered set of platforms for technology maturation and raising TRL's. The continued use of all of these platforms for technology development is essential to realize the long-term ambitions of the far-infrared community for large, actively cooled, space-based infrared telescopes. A potentially valuable addition to this technology maturation tier is the International Space Station, which offers a long-term, stable orbital platform with abundant power. 
\vspace{0.2cm}

\noindent {\bfseries Software and data archiving:} In the post-{\itshape Herschel} era, SOFIA and other sub-orbital platforms will play a critical role in mining the information-rich far-infrared spectral range, and in keeping the community moving forward. For example, the instruments flying on SOFIA and currently under development did not exist when {\itshape Herschel} instrumentation was defined. During this time, and henceforth, there is an urgent need to ensure community best-practices in software design, code sharing, and open sourcing via community-wide mechanisms. It is also important to maintain and enhance data archiving schemes that effectively bridge multiple complex platforms in a transparent way, and which enable access to the broadest possible spectrum of the community.

\section{Acknowledgements}\label{sect:ack}
We thank George Nelson and Kenol Jules for help on the capabilities of the International Space Station, and Jochem Baselmans for insights into KIDs. We also thank all speakers who took part in the FIR SIG Webinar series. This report developed in part from the presentations and discussions at the Far-Infrared Next Generation Instrumentation Community Workshop, held in Pasadena, California in March 2017. It is written as part of the activities of the Far-Infrared Science Interest Group. This work was supported by CNES.

\begin{landscape}

\begin{table}[ht]
\begin{threeparttable}[b]
\caption{A summary of enabling and enhancing technologies for far-infrared observing platforms.} \label{tab:summary}
{\footnotesize    
 \begin{tabular}{@{}|crr|p{1.4cm}p{1.4cm}p{1.4cm}p{1.4cm}p{1.4cm}p{1.4cm}p{1.4cm}|p{1.4cm}p{1.4cm}p{1.4cm}p{1.4cm}|@{}}
 \hline
  \hline
 \multicolumn{3}{|l|}{}                                                           & \multicolumn{7}{c|}{\cellcolor{blue!30}{\bfseries SPACE BASED}}                                                                                                                                                                  & \multicolumn{4}{c|}{\cellcolor{blue!30}{\bfseries ATMOSPHERE BASED}} \\
 \hline             
 \multicolumn{3}{|l|}{}                                                           & \cellcolor{yellow!35}OST\tnote{(a)}    & \cellcolor{yellow!15}SPICA     & \cellcolor{yellow!35}Probe & \cellcolor{yellow!15}CubeSats & \cellcolor{yellow!35}ISS     & \cellcolor{yellow!15}Interfer- & \cellcolor{yellow!35}Sounding & \cellcolor{lime!25}SOFIA   & \multicolumn{2}{c}{\cellcolor{lime!15}Balloons\tnote{(b)}}                    & \cellcolor{lime!25}Ground   \\
 \multicolumn{3}{|l|}{}                                                           & \cellcolor{yellow!35}                  & \cellcolor{yellow!15}          & \cellcolor{yellow!35}Class & \cellcolor{yellow!15}         & \cellcolor{yellow!35}        & \cellcolor{yellow!15}ometry    & \cellcolor{yellow!35}Rockets  & \cellcolor{lime!25}        & \cellcolor{lime!15}ULDB              & \cellcolor{lime!15}LDB                 & \cellcolor{lime!25}Based    \\
 \hline                
 \hline                
 \multicolumn{3}{|l|}{\cellcolor{gray!50}{\bfseries Direct Detectors\tnote{(c)}}} & \cellcolor{gray!50}                    & \cellcolor{gray!50}            & \cellcolor{gray!50}        & \cellcolor{gray!50}           & \cellcolor{gray!50}          & \cellcolor{gray!50}            & \cellcolor{gray!50}         & \cellcolor{gray!50}        & \cellcolor{gray!50}                    & \cellcolor{gray!50}                    & \cellcolor{gray!50}         \\
 \multicolumn{3}{|r|}{\cellcolor{green!25}Array size ($10^{4+}$pix)}              & \cellcolor{Sienna2}                    & \cellcolor{Sienna2}            & \cellcolor{Sienna2}        & \cellcolor{gray!50}           & \cellcolor{Sienna2}          & \cellcolor{SteelBlue1}         & \cellcolor{gray!50}         & \cellcolor{Sienna2}        & \cellcolor{Sienna2}                    & \cellcolor{Sienna2}                    & \cellcolor{Sienna2}  \\
 \multicolumn{3}{|r|}{\cellcolor{green!15}Sensitivity}                            & \cellcolor{Sienna2}                    & \cellcolor{Sienna2}            & \cellcolor{Sienna2}        & \cellcolor{Sienna2}           & \cellcolor{gray!50}          & \cellcolor{SteelBlue1}         & \cellcolor{Sienna2}         & \cellcolor{Sienna2}        & \cellcolor{Sienna2}                    & \cellcolor{SteelBlue1}                 & \cellcolor{SteelBlue1}          \\
 \multicolumn{3}{|r|}{\cellcolor{green!25}Speed}                                  & \cellcolor{SteelBlue1}                 & \cellcolor{gray!50}            & \cellcolor{SteelBlue1}     & \cellcolor{gray!50}           & \cellcolor{SteelBlue1}       & \cellcolor{SteelBlue1}         & \cellcolor{SteelBlue1}      & \cellcolor{SteelBlue1}     & \cellcolor{SteelBlue1}                 & \cellcolor{gray!50}                    & \cellcolor{SteelBlue1}          \\
 \multicolumn{3}{|r|}{\cellcolor{green!15}Dynamic Range}                          & \cellcolor{SteelBlue1}                 & \cellcolor{SteelBlue1}         & \cellcolor{SteelBlue1}     & \cellcolor{SteelBlue1}        & \cellcolor{SteelBlue1}       & \cellcolor{gray!50}            & \cellcolor{gray!50}         & \cellcolor{gray!50}        & \cellcolor{gray!50}                    & \cellcolor{gray!50}                    & \cellcolor{gray!50}         \\
 \multicolumn{3}{|r|}{\cellcolor{green!25}Readout: $10^{4}$pix}                   & \cellcolor{Sienna2}                    & \cellcolor{Sienna2}            & \cellcolor{Sienna2}        & \cellcolor{SteelBlue1}        & \cellcolor{Sienna2}          & \cellcolor{Sienna2}            & \cellcolor{gray!50}         & \cellcolor{Sienna2}        & \cellcolor{Sienna2}                    & \cellcolor{Sienna2}                    & \cellcolor{SteelBlue1}          \\
 \multicolumn{3}{|r|}{\cellcolor{green!15}Readout: $10^{5}$pix}                   & \cellcolor{Sienna2}                    & \cellcolor{gray!50}            & \cellcolor{Sienna2}        & \cellcolor{gray!50}           & \cellcolor{Sienna2}          & \cellcolor{gray!50}            & \cellcolor{gray!50}         & \cellcolor{Sienna2}        & \cellcolor{Sienna2}                    & \cellcolor{Sienna2}                    & \cellcolor{SteelBlue1}          \\
 \hline                                               
 \multicolumn{3}{|l|}{\cellcolor{gray!50}{\bfseries Heterodyne Detectors}\tnote{(d)}} & \cellcolor{gray!50}                    & \cellcolor{gray!50}            & \cellcolor{gray!50}        & \cellcolor{gray!50}           & \cellcolor{gray!50}          & \cellcolor{gray!50}            & \cellcolor{gray!50}         & \cellcolor{gray!50}        & \cellcolor{gray!50}                    & \cellcolor{gray!50}                    & \cellcolor{gray!50}         \\
 \multicolumn{3}{|r|}{\cellcolor{green!25}Array size ($10^{2+}$pix)}              & \cellcolor{SteelBlue1}                 & \cellcolor{gray!50}            & \cellcolor{SteelBlue1}     & \cellcolor{gray!50}           & \cellcolor{SteelBlue1}       & \cellcolor{gray!50}            & \cellcolor{gray!50}         & \cellcolor{Sienna2}        & \cellcolor{Sienna2}                    & \cellcolor{Sienna2}                    & \cellcolor{Sienna2}  \\
 \multicolumn{3}{|r|}{\cellcolor{green!15}LO bandwidth\tnote{(e)}}                & \cellcolor{SteelBlue1}                 & \cellcolor{gray!50}            & \cellcolor{SteelBlue1}     & \cellcolor{SteelBlue1}        & \cellcolor{SteelBlue1}       & \cellcolor{gray!50}            & \cellcolor{gray!50}         & \cellcolor{Sienna2}        & \cellcolor{Sienna2}                    & \cellcolor{Sienna2}                    & \cellcolor{Sienna2}  \\
 \multicolumn{3}{|r|}{\cellcolor{green!25}LO mass}                                & \cellcolor{gray!50}                    & \cellcolor{gray!50}            & \cellcolor{SteelBlue1}     & \cellcolor{Sienna2}           & \cellcolor{gray!50}          & \cellcolor{gray!50}            & \cellcolor{gray!50}         & \cellcolor{gray!50}        & \cellcolor{Sienna2}                    & \cellcolor{SteelBlue1}                 & \cellcolor{gray!50}         \\
 \multicolumn{3}{|r|}{\cellcolor{green!15}LO power draw}                          & \cellcolor{SteelBlue1}                 & \cellcolor{gray!50}            & \cellcolor{SteelBlue1}     & \cellcolor{Sienna2}           & \cellcolor{gray!50}          & \cellcolor{gray!50}            & \cellcolor{gray!50}         & \cellcolor{gray!50}        & \cellcolor{SteelBlue1}                 & \cellcolor{SteelBlue1}                 & \cellcolor{gray!50}         \\
 \multicolumn{3}{|r|}{\cellcolor{green!25}Mixer bandwidth}                        & \cellcolor{SteelBlue1}                 & \cellcolor{gray!50}            & \cellcolor{Sienna2}        & \cellcolor{SteelBlue1}        & \cellcolor{SteelBlue1}       & \cellcolor{gray!50}            & \cellcolor{gray!50}         & \cellcolor{Sienna2}        & \cellcolor{Sienna2}                    & \cellcolor{Sienna2}                    & \cellcolor{Sienna2}  \\
 \multicolumn{3}{|r|}{\cellcolor{green!15}Mixer sensitivity}                      & \cellcolor{Sienna2}                    & \cellcolor{gray!50}            & \cellcolor{Sienna2}        & \cellcolor{Sienna2}           & \cellcolor{SteelBlue1}       & \cellcolor{gray!50}            & \cellcolor{gray!50}         & \cellcolor{SteelBlue1}     & \cellcolor{SteelBlue1}                 & \cellcolor{SteelBlue1}                 & \cellcolor{gray!50}         \\
 \hline                                                            
 \multicolumn{3}{|l|}{\cellcolor{gray!50}{\bfseries Cryocoolers\tnote{(f)}}}      & \cellcolor{gray!50}                    & \cellcolor{gray!50}            & \cellcolor{gray!50}        & \cellcolor{gray!50}           & \cellcolor{gray!50}          & \cellcolor{gray!50}            & \cellcolor{gray!50}         & \cellcolor{gray!50}        & \cellcolor{gray!50}                    & \cellcolor{gray!50}                    & \cellcolor{gray!50}         \\
 \multicolumn{3}{|r|}{\cellcolor{green!25}Low-Power}                              & \cellcolor{SteelBlue1}                 & \cellcolor{gray!50}            & \cellcolor{SteelBlue1}     & \cellcolor{Sienna2}           & \cellcolor{gray!50}          & \cellcolor{gray!50}            & \cellcolor{SteelBlue1}      & \cellcolor{gray!50}        & \cellcolor{Sienna2}                    & \cellcolor{SteelBlue1}                 & \cellcolor{gray!50}         \\
 \multicolumn{3}{|r|}{\cellcolor{green!15}Low-Mass}                               & \cellcolor{SteelBlue1}                 & \cellcolor{gray!50}            & \cellcolor{SteelBlue1}     & \cellcolor{Sienna2}           & \cellcolor{gray!50}          & \cellcolor{gray!50}            & \cellcolor{SteelBlue1}      & \cellcolor{gray!50}        & \cellcolor{Sienna2}                    & \cellcolor{SteelBlue1}                 & \cellcolor{gray!50}         \\
 \hline                                                                                      
 \multicolumn{3}{|l|}{\cellcolor{gray!50}{\bfseries Mirrors/optics}}              & \cellcolor{gray!50}                    & \cellcolor{gray!50}            & \cellcolor{gray!50}        & \cellcolor{gray!50}           & \cellcolor{gray!50}          & \cellcolor{gray!50}            & \cellcolor{gray!50}         & \cellcolor{gray!50}        & \cellcolor{gray!50}                    & \cellcolor{gray!50}                    & \cellcolor{gray!50}         \\
 \multicolumn{3}{|r|}{\cellcolor{green!25}Low areal density}                      & \cellcolor{SteelBlue1}                 & \cellcolor{SteelBlue1}         & \cellcolor{gray!50}        & \cellcolor{SteelBlue1}        & \cellcolor{gray!50}          & \cellcolor{gray!50}            & \cellcolor{SteelBlue1}      & \cellcolor{gray!50}        & \cellcolor{Sienna2}                    & \cellcolor{SteelBlue1}                 & \cellcolor{gray!50}         \\
 \multicolumn{3}{|r|}{\cellcolor{green!15}Large aperture}                         & \cellcolor{Sienna2}                    & \cellcolor{gray!50}            & \cellcolor{gray!50}        & \cellcolor{gray!50}           & \cellcolor{SteelBlue1}       & \cellcolor{gray!50}            & \cellcolor{gray!50}         & \cellcolor{gray!50}        & \cellcolor{Sienna2}                    & \cellcolor{SteelBlue1}                 & \cellcolor{gray!50}         \\
 \multicolumn{3}{|r|}{\cellcolor{green!25}Deployable}                             & \cellcolor{SteelBlue1}                 & \cellcolor{gray!50}            & \cellcolor{gray!50}        & \cellcolor{Sienna2}           & \cellcolor{SteelBlue1}       & \cellcolor{gray!50}            & \cellcolor{gray!50}         & \cellcolor{gray!50}        & \cellcolor{gray!50}                    & \cellcolor{gray!50}                    & \cellcolor{gray!50}         \\
 \hline                                                                                                                                                                                
 \multicolumn{3}{|l|}{\cellcolor{gray!50}{\bfseries Other}}                       & \cellcolor{gray!50}                    & \cellcolor{gray!50}            & \cellcolor{gray!50}        & \cellcolor{gray!50}           & \cellcolor{gray!50}          & \cellcolor{gray!50}            & \cellcolor{gray!50}         & \cellcolor{gray!50}        & \cellcolor{gray!50}                    & \cellcolor{gray!50}                    & \cellcolor{gray!50}         \\
 \multicolumn{3}{|r|}{\cellcolor{green!25}Backend electronics}                    & \cellcolor{SteelBlue1}                 & \cellcolor{SteelBlue1}         & \cellcolor{SteelBlue1}     & \cellcolor{Sienna2}           & \cellcolor{gray!50}          & \cellcolor{Sienna2}            & \cellcolor{SteelBlue1}      & \cellcolor{SteelBlue1}     & \cellcolor{SteelBlue1}                 & \cellcolor{SteelBlue1}                 & \cellcolor{SteelBlue1}          \\
 \multicolumn{3}{|r|}{\cellcolor{green!15}Downlink systems}                       & \cellcolor{SteelBlue1}                 & \cellcolor{gray!50}            & \cellcolor{gray!50}        & \cellcolor{SteelBlue1}        & \cellcolor{gray!50}          & \cellcolor{gray!50}            & \cellcolor{gray!50}         & \cellcolor{gray!50}        & \cellcolor{gray!50}                    & \cellcolor{gray!50}                    & \cellcolor{gray!50}         \\
 \hline
  \hline
 \multicolumn{5}{r}{}      & \multicolumn{2}{c}{\cellcolor{Sienna2}\textcolor{white}{{\bfseries Enabling}}}& \multicolumn{2}{c}{\cellcolor{SteelBlue1}\textcolor{white}{{\bfseries Enhancing}}}&      &     &    &  \\
 \hline
  \hline
 \end{tabular}
 }
 \begin{tablenotes}
 {\footnotesize 
 \item[(a)] For the OST (\S\ref{origins}), the table refers to ``concept 1'', the more ambitious of the concepts investigated, with greater dependence on technology development. 
 \item[(b)] For balloons (\S\ref{ssect:uldb}) ULD balloons have flight times of 100+ days and carry payloads up to $\sim1800\,$kg. The $<50$ day LD balloons can carry up to $\sim2700\,$kg.
 \item[(c)] Fiducial targets for direct detectors (\S\ref{directdetect}) used for space-based imaging are a NEP of $1\times 10^{-19}\,\mathrm{W}/\sqrt{\mathrm{Hz}}$ and a readout system with $<3\,$kW power dissipation. They should also be compatible with an observatory cryogenic system.
 \item[(d)] For heterodyne instruments (\S\ref{ssect:het}): none are planned for SPICA (\S\ref{sec:spica}). For interferometers (\S\ref{ssect:firint}); all those proposed by the US community are direct detection; heterodyne interferometer needs have however been studied in Europe.
 \item[(e)] The assumed operating frequency range is 1-5\,THz. 
 \item[(f)] For cryocoolers (\S\ref{ssect:cryoc}) we do not distinguish between 4\,K and 0.1\,K coolers, since the choice is detector dependent. 
 }
\end{tablenotes}
 \end{threeparttable}

\end{table} 

\end{landscape}

\end{document}